\documentclass[preprint,12pt]{elsarticle}

\usepackage{amsmath}
\usepackage{amssymb}
\usepackage{graphicx}% Include figure files
\usepackage{dcolumn}% Align table columns on decimal point
\usepackage{bm}% bold math
\begin{document}

\title{Emergence of Unique Steady Edge States in Trapped Ultracold Atom Systems}% Force line breaks with \\

\author[ust,rcnas,gs]{Roland Cristopher F. Caballar}
 %\altaffiliation[Also at ]{Department of Physics and Mathematics, College of Science, University of Santo Tomas}%Lines break automatically or can be forced with \\
\ead{rfcaballar@ust.edu.ph}
\affiliation[ust]{organization={Physics of Life and Stuff Group, Department of Mathematics and Physics, College of Science, University of Santo Tomas},
             addressline={Espana Boulevard, Sampaloc},
             city={Manila},
             postcode={1008},
             state={},
             country={Philippines}}
\affiliation[rcnas]{organization={Mathematics and Theoretical Physics Cluster, Research Center for the Natural and Applied Sciences, University of Santo Tomas},
             addressline={Espana Boulevard, Sampaloc},
             city={Manila},
             postcode={1008},
             state={},
             country={Philippines}}
\affiliation[gs]{organization={M. Sc. Applied Physics Major in Medical Physics Program, The Graduate School, University of Santo Tomas},
             addressline={Espana Boulevard, Sampaloc},
             city={Manila},
             postcode={1008},
             state={},
             country={Philippines}}
%\affiliation{%
% Department of Physics and Mathematics, College of Science, University of Santo Tomas
%
%\affiliation{Research Center for the Natural and Applied Sciences, University of Santo Tomas, Espana Blvd., Sampaloc, Manila, Philippines}


%\affiliation{
%\affiliation{
 %Third institution, the second for Charlie Author
%\affiliation{%


\date{\today}% It is always \today, today,

\begin{abstract}
We show that, for a one - dimensional open quantum system of ultracold atoms trapped in an array of harmonic potentials that is weakly coupled to a background Bose - Einstein Condensate (BEC), a unique steady state emerges at either of the two edges of the array due to the combined effects of excitation via lasers of these ultracold atoms and decay back to their initial energy levels via emission of excitations into the BEC, acting as an excitation reservoir. We then demonstrate the existence of steady states of the master equation that describes the dynamics of this open quantum system with respect to the expectation value of the position of the harmonic potential containing these states, and show that these steady states occur at the edges of the array of harmonic potentials trapping these atoms. Using the open quantum system's master equation to evolve initial states of the system that are trapped in other harmonic potentials in the array numerically over time, we verify the analytic calculations demonstrating the existence of these steady states at the edge of the system, and further demonstrate that these steady states will emerge regardless of the number of atoms trapped in each of the harmonic potentials in the array. We note that the emergence of these edge states irrespective of the initial state of the system is similar to the non-Hermitian skin effect seen in other open quantum systems, and are of fundamental importance in quantum technological applications such as quantum transport and quantum batteries.
%\begin{description}
%\end{description}
\end{abstract}

\maketitle

\section{Introduction}
Open quantum systems have proven themselves to be central to the second quantum revolution (\cite{BreuerPetruccione}). In particular, open quantum systems have made it possible to formulate dissipative quantum computing algorithms which make use of the environment as a resource in its implementation (\cite{kraus, verstraete, kliesch, kastoryano2, reiter, shtanko, zapusek, eder}), as well as dissipative quantum state preparation mechanisms (\cite{daley, diehl, kastoryano, watanabe, caballar, seetharam, harrington, ghasemian, stefanini, lin}) and dissipative quantum transport mechanisms (\cite{landi, han, damanet, gou, bandyopadhyay, aksenov}). Open quantum systems have also been used to create various types of quantum technologies such as quantum sensors (\cite{reiter, montenegro, degen}), quantum heat engines and refrigerators (\cite{DeffnerCampbell, chen, zhang, villegas, villegas2, albay}), and dissipative quantum time crystals (\cite{iemini, gong, gambetta, buca, kessler}).

A particular type of open quantum system that is currently gaining interest in quantum technology, and which has been used in some of the applications of open quantum systems mentioned in the previous paragraph, is driven open quantum matter (\cite{sieberer}), which consists of non-equilibrium quantum systems that undergo a combination of coherent, driven and dissipative dynamics as they evolve over time. This combination of driven and dissipative dynamics pushes driven open quantum matter out of equilibrium, but this system will eventually evolve towards a unique stationary state in thermodynamic equilibrium (\cite{daley, diehl, watanabe, caballar, villegas3}). It is this ability of driven open quantum matter to attain a unique steady state despite evolving out of its equilibrium state due to the combination of driven and dissipative dynamics that permits dissipation to be harnessed as a resource in quantum computing, quantum information and quantum technological applications (\cite{diehl, kastoryano, kastoryano2}), specifically for the dissipative preparation of quantum many-body states that are of interest in these applications (\cite{diehl, kastoryano, kastoryano2, caballar, sweke1, sweke2, ghasemian, wang, caballar2}). It is of interest to note that one may use either Markovian (\cite{BreuerPetruccione, li, sweke3}) or non-Markovian methods (\cite{BreuerPetruccione, breuer, semin, semina, sweke4, devega, butanas}) to analyze the dynamics of driven open quantum matter, with each set of methods suitable for specific conditions under which the quantum matter experiences both driven and dissipative dynamics, such as the timescale over which the correlations in the bath coupled to the system decay.

Another reason why driven open quantum matter, and dissipative quantum systems in general, are currently systems of interest is due to the emergence of robust edge states in non-Hermitian quantum systems (\cite{lee, gong2, yao, alvarez, song, haga, gohsrich, hu}). The emergence of these edge states, also known as the non-Hermitian skin effect, is characterized by the localization of a majority of the eigenstates of a non-Hermitian operator at the boundaries of the system as it evolves over time. This effect is especially significant for open quantum systems, whose time evolution is governed by the master equation which has both Hermitian (the Hamiltonian operator denoting the coherent dynamics of the system) and non - Hermitian (the Lindblad operator denoting the dissipative dynamics of the system) components. At the same time, emergence of these edge states is a hallmark of the bulk-boundary correspondence principle, which states that the existence of these edge states are dictated by topological invariants in the bulk of the material. Considering that the Lindblad operator of an open quantum system is non-Hermitian, the topological invariants of an open quantum system are then characterized by the spectrum of the system's non-Hermitian master equation, with these topological invariants indicating the existence of robust, unique edge states accumulating at the boundaries of the open quantum system as it evolves over time. Examples of these topological invariants are the winding numbers of the energy eigenvalues of non-Hermitian Hamiltonians (\cite{zhang2}).

In this paper, we investigate the emergence of such unique edge states as steady states of a particular many-body open quantum system, namely a gas of ultracold atoms trapped in an array of harmonic potentials which undergo a combination of driven and dissipative dynamics, in the form of successive excitations to higher energy levels in harmonic traps away from their initial trapping location and dissipative emission of Bogoliubov excitations by these excited atoms into a Bose - Einstein Condensate (BEC), which is weakly coupled to this trapped ultracold atom system and acts as a reservoir of these excitations, to enable their return to their original energy levels but at a trapping location which may be different from their original one. We show, using numerical methods and after deriving the interaction Hamiltonian for this open quantum system and the master equation describing its driven-dissipative dynamics, that unique steady states localized at the edges of this system will emerge as the system evolves over time, with these steady edge states characterized by their particle number. The emergence of these unique steady edge states indicates that the system can be used for various quantum technologies which involve the use of these edge states, such as efficient quantum transport of states and quantum energy storage devices requiring ready access to states located at the edge of the system. 

\section{General Description of the Trapped Ultracold Open Quantum System}

The open quantum system to be considered in this paper is a gas of $N$ ultracold atoms trapped in an array of $2L+1, L=1,2,3,...$ one-dimensional harmonic potentials, and immersed in a background BEC which serves as a reservoir of excitations emitted by the trapped ultracold atoms. The open quantum system is similar to one proposed in Refs. \cite{daley,diehl,caballar,caballar2} while the trapping potential is a generalization of one proposed in Ref. \cite{caballar}. The figure below is a schematic diagram showing three of the harmonic traps in the array, and how the trapped ultracold atom system evolves over time by a combination of driven and dissipative dynamics. 
\begin{figure}[htb]
\includegraphics[width=1.0\columnwidth, height=0.3\textheight]{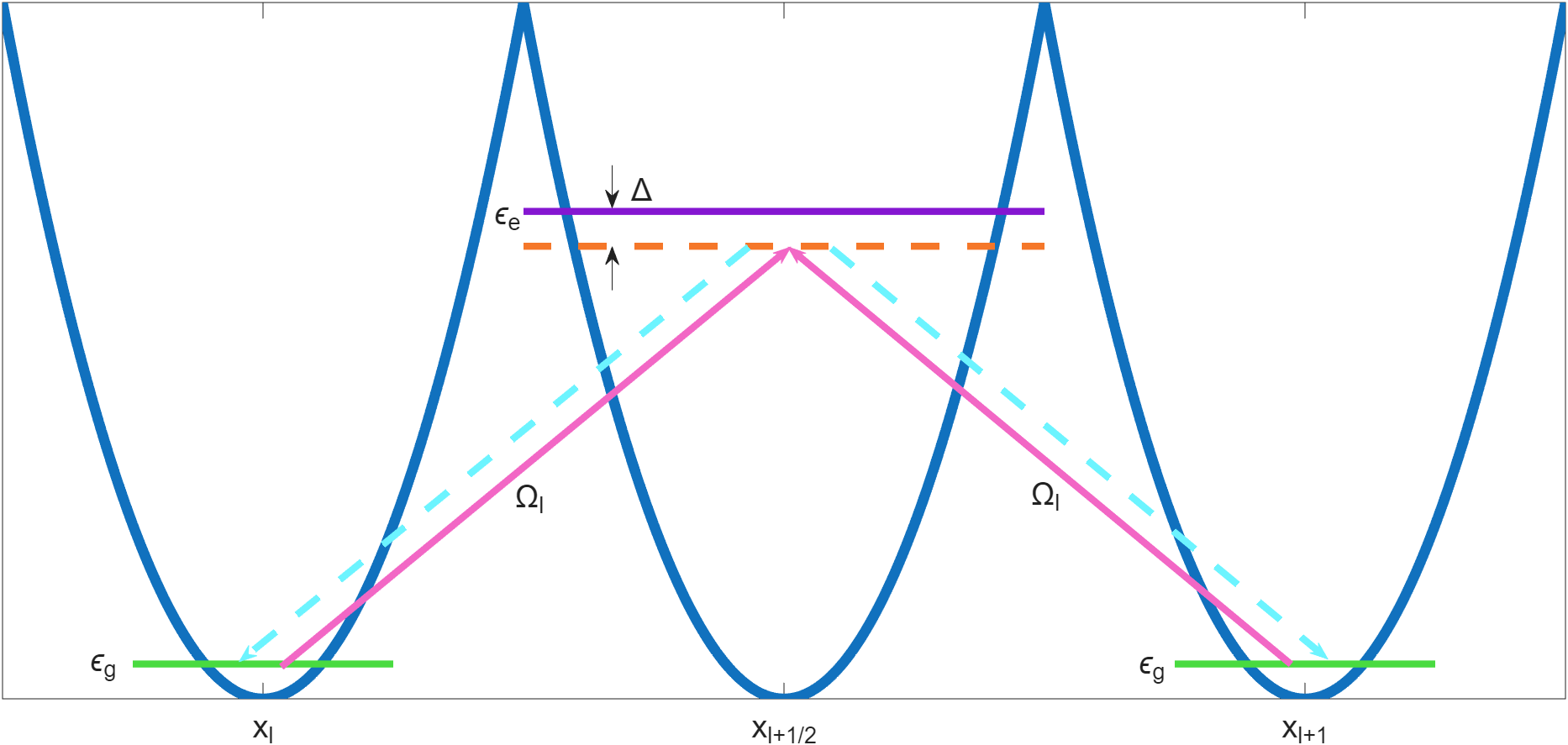}
\caption{\label{fig:harmonictraparrayschem} Schematic diagram of a portion of the harmonic oscillator array used to trap the ultracold atoms in this open quantum system. The atoms are initially in the ground state energy level $\epsilon_g$ of the harmonic oscillators centered at $x_l$ and $x_{l+1}$ (green lines), and are excited by Rabi lasers (pink lines) with Rabi frequencies $\Omega_l$ to the excited energy level (dashed orange line) located below the higher energy level $\epsilon_e$ of the harmonic potential centered at $x_{l+1/2}$ (solid purple line). Emission of an excitation with energy $E_\mathbf{k}$ (dashed sky blue line) into the background BEC will cause the excited atoms to return to the ground state energy level in either of the two neighboring harmonic oscillators centered at $x_l$ and $x_{l+1}$.}
\end{figure}

As shown in the figure, the ultracold atoms are initially trapped in the harmonic potentials located at $x_j = (2j-L-2)x_a, j=1,2,...,L+1, x_a\in\Re,$ in the array, at the ground state energy levels $\epsilon_g$ of these harmonic traps. Each of these traps have the same ground state energy levels, and  contain $n_j$ atoms, with the initial number of atoms in each trap satisfying the following condition:

\begin{equation}
\sum_{j=1}^{L+1}n_j = N
\label{initpartnumcond}
\end{equation}
The trapped ultracold atoms are then excited, using Raman lasers with Rabi frequencies $\Omega_j$, to an energy level close to one of the higher energy levels $\epsilon_e$ in one of the neighboring harmonic potentials. We note that while the laser frequencies $\Omega_j$ will be different from each other, each of the energy levels to which the trapped atoms are excited are close to, but not equal to, the same excited energy level $\epsilon_e$ in the neighboring harmonic potentials to which these atoms are excited. At the same time, the location of the harmonic trap to which they will be excited to will vary based on the trapped atoms' initial locations. In particular, atoms initially trapped in the harmonic potential located at $x_l = (2l-L-2)x_a$ at the energy level $\epsilon_g$ will be excited by Raman lasers to an energy level close to $\epsilon_e$ at the harmonic potential located either at $x_{l-1/2} = (2l-L-3)x_a$ or $x_{l+1/2} = (2l-L-1)x_a$, both of which are located beside the harmonic potential located at $x_l$. Now the reason why the energy level to which the atoms will be excited is close to but not equal to the higher energy level $\epsilon_e$ of the neighboring harmonic potentials is because the Rabi frequency of the laser exciting these atoms are not equal to the frequency $\omega_e = \epsilon_e / \hbar$ corresponding to this energy level of the neighboring harmonic potential. In particular, the Rabi lasers used to excite the trapped atoms from their initial location and energy level to their subsequent location and energy level is detuned from the frequency corresponding to the energy level $\epsilon_e$ of the neighboring harmonic potentials by an amount $\Delta_j = \omega_e - \Omega_j$, with this detuning together with the Rabi frequency of the laser being significant in deriving the master equation governing the dynamics of these trapped ultracold atoms.

After excitation to the higher energy levels in the neighboring harmonic potentials located at $x_{j\pm 1/2}$, the excited ultracold atoms will then, due to the instability of the quantum state corresponding to the energy level to which they are excited, decay back to their initial energy level, which is the ground state energy level of the harmonic potential in the trap array $\epsilon_g$. In doing so, the trapped ultracold atoms will emit a Bogoliubov excitation into the background BEC in which the system is immersed. Now the coupling between the system and the background BEC is formulated in such a way that the entire system of trapped ultracold atoms are weakly interacting with the background BEC. As such, the BEC then serves as a reservoir for the excitations emitted by the excited ultracold atoms as they decay from their excited energy states back to their initial energy states, with the excitations having minimal effect on the BEC. However, since the ultracold atoms have been excited to one of two possible harmonic potentials adjacent to the harmonic potential in which they have initially been trapped, these excited atoms may decay into the ground state energy level in one of two possible harmonic potentials adjacent to the harmonic potential to which they have been excited. This signifies that a trapped ultracold atom originally at energy level $\epsilon_g$ at the harmonic trap located at $x_l = (2l-L-2)x_a$ and excited by a Raman laser to a higher energy level in the harmonic potential located at $x_{l+1/2} = (2l-L-1)x_a$ will then decay, by emission of a Bogoliubov excitation into the background BEC, back to its initial energy level $\epsilon_g$, but to either the harmonic potential located at $x_l = (2l-L-2)x_a$ where it was originally located, or the harmonic potential located at $x_{l+1} = (2l-L)x_a$, both of which are adjacent to the harmonic potential located at $x_{l+1/2}$. 

This combination of driven (excitation via Raman lasers with Rabi frequencies $\Omega_j$ of trapped ultracold atoms initially at energy level $\epsilon_g$ in harmonic potentials located at $x_j = (2j-L)x_a, 0\leq j \leq L, x_a\in\Re$ to a neighboring harmonic potential located at $x_{j\pm 1/2}$ at an energy level close to the harmonic potential's energy level $\epsilon_e > \epsilon_g$) and dissipative (decay of excited atoms excited to an energy level close to the energy level $\epsilon_e > \epsilon_g$ of the harmonic potential located at $x_{j\pm 1/2}$ via emission of a Bogoliubov excitation into the background BEC serving as an excitation reservoir to the ground state energy level $\epsilon_g$ of a harmonic potential located at $x_{j}$ or $x_{j\pm 1}$) will then result in an open quantum trapped ultracold atom system with a reservoir of Bogoliubov excitations that can facilitate transport of trapped ultracold atoms from one harmonic potential in the array to another.

\section{Derivation of the Interaction Hamiltonian of the Trapped Ultracold System and the BEC Excitation Reservoir} 

In order to describe the dynamics of this open quantum system, we first need to derive the interaction Hamiltonian $\hat{H}_{SB}$ between the trapped ultracold atoms and the background BEC that form this open quantum system. In terms of the field operators of the trapped ultracold atoms and the background BEC, this interaction Hamiltonian has the following form:

\begin{equation}
\hat{H}_{SB}=\frac{2\pi a_{SB}}{\mu}\int dx\hat{\psi}^{\dagger}_{S}\hat{\psi}_{S}\hat{\psi}^{\dagger}_{B}\hat{\psi}_{B}
\label{inthamgenform}
\end{equation}
In this expression, $a_{SB}$ is the scattering length between the atoms in the trapped ultracold atom gas and the BEC, $\mu$ is the reduced mass of the trapped ultracold atoms and the atoms of the BEC, and $\hat{\psi}_S$ and $\hat{\psi}_B$ are the field operators for the system and the reservoir respectively. For $\hat{\psi}_{S}$, its explicit form is given by
\begin{equation}
\hat{\psi}_{S}=\sum_{j=1}^{L+1}\phi_{g,j}(x)\hat{a}_{g,j}+\sum_{j=1}^{L}\phi_{e,j+1/2}(x)\hat{a}_{e,j+1/2}
\label{fieldopsys}
\end{equation}
Here, $\hat{a}_{g,j}$ and $\hat{a}_{e,j+1/2}$ are the many-body particle annihilation operators for the ground state of the harmonic oscillator potential centered at $x_j = 2j-L$ and the excited state of the harmonic oscillator potential centered at $x_{j+1/2} = 2j+1-L$, respectively, and $\phi_{g,j}(x)$ and $\phi_{e,j+1/2}(x)$ are the ground state and excited state wavefunctions of the harmonic oscillator potentials centered at $x_j = 2j-L$ and $x_{j+1/2} = 2j+1-L$, respectively, with the explicit forms of these wavefunctions given by

\begin{eqnarray}
&&\phi_{g,j}(x)=\left(\frac{m_s \omega_g}{\pi\hbar}\right)^{1/4}\exp\left(-\frac{m_s \omega_g}{2\hbar}(x-x_j)^2\right)\underset{\sigma_g \rightarrow 0}\longrightarrow\sqrt{2}\pi^{1/4}\sigma_{g}^{1/2}\delta(x-x_j)\nonumber\\
&&\phi_{e,j+1/2}(x)=\left(\frac{m_s \omega_e}{\pi\hbar}\right)^{1/4}\sqrt{\frac{2m_s \omega_e}{\hbar}}(x-x_{j+1/2})\exp\left(-\frac{m_s \omega_e}{2\hbar}(x-x_{j+1/2})^2\right)\nonumber\\
&&=\sqrt{2}\pi^{-1/4}\sigma_{e}^{-3/4}(x-x_{j+1/2})\exp\left(-\frac{(x-x_{j+1/2})^2}{2\sigma_e^2}\right)\nonumber\\
\label{syswavfcns}
\end{eqnarray}

On the other hand, for $\hat{\psi}_{B}$, its explicit form is given by
\begin{equation}
\hat{\psi}_{B}(x)=\sqrt{\rho_B}+\delta\hat{\psi}_B (x)
\label{fieldopres}
\end{equation}
Breaking down the terms in this field operator, we find that $\rho_B$ is the condensate density, and
\begin{equation}
\delta\hat{\psi}_B = \frac{1}{\sqrt{L}}\sum_{k}\left(u_k e^{ikx}\hat{b}_k +v_{k}e^{-ikx}\hat{b}^{\dagger}_k \right),
\label{becexcifieldop}
\end{equation}
is the term corresponding to excitations in the condensate. In this excitation term, $L$ is the length of the BEC, $u_k =(1-L^{2}_k )^{-1/2}$, $v_k =L_k (1-L^{2}_k )^{-1/2}$, $L_k =(E_k -(k^2/2m_{B})-m_{B}c^2)/m_{B}c^2$, and $E_k$ is the energy of the excitations emitted into the condensate given by

\begin{equation}
E_k = ck\sqrt{1+\left(\frac{k}{2m_B c}\right)^2}
\end{equation}
with $c=\sqrt{g_{BB}\rho_B /m_B}$ being the velocity of sound in the condensate and $g_{BB}$ the interaction strength among the condensate atoms. Here, we make use of the assumption that the excitatons emitted into and from the BEC are sound - like, so that $E_k\approx ck$. However, this approximation implies that $k^2 / 4(m_B c)^2 <<1$. Now these simplifications will enable us to determine the explicit form of $\hat{\psi}_{B}^{\dagger}\hat{\psi}_B$:

\begin{eqnarray}
&&\hat{\psi}^{\dagger}_{B}\hat{\psi}_{B}=\left(\sqrt{\rho_B}+\delta\hat{\psi}^{\dagger}_{B}\right)\left(\sqrt{\rho_B}+\delta\hat{\psi}_{B}\right)\nonumber\\
&&\approx\rho_{B}+\sqrt{\rho_{B}}\left(\delta\hat{\psi}^{\dagger}_{B}+\delta\hat{\psi}_{B}\right)\nonumber\\
&&=\rho_{B}+\sqrt{\frac{\rho_{B}}{L_k}}\sum_{k}(u_k +v_k)(e^{ikx}\hat{b}_{k}+e^{-ikx}\hat{b}^{\dagger}_{k})
\end{eqnarray}

In determining this product, we neglect second order terms in $\hat{b}_k$ and $\hat{b}_{k}^{\dagger}$, on the basis that the trace of these second order terms will vanish, based on our earlier approximation that second - order terms in $k$ are negligible. Thus, using the earlier approximations made, we obtain the following terms in $\hat{\psi}^{\dagger}_{B}\hat{\psi}_{B}$: 
\begin{eqnarray}
&&L_k =\frac{1}{m_B c^2}\left(E_k -\frac{k^2}{2m_{B}}-m_{B}c^2\right)\approx \frac{k}{m_B c}-1\nonumber\\
&&u_k+v_k=\frac{1+L_{k}}{\sqrt{1-L^2_k}}=\sqrt{\frac{1+L_{k}}{1-L_{k}}}\approx\sqrt{\frac{k/m_B c}{2-k/m_B c}}\nonumber\\
&&=\sqrt{\frac{k}{2m_B c}}\left(1-\frac{k}{2m_B c}\right)^{-1/2}\approx\sqrt{\frac{k}{2m_B c}}\nonumber\\
\end{eqnarray}
These terms, in turn, will lead to the following form of $\hat{\psi}^{\dagger}_{B}\hat{\psi}_B$:

\begin{eqnarray}
&&\hat{\psi}^{\dagger}_B\hat{\psi}_B = \left(\sqrt{\rho}_B+\frac{1}{\sqrt{L}}\sum_{k}(u_k e^{-ikx}\hat{b}^\dagger_k +v_k e^{ikx}\hat{b}_k )\right)\left(\sqrt{\rho}_B+\frac{1}{\sqrt{L}}\sum_{k}(u_k e^{ikx}\hat{b}_k + v_k e^{-ikx}\hat{b}^\dagger_k )\right)\nonumber\\
&&\approx\sqrt{\frac{\rho_B}{L}}\sum_{k}(u_k +v_k )(e^{ikx}\hat{b}_k+e^{-ikx}\hat{b}^\dagger_k )=\sqrt{\frac{\rho_B}{2m_B cV}}\sum_{k}\sqrt{k}(e^{ikx}\hat{b}_k +e^{-ikx}\hat{b}^\dagger_k )\nonumber\\
\label{becfieldopprod}
\end{eqnarray}

Substituting Eqs. (\ref{becfieldopprod}) and (\ref{fieldopsys}) into Eq. (\ref{inthamgenform}), and carrying out the integration, we find that

\begin{eqnarray}
&&\hat{H}_{SB}=\frac{2\pi a_{SB}}{\mu}\sqrt{\frac{\rho_B}{2m_B cL}}\int dx \left(\sum_{j=1}^{L+1}\phi_{g,j}(x)\hat{a}^{\dagger}_{g,j}+\sum_{j=1}^{L}\phi_{e,j+1/2}(x)\hat{a}^{\dagger}_{e,j+1/2}\right)\nonumber\\
&&\times\left(\sum_{j'=1}^{L+1}\phi_{g,j'}(x)\hat{a}_{g,j'}+\sum_{j'=1}^{L}\phi_{e,j'+1/2}(x)\hat{a}_{e,j'+1/2}\right)\sum_{k}\sqrt{k}(e^{ikx}\hat{b}_k +e^{-ikx}\hat{b}^\dagger_k )\nonumber\\
\label{inthamexp1}
\end{eqnarray}

In evaluating this integral, we neglect terms proportional to $\hat{a}^{\dagger}_{g,j'}\hat{a}_{g,j}$ and $\hat{a}^{\dagger}_{e,j'+1/2}\hat{a}_{e,j+1/2}$, since these terms will vanish once we evaluate the double commutator $\left[\hat{H}_{SB}(t),\left[\hat{H}_{SB}(t-t'),\hat{\rho}_{S}(t')\otimes\hat{\rho}_{B}\right]\right]$ in Eq.
(\ref{bornmarkovmasteq}). Thus, we consider only the following terms in the evaluation of the integral in Eq. (\ref{inthamexp1}):

\begin{eqnarray}
&&\int dx e^{\pm ikx}\phi_{g,j}(x)\phi_{e,j'+1/2}(x)=\left(\frac{m_s \omega_g}{\pi\hbar}\right)^{1/4}\left(\frac{m_s \omega_e}{\pi\hbar}\right)^{1/4}\sqrt{\frac{2m_s \omega_e}{\hbar}}\nonumber\\
&&\times\int dx\; e^{\pm ikx}\exp\left(-\frac{m_s \omega_g}{2\hbar}(x-x_j)^2\right)(x-x_{j'+1/2})\exp\left(-\frac{m_s \omega_e}{2\hbar}(x-x_{j'+1/2})^2\right)\nonumber\\
&&\underset{\sigma_g \rightarrow 0}\longrightarrow 2\sigma_{g}^{1/2}\sigma_{e}^{-3/4}\int dx\;e^{\pm ikx}\delta(x-x_j)(x-x_{j'+1/2})\exp\left(-\frac{(x-x_{j'+1/2})^2}{2\sigma_e^2}\right)\nonumber\\
&&=2\sigma_{g}^{1/2}\sigma_{e}^{-3/4}e^{\pm ikx_j}(x_j-x_{j'+1/2})\exp\left(-\frac{(x_j-x_{j'+1/2})^2}{2\sigma_e^2}\right)
\label{overlapint}
\end{eqnarray}
The result of the integration will, due to the exponential term $\exp\left(-\frac{(x_j-x_{j'+1/2})^2}{2\sigma_e^2}\right)$, vanish for large $x_{j}-x_{j'+1/2}$. As such, the dominant terms from this integral will be terms for which either $j'=j$ or $j'=j-1$. However, these values of $j'$ and $j$ will correspond to values of $x_j$ and $x_{j+1/2}$ which denote the centroids of harmonic potentials that are adjacent to each other, i. e. harmonic potentials centered at $x_j = (2j-L-2)x_a$ and $x_{j+1/2} = (2j-L-1)x_a$ or harmonic potentials centered at $x_{j+1} = (2j+2-L)x_a$ and $x_{j+1/2} = (2j+1-L)x_a$. This further implies that the terms $x_j-x_{j'+1/2}$ appearing in the result of the integration given by Eq. (\ref{overlapint}) will have the form
\begin{eqnarray}
&&x_{j}-x_{j+1/2}=(2j-L-2)x_a-(2j-L-1)x_a=-x_a\nonumber\\
&&x_{j+1}-x_{j+1/2}=(2j-L)x_a-(2j-L-1)x_a=x_a\nonumber\\
\end{eqnarray}
At the same time, in order to minimize the probability that atoms trapped in the harmonic potentials of the array will tunnel without undergoing the driven and dissipative processes described earlier from one harmonic potential to another, we can set the locations of the centers of the harmonic potentials to be very far apart from each other relative to the condensate atoms' wavenumber $k$, such that $kx_j<<1 \forall j$. With this approximation, $e^{\pm ikx_j}\approx 1$, and so the surviving terms in Eq. (\ref{inthamexp1}) will be

\begin{eqnarray}
&&\hat{H}_{SB}=\frac{4\pi a_{SB}}{\mu\sigma_e^{1/4}}\sqrt{\frac{\rho_B \sigma_g}{2m_B cL\sigma_e}}x_a\exp\left(-\frac{x_a^2}{2\sigma_e^2}\right)\sum_{j=1}^{L}\sum_{k}\sqrt{k}\left(-\left(\hat{a}^{\dagger}_{e,j+1/2}\hat{a}_{g,j}+\hat{a}^{\dagger}_{g,j}\hat{a}_{e,j+1/2}\right)\right. \nonumber\\
&&\left.+\left(\hat{a}^{\dagger}_{e,j+1/2}\hat{a}_{g,j+1}+\hat{a}^{\dagger}_{g,j+1}\hat{a}_{e,j+1/2}\right)\right)\left(\hat{b}_{k}+\hat{b}^{\dagger}_{k}\right)\nonumber\\
\label{inthamexp2}
\end{eqnarray}

\section{Master Equation for the Driven - Dissipative Dynamics of the Open Quantum Trapped Ultracold Atom System}
To derive the master equation describing the time evolution of the open quantum trapped ultracold atom system we are considering in this paper, we make use of the Born - Markov approximation for open quantum systems, which assumes that the quantum system is weakly coupled to another quantum system that serves as a bath, or reservoir, for quantities that are emitted by the first quantum system (such as heat, electromagnetic radiation, or Bogoliubov excitations) and which assumes that the relaxation time for the system is much greater than the timescale over which correlations between the system and the bath decay. Under this approximation, the resulting master equation will have the form

\begin{equation}
\frac{d}{dt}\hat{\rho}_{S}(t)=-\int_{0}^{+\infty}dt' Tr_{B}\left[\hat{H}_{SB}(t),\left[\hat{H}_{SB}(t-t'),\hat{\rho}_{S}(t)\otimes\hat{\rho}_{B}\right]\right]
\label{bornmarkovmasteq}
\end{equation}
In this equation, the density operators $\hat{\rho}_{S}(t)$ and $\hat{\rho}_{B}$ are the density operators corresponding to the trapped ultracold atoms (the system) and the background BEC (the bath, or reservoir, of excitations) respectively, and $\hat{H}_{SB}(t)$ is the time-evolved interaction Hamiltonian for the trapped ultracold atom open quantum system considered in this paper. Now to determine the explicit form of $\hat{H}_{SB}(t)$, we evolve the interaction Hamiltonian given by Eq. (\ref{inthamexp2}) over time using the following time evolution rule in the Heisenberg picture:

\begin{equation}
\hat{H}_{SB}(t)=exp\left(\frac{i}{\hbar}(\hat{H}_{S}+\hat{H}_{B})t\right)\hat{H}_{SB}exp\left(-\frac{i}{\hbar}(\hat{H}_{S}+\hat{H}_{B})t\right)
\label{heisenevointeracthamil}
\end{equation}
In using the time evolution rule in the Heisenberg picture, the Hamiltonians used to evolve the interaction Hamiltonian for the open quantum system are those for the trapped ultracold atom system and the BEC excitation reservoir, which have the explicit forms

\begin{equation}
\hat{H}_{S}=\sum_{j=1}^{L}\left(\epsilon_{g}\hat{a}^{\dagger}_{g,j}\hat{a}_{g,j}+\epsilon_{e}\hat{a}^{\dagger}_{e,j}\hat{a}_{e,j}\right)
\label{sysham}
\end{equation}
\begin{equation}
\hat{H}_{B}=\sum_{k}E_{k}\hat{b}^{\dagger}_{k}\hat{b}_{k}
\label{becham}
\end{equation}
In these Hamiltonians, the many-body creation (annihilation) operators for the trapped ultracold atoms in the ground and excited states, $\hat{a}^{\dagger}_{g,j}\; (\hat{a}_{g,j})$ and $\hat{a}^{\dagger}_{e,j}\; (\hat{a}_{e,j})$ respectively, and the particle creation (annihilation) operators for the atoms in the BEC reservoir, $\hat{b}^{\dagger}_{k}\; (\hat{b}_{k})$, satisfy the following commutation relations:
\begin{equation}
\left[\hat{a}_{g,j},\hat{a}^{\dagger}_{g,j'}\right]=\left[\hat{a}_{e,j},\hat{a}^{\dagger}_{e,j'}\right]=\delta_{j,j'},\;\left[\hat{a}_{g,j},\hat{a}^{\dagger}_{e,j'}\right]=\left[\hat{a}_{e,j},\hat{a}^{\dagger}_{g,j'}\right]=0\nonumber\\
\label{syscommrel}
\end{equation}
\begin{equation}
\left[\hat{b}_{k},\hat{b}^{\dagger}_{k'}\right]=\delta_{k,k'},\;\left[\hat{b}_{k},\hat{b}_{k'}\right]=\left[\hat{b}^{\dagger}_{k},\hat{b}^{\dagger}_{k'}\right]=0
\label{beccommrel}
\end{equation}
In performing the time evolution, we make use of the Baker-Campbell-Hausdorff identity for time - evolved quantum observables in the Heisenberg picture:
\begin{equation}
\hat{A}(t)=\exp\left(\frac{i}{\hbar}t\hat{H}\right)\hat{A}\exp\left(-\frac{i}{\hbar}t\hat{H}\right)=\hat{A}+\frac{i}{\hbar}t\left[\hat{H},\hat{A}\right]+\frac{1}{2!}\left(\frac{i}{\hbar}t\right)^2 \left[\hat{H},\left[\hat{H},\hat{A}\right]\right]+...
\label{bch}
\end{equation}
Using this identity in the time evolution of the interaction Hamiltonian of the trapped ultracold atom open quantum system, together with the commutation relations for the system and reservoir creation and annihilation operators given by Eqs. (\ref{syscommrel}) and (\ref{beccommrel}), we then obtain the following expression:
\begin{eqnarray}
&&\hat{H}_{SB}(t)=\frac{4\pi a_{SB}}{\mu\sigma_e^{1/4}}\sqrt{\frac{\rho_B \sigma_g}{2m_B cL\sigma_e}}x_a\exp\left(-\frac{x_a^2}{2\sigma_e^2}\right)\nonumber\\
&&\times\sum_{j=1}^{L-1}\sum_{k}\sqrt{k}\left(-\left(e^{\frac{i}{\hbar}(\epsilon_{e}-\epsilon_{g})t}\hat{a}^{\dagger}_{e,j+1/2}\hat{a}_{g,j}+e^{-\frac{i}{\hbar}(\epsilon_{e}-\epsilon_{g})t}\hat{a}^{\dagger}_{g,j}\hat{a}_{e,j+1/2}\right)\right. \nonumber\\
&&\left.+\left(e^{\frac{i}{\hbar}(\epsilon_{e}-\epsilon_{g})t}\hat{a}^{\dagger}_{e,j+1/2}\hat{a}_{g,j+1}+e^{-\frac{i}{\hbar}(\epsilon_{e}-\epsilon_{g})t}\hat{a}^{\dagger}_{g,j+1}\hat{a}_{e,j+1/2}\right)\right)\left(e^{-\frac{i}{\hbar}ck t}\hat{b}_{k}+e^{\frac{i}{\hbar}ck t}\hat{b}^{\dagger}_{k}\right)\nonumber\\
\label{inthamtimeevo}
\end{eqnarray}

In evolving the interaction Hamiltonian over time, we make the substitution $E_k = ck$ in the reservoir Hamiltonian $\hat{H}_B$, in accordance with the assumptions that we made regarding the excitation energies earlier. We then substitute this time-evolved interaction Hamiltonian into the Born-Markov form of the quantum master equation as given in Eq. (\ref{bornmarkovmasteq}) and evaluate the resulting trace over the reservoir variables $\hat{b}_{k},\;\hat{b}^{\dagger}_{k}$ as well as the integrals over time. In taking the trace over the reservoir variables, we follow the convention in Ref. (\cite{caballar}) which states that 

\begin{equation}
Tr_{B}(\hat{b}_{k}\hat{b}_{k'}\hat{\rho}_B)=Tr_{B}(\hat{b}^{\dagger}_{k}\hat{b}^{\dagger}_{k'}\hat{\rho}_B)=Tr_{B}(\hat{b}^{\dagger}_{k}\hat{b}_{k'}\hat{\rho}_B)\approx 0,\;Tr_{B}(\hat{b}_{k}\hat{b}^{\dagger}_{k'}\hat{\rho}_B)\approx \delta_{k,k'}
\label{bathvartrace}
\end{equation}

Substituting the time-evolved interaction Hamiltonian given by Eq. (\ref{inthamtimeevo}) into the Born - Markov master equation, Eq. (\ref{bornmarkovmasteq}), we then carry out the double commutation, trace over the bath variables, and time integration operations to obtain the final form of the master equation for this open quantum system. The trace over the bath variables is accomplished with the help of the identities given by Eq. (\ref{bathvartrace}), such that only terms of the form $Tr_{B}(\hat{b}_{k}\hat{b}^{\dagger}_{k}\hat{\rho}_{B}), Tr_{B}(\hat{\rho}_{B}\hat{b}_{k}\hat{b}^{\dagger}_{k})$ and $Tr_{B}(\hat{b}^{\dagger}_{k}\hat{\rho_{B}}\hat{b}_{k})$ remain in the master equation. In doing so, we obtain the following term:

\begin{eqnarray}
&&\int_{0}^{+\infty}dt' Tr_{B}\left[\hat{H}_{SB}(t),\left[\hat{H}_{SB}(t-t'),\hat{\rho}_{S}(t)\otimes\hat{\rho}_{B}\right]\right]=\frac{16\pi^2 a^{2}_{SB}}{\mu^2 \sigma_e^{1/2}}\left(\frac{\rho_B \sigma_g}{2m_B cL\sigma_e}x^{2}_a\right)\exp\left(-\frac{x_a^2}{\sigma_e^2}\right)\nonumber\\
&&\times\int_{0}^{+\infty}dt'Tr_{B}\left[\sum_{j=1}^{L-1}\sum_{k}\sqrt{k}\left(-\left(e^{\frac{i}{\hbar}(\epsilon_{e}-\epsilon_{g})t}\hat{a}^{\dagger}_{e,j+1/2}\hat{a}_{g,j}+e^{-\frac{i}{\hbar}(\epsilon_{e}-\epsilon_{g})t}\hat{a}^{\dagger}_{g,j}\hat{a}_{e,j+1/2}\right)\right.\right.\nonumber\\
&&\left.+\left(e^{\frac{i}{\hbar}(\epsilon_{e}-\epsilon_{g})t}\hat{a}^{\dagger}_{e,j+1/2}\hat{a}_{g,j+1}+e^{-\frac{i}{\hbar}(\epsilon_{e}-\epsilon_{g})t}\hat{a}^{\dagger}_{g,j+1}\hat{a}_{e,j+1/2}\right)\right)\left(e^{-\frac{i}{\hbar}ck t}\hat{b}_{k}+e^{\frac{i}{\hbar}ck t}\hat{b}^{\dagger}_{k}\right),\nonumber\\
&&\left[\sum_{j'=1}^{L-1}\sum_{k'}\sqrt{k'}\left(-\left(e^{\frac{i}{\hbar}(\epsilon_{e}-\epsilon_{g})(t-t')}\hat{a}^{\dagger}_{e,j'+1/2}\hat{a}_{g,j'}+e^{-\frac{i}{\hbar}(\epsilon_{e}-\epsilon_{g})(t-t')}\hat{a}^{\dagger}_{g,j'}\hat{a}_{e,j'+1/2}\right)\right.\right.\nonumber\\
&&\left.+\left(e^{\frac{i}{\hbar}(\epsilon_{e}-\epsilon_{g})(t-t')}\hat{a}^{\dagger}_{e,j'+1/2}\hat{a}_{g,j'+1}+e^{-\frac{i}{\hbar}(\epsilon_{e}-\epsilon_{g})(t-t')}\hat{a}^{\dagger}_{g,j'+1}\hat{a}_{e,j'+1/2}\right)\right)\left(e^{-\frac{i}{\hbar}ck'(t-t')}\hat{b}_{k'}+e^{\frac{i}{\hbar}ck'(t-t')}\hat{b}^{\dagger}_{k'}\right),\nonumber\\
&&\left.\left.\hat{\rho}_{S}(t')\otimes\hat{\rho}_{B}\right]\right]\nonumber\\
&&=\frac{16\pi^2 a^{2}_{SB}}{\mu^2 \sigma_e^{1/2}}\left(\frac{\rho_B \sigma_g}{2m_B cL\sigma_e}x^{2}_a\right)\exp\left(-\frac{x_a^2}{\sigma_e^2}\right)\sum_{j,j'=1}^{L-1}\sum_{k}k\int_{0}^{+\infty}dt'\left[\left(\left(e^{\frac{i}{\hbar}(\epsilon_{e}-\epsilon_{g})t}\hat{a}^{\dagger}_{e,j+1/2}\hat{a}_{g,j}\right.\right.\right.\nonumber\\
&&\left.\left.+e^{-\frac{i}{\hbar}(\epsilon_{e}-\epsilon_{g})t}\hat{a}^{\dagger}_{g,j}\hat{a}_{e,j+1/2}\right)-\left(e^{\frac{i}{\hbar}(\epsilon_{e}-\epsilon_{g})t}\hat{a}^{\dagger}_{e,j+1/2}\hat{a}_{g,j+1}+e^{-\frac{i}{\hbar}(\epsilon_{e}-\epsilon_{g})t}\hat{a}^{\dagger}_{g,j+1}\hat{a}_{e,j+1/2}\right)\right)\nonumber\\
&&\times\left(\left(e^{\frac{i}{\hbar}(\epsilon_{e}-\epsilon_{g})(t-t')}\hat{a}^{\dagger}_{e,j'+1/2}\hat{a}_{g,j'}+e^{-\frac{i}{\hbar}(\epsilon_{e}-\epsilon_{g})(t-t')}\hat{a}^{\dagger}_{g,j'}\hat{a}_{e,j'+1/2}\right)\right.\nonumber\\
&&\left.-\left(e^{\frac{i}{\hbar}(\epsilon_{e}-\epsilon_{g})(t-t')}\hat{a}^{\dagger}_{e,j'+1/2}\hat{a}_{g,j'+1}+e^{-\frac{i}{\hbar}(\epsilon_{e}-\epsilon_{g})(t-t')}\hat{a}^{\dagger}_{g,j'+1}\hat{a}_{e,j'+1/2}\right)\right)e^{-\frac{i}{\hbar}ckt'}\hat{\rho}_{S}(t)\nonumber\\
&&-\left(\left(e^{\frac{i}{\hbar}(\epsilon_{e}-\epsilon_{g})t}\hat{a}^{\dagger}_{e,j+1/2}\hat{a}_{g,j}+e^{-\frac{i}{\hbar}(\epsilon_{e}-\epsilon_{g})t}\hat{a}^{\dagger}_{g,j}\hat{a}_{e,j+1/2}\right)-\left(e^{\frac{i}{\hbar}(\epsilon_{e}-\epsilon_{g})t}\hat{a}^{\dagger}_{e,j+1/2}\hat{a}_{g,j+1}\right.\right.\nonumber\\
&&\left.\left.+e^{-\frac{i}{\hbar}(\epsilon_{e}-\epsilon_{g})t}\hat{a}^{\dagger}_{g,j+1}\hat{a}_{e,j+1/2}\right)\right)e^{\frac{i}{\hbar}ckt'}\hat{\rho}_{S}(t)\left(\left(e^{\frac{i}{\hbar}(\epsilon_{e}-\epsilon_{g})(t-t')}\hat{a}^{\dagger}_{e,j'+1/2}\hat{a}_{g,j'}+e^{-\frac{i}{\hbar}(\epsilon_{e}-\epsilon_{g})(t-t')}\hat{a}^{\dagger}_{g,j'}\hat{a}_{e,j'+1/2}\right)\right.\nonumber\\
&&\left.-\left(e^{\frac{i}{\hbar}(\epsilon_{e}-\epsilon_{g})(t-t')}\hat{a}^{\dagger}_{e,j'+1/2}\hat{a}_{g,j'+1}+e^{-\frac{i}{\hbar}(\epsilon_{e}-\epsilon_{g})(t-t')}\hat{a}^{\dagger}_{g,j'+1}\hat{a}_{e,j'+1/2}\right)\right)-\left(\left(e^{\frac{i}{\hbar}(\epsilon_{e}-\epsilon_{g})(t-t')}\hat{a}^{\dagger}_{e,j'+1/2}\hat{a}_{g,j'}\right.\right.\nonumber\\
&&\left.\left.+e^{-\frac{i}{\hbar}(\epsilon_{e}-\epsilon_{g})(t-t')}\hat{a}^{\dagger}_{g,j'}\hat{a}_{e,j'+1/2}\right)-\left(e^{\frac{i}{\hbar}(\epsilon_{e}-\epsilon_{g})(t-t')}\hat{a}^{\dagger}_{e,j'+1/2}\hat{a}_{g,j'+1}+e^{-\frac{i}{\hbar}(\epsilon_{e}-\epsilon_{g})(t-t')}\hat{a}^{\dagger}_{g,j'+1}\hat{a}_{e,j'+1/2}\right)\right)\nonumber\\
&&\times e^{-\frac{i}{\hbar}ckt'}\hat{\rho}_{S}(t)\left(\left(e^{\frac{i}{\hbar}(\epsilon_{e}-\epsilon_{g})t}\hat{a}^{\dagger}_{e,j+1/2}\hat{a}_{g,j}+e^{-\frac{i}{\hbar}(\epsilon_{e}-\epsilon_{g})t}\hat{a}^{\dagger}_{g,j}\hat{a}_{e,j+1/2}\right)-\left(e^{\frac{i}{\hbar}(\epsilon_{e}-\epsilon_{g})t}\hat{a}^{\dagger}_{e,j+1/2}\hat{a}_{g,j+1}\right.\right.\nonumber\\
&&+\left.\left.e^{-\frac{i}{\hbar}(\epsilon_{e}-\epsilon_{g})t}\hat{a}^{\dagger}_{g,j+1}\hat{a}_{e,j+1/2}\right)\right)+e^{\frac{i}{\hbar}ckt'}\hat{\rho}_{S}(t)\left(\left(e^{\frac{i}{\hbar}(\epsilon_{e}-\epsilon_{g})(t-t')}\hat{a}^{\dagger}_{e,j'+1/2}\hat{a}_{g,j'}\right.\right.\nonumber\\
&&\left.\left.+e^{-\frac{i}{\hbar}(\epsilon_{e}-\epsilon_{g})(t-t')}\hat{a}^{\dagger}_{g,j'}\hat{a}_{e,j'+1/2}\right)-\left(e^{\frac{i}{\hbar}(\epsilon_{e}-\epsilon_{g})(t-t')}\hat{a}^{\dagger}_{e,j'+1/2}\hat{a}_{g,j'+1}+e^{-\frac{i}{\hbar}(\epsilon_{e}-\epsilon_{g})(t-t')}\hat{a}^{\dagger}_{g,j'+1}\hat{a}_{e,j'+1/2}\right)\right)\nonumber\\
&&\times\left(\left(e^{\frac{i}{\hbar}(\epsilon_{e}-\epsilon_{g})t}\hat{a}^{\dagger}_{e,j+1/2}\hat{a}_{g,j}+e^{-\frac{i}{\hbar}(\epsilon_{e}-\epsilon_{g})t}\hat{a}^{\dagger}_{g,j}\hat{a}_{e,j+1/2}\right)-\left(e^{\frac{i}{\hbar}(\epsilon_{e}-\epsilon_{g})t}\hat{a}^{\dagger}_{e,j+1/2}\hat{a}_{g,j+1}\right.\right.\nonumber\\
&&\left.\left.\left.+e^{-\frac{i}{\hbar}(\epsilon_{e}-\epsilon_{g})t}\hat{a}^{\dagger}_{g,j+1}\hat{a}_{e,j+1/2}\right)\right)\right]\nonumber\\
\label{masteqrhs1}
\end{eqnarray}

Eq. (\ref{masteqrhs1}) has integrals of the form $\int_{0}^{\infty}dt'\;e^{\pm\frac{i}{\hbar}((\epsilon_e -\epsilon_g \pm ck))t'}$, which, using the definition of the Dirac delta distribution for oscillatory integrals, can be written as

\begin{equation}
\int_{0}^{\infty}dt'\;e^{\pm\frac{i}{\hbar}((\epsilon_e -\epsilon_g \pm ck))t'}=\pm\frac{\hbar}{\sqrt{2\pi}}\delta((\epsilon_e -\epsilon_g \pm ck))
\label{diracdeltaintegral}
\end{equation}
Now we substitute these delta distribution forms of the integrals back into Eq. (\ref{masteqrhs1}), and under the assumption that $k$ can be approximated as a continuous variable, the summation over the discrete wavenumber $k$ is transformed into an integration over the continuous variable $k$, $\sum_{k}\rightarrow\frac{1}{2\pi}\int_{-\infty}^{\infty}dk'$, so that Eq. (\ref{masteqrhs1}) becomes

\begin{eqnarray}
&&\int_{0}^{+\infty}dt' Tr_{B}\left[\hat{H}_{SB}(t),\left[\hat{H}_{SB}(t-t'),\hat{\rho}_{S}(t)\otimes\hat{\rho}_{B}\right]\right]\nonumber\\
&&=\frac{8\pi\hbar a^{2}_{SB}}{\mu^2 \sqrt{2\pi\sigma_e}}\left(\frac{\rho_B \sigma_g (\epsilon_e - \epsilon_g)}{2m_B c^2 L\sigma_e}x^{2}_a\right)e^{-\frac{x_a^2}{\sigma_e^2}}\sum_{j,j'=1}^{L-1}\left[-\left(\left(e^{\frac{i}{\hbar}(\epsilon_{e}-\epsilon_{g})t}\hat{a}^{\dagger}_{e,j+1/2}\hat{a}_{g,j}\right.\right.\right.\nonumber\\
&&\left.\left.+e^{-\frac{i}{\hbar}(\epsilon_{e}-\epsilon_{g})t}\hat{a}^{\dagger}_{g,j}\hat{a}_{e,j+1/2}\right)-\left(e^{\frac{i}{\hbar}(\epsilon_{e}-\epsilon_{g})t}\hat{a}^{\dagger}_{e,j+1/2}\hat{a}_{g,j+1}+e^{-\frac{i}{\hbar}(\epsilon_{e}-\epsilon_{g})t}\hat{a}^{\dagger}_{g,j+1}\hat{a}_{e,j+1/2}\right)\right)\nonumber\\
&&\times\left(\left(e^{\frac{i}{\hbar}(\epsilon_{e}-\epsilon_{g})t}\hat{a}^{\dagger}_{e,j'+1/2}\hat{a}_{g,j'}-e^{-\frac{i}{\hbar}(\epsilon_{e}-\epsilon_{g})t}\hat{a}^{\dagger}_{g,j'}\hat{a}_{e,j'+1/2}\right)\right.\nonumber\\
&&\left.-\left(e^{\frac{i}{\hbar}(\epsilon_{e}-\epsilon_{g})t}\hat{a}^{\dagger}_{e,j'+1/2}\hat{a}_{g,j'+1}-e^{-\frac{i}{\hbar}(\epsilon_{e}-\epsilon_{g})t}\hat{a}^{\dagger}_{g,j'+1}\hat{a}_{e,j'+1/2}\right)\right)\hat{\rho}_{S}(t)\nonumber\\
&&-\left(\left(e^{\frac{i}{\hbar}(\epsilon_{e}-\epsilon_{g})t}\hat{a}^{\dagger}_{e,j+1/2}\hat{a}_{g,j}+e^{-\frac{i}{\hbar}(\epsilon_{e}-\epsilon_{g})t}\hat{a}^{\dagger}_{g,j}\hat{a}_{e,j+1/2}\right)-\left(e^{\frac{i}{\hbar}(\epsilon_{e}-\epsilon_{g})t}\hat{a}^{\dagger}_{e,j+1/2}\hat{a}_{g,j+1}\right.\right.\nonumber\\
&&\left.\left.+e^{-\frac{i}{\hbar}(\epsilon_{e}-\epsilon_{g})t}\hat{a}^{\dagger}_{g,j+1}\hat{a}_{e,j+1/2}\right)\right)\hat{\rho}_{S}(t)\left(\left(e^{\frac{i}{\hbar}(\epsilon_{e}-\epsilon_{g})t}\hat{a}^{\dagger}_{e,j'+1/2}\hat{a}_{g,j'}-e^{-\frac{i}{\hbar}(\epsilon_{e}-\epsilon_{g})t}\hat{a}^{\dagger}_{g,j'}\hat{a}_{e,j'+1/2}\right)\right.\nonumber\\
&&\left.-\left(e^{\frac{i}{\hbar}(\epsilon_{e}-\epsilon_{g})t}\hat{a}^{\dagger}_{e,j'+1/2}\hat{a}_{g,j'+1}-e^{-\frac{i}{\hbar}(\epsilon_{e}-\epsilon_{g})t}\hat{a}^{\dagger}_{g,j'+1}\hat{a}_{e,j'+1/2}\right)\right)+\left(\left(e^{\frac{i}{\hbar}(\epsilon_{e}-\epsilon_{g})t}\hat{a}^{\dagger}_{e,j'+1/2}\hat{a}_{g,j'}\right.\right.\nonumber\\
&&\left.\left.-e^{-\frac{i}{\hbar}(\epsilon_{e}-\epsilon_{g})t}\hat{a}^{\dagger}_{g,j'}\hat{a}_{e,j'+1/2}\right)-\left(e^{\frac{i}{\hbar}(\epsilon_{e}-\epsilon_{g})t}\hat{a}^{\dagger}_{e,j'+1/2}\hat{a}_{g,j'+1}-e^{-\frac{i}{\hbar}(\epsilon_{e}-\epsilon_{g})t}\hat{a}^{\dagger}_{g,j'+1}\hat{a}_{e,j'+1/2}\right)\right)\nonumber\\
&&\times\hat{\rho}_{S}(t)\left(\left(e^{\frac{i}{\hbar}(\epsilon_{e}-\epsilon_{g})t}\hat{a}^{\dagger}_{e,j+1/2}\hat{a}_{g,j}+e^{-\frac{i}{\hbar}(\epsilon_{e}-\epsilon_{g})t}\hat{a}^{\dagger}_{g,j}\hat{a}_{e,j+1/2}\right)-\left(e^{\frac{i}{\hbar}(\epsilon_{e}-\epsilon_{g})t}\hat{a}^{\dagger}_{e,j+1/2}\hat{a}_{g,j+1}\right.\right.\nonumber\\
&&+\left.\left.e^{-\frac{i}{\hbar}(\epsilon_{e}-\epsilon_{g})t}\hat{a}^{\dagger}_{g,j+1}\hat{a}_{e,j+1/2}\right)\right)+\hat{\rho}_{S}(t)\left(\left(e^{\frac{i}{\hbar}(\epsilon_{e}-\epsilon_{g})t}\hat{a}^{\dagger}_{e,j'+1/2}\hat{a}_{g,j'}\right.\right.\nonumber\\
&&\left.\left.-e^{-\frac{i}{\hbar}(\epsilon_{e}-\epsilon_{g})t}\hat{a}^{\dagger}_{g,j'}\hat{a}_{e,j'+1/2}\right)-\left(e^{\frac{i}{\hbar}(\epsilon_{e}-\epsilon_{g})t}\hat{a}^{\dagger}_{e,j'+1/2}\hat{a}_{g,j'+1}-e^{-\frac{i}{\hbar}(\epsilon_{e}-\epsilon_{g})t}\hat{a}^{\dagger}_{g,j'+1}\hat{a}_{e,j'+1/2}\right)\right)\nonumber\\
&&\times\left(\left(e^{\frac{i}{\hbar}(\epsilon_{e}-\epsilon_{g})t}\hat{a}^{\dagger}_{e,j+1/2}\hat{a}_{g,j}+e^{-\frac{i}{\hbar}(\epsilon_{e}-\epsilon_{g})t}\hat{a}^{\dagger}_{g,j}\hat{a}_{e,j+1/2}\right)-\left(e^{\frac{i}{\hbar}(\epsilon_{e}-\epsilon_{g})t}\hat{a}^{\dagger}_{e,j+1/2}\hat{a}_{g,j+1}\right.\right.\nonumber\\
&&\left.\left.\left.+e^{-\frac{i}{\hbar}(\epsilon_{e}-\epsilon_{g})t}\hat{a}^{\dagger}_{g,j+1}\hat{a}_{e,j+1/2}\right)\right)\right]\nonumber\\
\label{masteqrhs2}
\end{eqnarray}
This right hand side of the master equation can be further simplified by gathering like terms:

\begin{eqnarray}
&&\int_{0}^{+\infty}dt' Tr_{B}\left[\hat{H}_{SB}(t),\left[\hat{H}_{SB}(t-t'),\hat{\rho}_{S}(t)\otimes\hat{\rho}_{B}\right]\right]\nonumber\\
&&=\frac{8\pi\hbar a^{2}_{SB}}{\mu^2 \sqrt{2\pi\sigma_e}}\left(\frac{\rho_B \sigma_g (\epsilon_e - \epsilon_g)}{2m_B c^2 L\sigma_e}x^{2}_a\right)e^{-\frac{x_a^2}{\sigma_e^2}}\sum_{j,j'=1}^{L-1}\left[-\left(e^{\frac{i}{\hbar}(\epsilon_{e}-\epsilon_{g})t}\hat{a}^{\dagger}_{e,j+1/2}\left(\hat{a}_{g,j}-\hat{a}_{g,j+1}\right)\right.\right.\nonumber\\
&&\left.+e^{-\frac{i}{\hbar}(\epsilon_{e}-\epsilon_{g})t}\left(\hat{a}^{\dagger}_{g,j}-\hat{a}^{\dagger}_{g,j+1}\right)\hat{a}_{e,j+1/2}\right)\left(e^{\frac{i}{\hbar}(\epsilon_{e}-\epsilon_{g})t}\hat{a}^{\dagger}_{e,j'+1/2}\left(\hat{a}_{g,j'}-\hat{a}_{g,j'+1}\right)\right.\nonumber\\
&&\left.-e^{-\frac{i}{\hbar}(\epsilon_{e}-\epsilon_{g})t}\left(\hat{a}^{\dagger}_{g,j'}-\hat{a}^{\dagger}_{g,j'+1}\right)\hat{a}_{e,j'+1/2}\right)\hat{\rho}_{S}(t)\nonumber\\
&&-\left(e^{\frac{i}{\hbar}(\epsilon_{e}-\epsilon_{g})t}\hat{a}^{\dagger}_{e,j+1/2}\left(\hat{a}_{g,j}-\hat{a}_{g,j+1}\right)+e^{-\frac{i}{\hbar}(\epsilon_{e}-\epsilon_{g})t}\left(\hat{a}^{\dagger}_{g,j}-\hat{a}^{\dagger}_{g,j+1}\right)\hat{a}_{e,j+1/2}\right)\nonumber\\
&&\times\hat{\rho}_{S}(t)\left(e^{\frac{i}{\hbar}(\epsilon_{e}-\epsilon_{g})t}\hat{a}^{\dagger}_{e,j'+1/2}\left(\hat{a}_{g,j'}-\hat{a}_{g,j'+1}\right)-e^{-\frac{i}{\hbar}(\epsilon_{e}-\epsilon_{g})t}\left(\hat{a}^{\dagger}_{g,j'}-\hat{a}^{\dagger}_{g,j'+1}\right)\hat{a}_{e,j'+1/2}\right)\nonumber\\
&&+\left(e^{\frac{i}{\hbar}(\epsilon_{e}-\epsilon_{g})t}\hat{a}^{\dagger}_{e,j'+1/2}\left(\hat{a}_{g,j'}-\hat{a}_{g,j'+1}\right)-e^{-\frac{i}{\hbar}(\epsilon_{e}-\epsilon_{g})t}\left(\hat{a}^{\dagger}_{g,j'}-\hat{a}^{\dagger}_{g,j'+1}\right)\hat{a}_{e,j'+1/2}\right)\nonumber\\
&&\times\hat{\rho}_{S}(t)\left(e^{\frac{i}{\hbar}(\epsilon_{e}-\epsilon_{g})t}\hat{a}^{\dagger}_{e,j+1/2}\left(\hat{a}_{g,j}-\hat{a}_{g,j+1}\right)+e^{-\frac{i}{\hbar}(\epsilon_{e}-\epsilon_{g})t}\left(\hat{a}^{\dagger}_{g,j}-\hat{a}^{\dagger}_{g,j+1}\right)\hat{a}_{e,j+1/2}\right)\nonumber\\
&&+\hat{\rho}_{S}(t)\left(e^{\frac{i}{\hbar}(\epsilon_{e}-\epsilon_{g})t}\hat{a}^{\dagger}_{e,j'+1/2}\left(\hat{a}_{g,j'}-\hat{a}_{g,j'+1}\right)-e^{-\frac{i}{\hbar}(\epsilon_{e}-\epsilon_{g})t}\left(\hat{a}^{\dagger}_{g,j'}-\hat{a}^{\dagger}_{g,j'+1}\right)\hat{a}_{e,j'+1/2}\right)\nonumber\\
&&\times\left.\left(e^{\frac{i}{\hbar}(\epsilon_{e}-\epsilon_{g})t}\hat{a}^{\dagger}_{e,j+1/2}\left(\hat{a}_{g,j}-\hat{a}_{g,j+1}\right)+e^{-\frac{i}{\hbar}(\epsilon_{e}-\epsilon_{g})t}\left(\hat{a}^{\dagger}_{g,j}-\hat{a}^{\dagger}_{g,j+1}\right)\hat{a}_{e,j+1/2}\right)\right]\nonumber\\
\label{mastereqrhs3}
\end{eqnarray}
Next, we distribute the sum $\sum_{j,j'=1}^{L-1}=\sum{j=1}^{L-1}\sum_{j'=1}^{L-1}$ to those components of this expression with indices $j$ and $j'$, and factor out exponential time terms $\exp\left(\pm\frac{i}{\hbar}(\epsilon_e -\epsilon_g)t\right)$ in such a way that either of the following factorizations occur,

\begin{eqnarray}
&&e^{\frac{i}{\hbar}(\epsilon_{e}-\epsilon_{g})t}\hat{a}^{\dagger}_{e,j+1/2}\left(\hat{a}_{g,j}-\hat{a}_{g,j+1}\right)\pm e^{-\frac{i}{\hbar}(\epsilon_{e}-\epsilon_{g})t}\left(\hat{a}^{\dagger}_{g,j}-\hat{a}^{\dagger}_{g,j+1}\right)\hat{a}_{e,j+1/2}\nonumber\\
&&=e^{\frac{i}{\hbar}(\epsilon_{e}-\epsilon_{g})t}\left(\hat{a}^{\dagger}_{e,j+1/2}\left(\hat{a}_{g,j}-\hat{a}_{g,j+1}\right)\pm e^{-\frac{2i}{\hbar}(\epsilon_{e}-\epsilon_{g})t}\left(\hat{a}^{\dagger}_{g,j}-\hat{a}^{\dagger}_{g,j+1}\right)\hat{a}_{e,j+1/2}\right),\nonumber\\
&&e^{\frac{i}{\hbar}(\epsilon_{e}-\epsilon_{g})t}\hat{a}^{\dagger}_{e,j+1/2}\left(\hat{a}_{g,j}-\hat{a}_{g,j+1}\right)\pm e^{-\frac{i}{\hbar}(\epsilon_{e}-\epsilon_{g})t}\left(\hat{a}^{\dagger}_{g,j}-\hat{a}^{\dagger}_{g,j+1}\right)\hat{a}_{e,j+1/2}\nonumber\\
&&=e^{-\frac{i}{\hbar}(\epsilon_{e}-\epsilon_{g})t}\left(e^{\frac{2i}{\hbar}(\epsilon_{e}-\epsilon_{g})t}\hat{a}^{\dagger}_{e,j+1/2}\left(\hat{a}_{g,j}-\hat{a}_{g,j+1}\right)\pm\left(\hat{a}^{\dagger}_{g,j}-\hat{a}^{\dagger}_{g,j+1}\right)\hat{a}_{e,j+1/2}\right)\nonumber\\
\end{eqnarray}
with the exponential time term factored out, $e^{\pm\frac{i}{\hbar}(\epsilon_e -\epsilon_g)t}$, cancelling out by multiplication with the adjacent term whose exponential time term factored out has an opposite sign from the first exponential time term factored out. In doing so, we obtain the following term:
\begin{eqnarray}
&&\int_{0}^{+\infty}dt' Tr_{B}\left[\hat{H}_{SB}(t),\left[\hat{H}_{SB}(t-t'),\hat{\rho}_{S}(t)\otimes\hat{\rho}_{B}\right]\right]\nonumber\\ 
&&=\frac{8\pi\hbar a^{2}_{SB}}{\mu^2 \sqrt{2\pi\sigma_e}}\left(\frac{\rho_B \sigma_g (\epsilon_e - \epsilon_g)}{2m_B c^2 L\sigma_e}x^{2}_a\right)e^{-\frac{x_a^2}{\sigma_e^2}}\left[-\left(e^{\frac{2i}{\hbar}(\epsilon_{e}-\epsilon_{g})t}\sum_{j=1}^{L-1}\hat{a}^{\dagger}_{e,j+1/2}\left(\hat{a}_{g,j}-\hat{a}_{g,j+1}\right)\right.\right.\nonumber\\
&&\left.+\sum_{j=1}^{L-1}\left(\hat{a}^{\dagger}_{g,j}-\hat{a}^{\dagger}_{g,j+1}\right)\hat{a}_{e,j+1/2}\right)\left(\sum_{j'=1}^{L-1}\hat{a}^{\dagger}_{e,j'+1/2}\left(\hat{a}_{g,j'}-\hat{a}_{g,j'+1}\right)\right.\nonumber\\
&&\left.-e^{-\frac{2i}{\hbar}(\epsilon_{e}-\epsilon_{g})t}\sum_{j'=1}^{L}\left(\hat{a}^{\dagger}_{g,j'}-\hat{a}^{\dagger}_{g,j'+1}\right)\hat{a}_{e,j'+1/2}\right)\hat{\rho}_{S}(t)\nonumber\\
&&-\left(\sum_{j=1}^{L-1}\hat{a}^{\dagger}_{e,j+1/2}\left(\hat{a}_{g,j}-\hat{a}_{g,j+1}\right)+e^{-\frac{2i}{\hbar}(\epsilon_{e}-\epsilon_{g})t}\sum_{j=1}^{L-1}\left(\hat{a}^{\dagger}_{g,j}-\hat{a}^{\dagger}_{g,j+1}\right)\hat{a}_{e,j+1/2}\right)\nonumber\\
&&\times\hat{\rho}_{S}(t)\left(e^{\frac{2i}{\hbar}(\epsilon_e -\epsilon_g)t}\sum_{j'=1}^{L-1}\hat{a}^{\dagger}_{e,j'+1/2}\left(\hat{a}_{g,j'}-\hat{a}_{g,j'+1}\right)-\sum_{j'=1}^{L-1}\left(\hat{a}^{\dagger}_{g,j'}-\hat{a}^{\dagger}_{g,j'+1}\right)\hat{a}_{e,j'+1/2}\right)\nonumber\\
&&+\left(\sum_{j'=1}^{L-1}\hat{a}^{\dagger}_{e,j'+1/2}\left(\hat{a}_{g,j'}-\hat{a}_{g,j'+1}\right)-e^{-\frac{2i}{\hbar}(\epsilon_{e}-\epsilon_{g})t}\sum_{j'=1}^{L-1}\left(\hat{a}^{\dagger}_{g,j'}-\hat{a}^{\dagger}_{g,j'+1}\right)\hat{a}_{e,j'+1/2}\right)\nonumber\\
&&\times\hat{\rho}_{S}(t)\left(e^{\frac{2i}{\hbar}(\epsilon_{e}-\epsilon_{g})t}\sum_{j=1}^{L-1}\hat{a}^{\dagger}_{e,j+1/2}\left(\hat{a}_{g,j}-\hat{a}_{g,j+1}\right)+\sum_{j=1}^{L-1}\left(\hat{a}^{\dagger}_{g,j}-\hat{a}^{\dagger}_{g,j+1}\right)\hat{a}_{e,j+1/2}\right)\nonumber\\
&&+\hat{\rho}_{S}(t)\left(e^{\frac{2i}{\hbar}(\epsilon_{e}-\epsilon_{g})t}\sum_{j'=1}^{L-1}\hat{a}^{\dagger}_{e,j'+1/2}\left(\hat{a}_{g,j'}-\hat{a}_{g,j'+1}\right)-\sum_{j'=1}^{L-1}\left(\hat{a}^{\dagger}_{g,j'}-\hat{a}^{\dagger}_{g,j'+1}\right)\hat{a}_{e,j'+1/2}\right)\nonumber\\
&&\times\left.\left(\sum_{j=1}^{L-1}\hat{a}^{\dagger}_{e,j+1/2}\left(\hat{a}_{g,j}-\hat{a}_{g,j+1}\right)+e^{-\frac{2i}{\hbar}(\epsilon_{e}-\epsilon_{g})t}\sum_{j=1}^{L-1}\left(\hat{a}^{\dagger}_{g,j}-\hat{a}^{\dagger}_{g,j+1}\right)\hat{a}_{e,j+1/2}\right)\right]\nonumber\\
\label{masteqrhs4}
\end{eqnarray}
Now let us use adiabatic elimination to express the creation and annihilation operators for the harmonic potential excited states at $x_{j+1/2}$ as linear combinations of creation and annihilation operators for the harmonic potential ground states at $x_j$ and $x_{j+1}$:

\begin{equation}
\hat{a}_{e,j+1/2}=\frac{\Omega_j}{\sqrt{2}\Delta_j}\left(\hat{a}_{g,j}+\hat{a}_{g,j+1}\right),\;\hat{a}^{\dagger}_{e,j+1/2}=\frac{\Omega_j}{\sqrt{2}\Delta_j}\left(\hat{a}^{\dagger}_{g,j}+\hat{a}^{\dagger}_{g,j+1}\right)
\label{adiabelemexcop}
\end{equation}
Substituting Eq. (\ref{adiabelemexcop}) back into Eq. (\ref{mastereqrhs3}), we find that we have terms of the following form emerging in this equation:

\begin{eqnarray}
&&\sum_{j=1}^{L-1}\hat{a}^{\dagger}_{e,j+1/2}\left(\hat{a}_{g,j}-\hat{a}_{g,j+1}\right)=\sum_{j=1}^{L-1}\frac{\Omega_j}{\sqrt{2}\Delta_j}\left(\hat{a}^{\dagger}_{g,j}+\hat{a}^{\dagger}_{g,j+1}\right)\left(\hat{a}_{g,j}-\hat{a}_{g,j+1}\right)\nonumber\\
&&\sum_{j=1}^{L-1}\left(\hat{a}^{\dagger}_{g,j}-\hat{a}^{\dagger}_{g,j+1}\right)\hat{a}_{e,j+1/2}=\sum_{j=1}^{L-1}\frac{\Omega_j}{\sqrt{2}\Delta_j}\left(\hat{a}^{\dagger}_{g,j}-\hat{a}^{\dagger}_{g,j+1}\right)\left(\hat{a}_{g,j}+\hat{a}_{g,j+1}\right)\nonumber\\
\label{adiabelemoppord}
\end{eqnarray}
For both of these terms, we factor out the terms $\frac{\Omega_1}{\Delta_1}$ which comes from the first term in the series, and the numerical factors $1/\sqrt{2}$ from each term in the series, giving us

\begin{eqnarray}
&&\sum_{j=1}^{L-1}\hat{a}^{\dagger}_{e,j+1/2}\left(\hat{a}_{g,j}-\hat{a}_{g,j+1}\right)=\frac{\Omega_1}{\sqrt{2}\Delta_1}\sum_{j=1}^{L-1}\epsilon_{j}\left(\hat{a}^{\dagger}_{g,j}+\hat{a}^{\dagger}_{g,j+1}\right)\left(\hat{a}_{g,j}-\hat{a}_{g,j+1}\right)\nonumber\\
&&\sum_{j=1}^{L-1}\left(\hat{a}^{\dagger}_{g,j}-\hat{a}^{\dagger}_{g,j+1}\right)\hat{a}_{e,j+1/2}=\frac{\Omega_1}{\sqrt{2}\Delta_1}\sum_{j=1}^{L-1}\epsilon_{j}\left(\hat{a}^{\dagger}_{g,j}-\hat{a}^{\dagger}_{g,j+1}\right)\left(\hat{a}_{g,j}+\hat{a}_{g,j+1}\right)\nonumber\\
\label{adiabelemoppord2}
\end{eqnarray}
Here, the coefficients $\epsilon_j$ have the form
\begin{equation}
\epsilon_j = \frac{\Omega_j}{\Omega_1}\frac{\Delta_1}{\Delta_j},
\label{epsilonj}
\end{equation}
with $\epsilon_1 = 1$ since we factored out $\frac{\Omega_1}{\Delta_1}$. Finally, we make the following identification for the sum of the products of these annihilation and creation operators:

\begin{equation}
\hat{c}=\sum_{j=1}^{L-1}\epsilon_{j}\left(\hat{a}^{\dagger}_{g,j}+\hat{a}^{\dagger}_{g,j+1}\right)\left(\hat{a}_{g,j}-\hat{a}_{g,j+1}\right),\;\hat{c}^{\dagger}=\sum_{j=1}^{L-1}\epsilon_{j}\left(\hat{a}^{\dagger}_{g,j}-\hat{a}^{\dagger}_{g,j+1}\right)\left(\hat{a}_{g,j}+\hat{a}_{g,j+1}\right)
\label{jumpopdef}
\end{equation}
These operators, known as the jump operators for this open quantum system, allows us to write Eq. (\ref{masteqrhs4}) in the following much simplified form:

\begin{eqnarray}
&&\int_{0}^{+\infty}dt' Tr_{B}\left[\hat{H}_{SB}(t),\left[\hat{H}_{SB}(t-t'),\hat{\rho}_{S}(t)\otimes\hat{\rho}_{B}\right]\right]\nonumber\\ 
&&=\frac{4\pi\hbar a^{2}_{SB}}{\mu^2 \sqrt{2\pi\sigma_e}}\left(\frac{\rho_B \sigma_g (\epsilon_e - \epsilon_g)}{2m_B c^2 L\sigma_e}x^{2}_a\right)\frac{\Omega_1^2}{\Delta_1^2}e^{-\frac{x_a^2}{\sigma_e^2}}\left[-\left(e^{\frac{2i}{\hbar}(\epsilon_{e}-\epsilon_{g})t}\hat{c}+\hat{c}^{\dagger}\right)\left(\hat{c}-e^{-\frac{2i}{\hbar}(\epsilon_{e}-\epsilon_{g})t}\hat{c}^{\dagger}\right)\hat{\rho}_{S}(t)\right.\nonumber\\
&&-\left(\hat{c}+e^{-\frac{2i}{\hbar}(\epsilon_{e}-\epsilon_{g})t}\hat{c}^{\dagger}\right)\hat{\rho}_{S}(t)\left(e^{\frac{2i}{\hbar}(\epsilon_e -\epsilon_g)t}\hat{c}-\hat{c}^{\dagger}\right)+\left(\hat{c}-e^{-\frac{2i}{\hbar}(\epsilon_{e}-\epsilon_{g})t}\hat{c}^{\dagger}\right)\hat{\rho}_{S}(t)\left(e^{\frac{2i}{\hbar}(\epsilon_{e}-\epsilon_{g})t}\hat{c}+\hat{c}^{\dagger}\right)\nonumber\\
&&\nonumber\\
&&\left.+\hat{\rho}_{S}(t)\left(e^{\frac{2i}{\hbar}(\epsilon_{e}-\epsilon_{g})t}\hat{c}-\hat{c}^{\dagger}\right)\left(\hat{c}+e^{-\frac{2i}{\hbar}(\epsilon_{e}-\epsilon_{g})t}\hat{c}^{\dagger}\right)\right]\nonumber\\
\label{masteqrhs5}
\end{eqnarray}

Now let us define the following constant that appears in Eq. (\ref{masteqrhs5}):

\begin{equation}
\gamma=\frac{4\pi\hbar a^{2}_{SB}}{\mu^2 \sqrt{2\pi\sigma_e}}\left(\frac{\rho_B \sigma_g (\epsilon_e - \epsilon_g)}{2m_B c^2 L\sigma_e}\right)\left(\frac{\Omega_1}{\Delta_1}x_a\right)^{2}\exp\left(-\frac{x_a^2}{\sigma_e^2}\right)
\label{gammaconst}
\end{equation}
Carrying out the operator multiplications in Eq. (\ref{masteqrhs5}) and substituting Eq. (\ref{gammaconst}) in Eq. (\ref{masteqrhs5}), we obtain the following equation:

\begin{eqnarray}
&&\int_{0}^{+\infty}dt' Tr_{B}\left[\hat{H}_{SB}(t),\left[\hat{H}_{SB}(t-t'),\hat{\rho}_{S}(t)\otimes\hat{\rho}_{B}\right]\right]\nonumber\\ 
&&=\gamma\left[-\left(\hat{c}^{\dagger}\hat{c}-\hat{c}\hat{c}^{\dagger}+e^{\frac{2i}{\hbar}(\epsilon_{e}-\epsilon_{g})t}\hat{c}\hat{c}-e^{-\frac{2i}{\hbar}(\epsilon_{e}-\epsilon_{g})t}\hat{c}^{\dagger}\hat{c}^{\dagger}\right)\hat{\rho}_{S}(t)\right.\nonumber\\
&&-\left(\hat{c}^{\dagger}\hat{\rho}_{S}(t)\hat{c}-\hat{c}\hat{\rho}_{S}(t)\hat{c}^{\dagger}+e^{\frac{2i}{\hbar}(\epsilon_e-\epsilon_g)t}\hat{c}\hat{\rho}_{S}(t)\hat{c}-e^{-\frac{2i}{\hbar}(\epsilon_e-\epsilon_g)t}\hat{c}^{\dagger}\hat{\rho}_{S}(t)\hat{c}^{\dagger}\right)\nonumber\\
&&+\left(\hat{c}\hat{\rho}_{S}(t)\hat{c}^{\dagger}-\hat{c}^{\dagger}\hat{\rho}_{S}(t)\hat{c}+e^{\frac{2i}{\hbar}(\epsilon_e -\epsilon_g)t}\hat{c}\hat{\rho}_{S}(t)\hat{c}-e^{\frac{-2i}{\hbar}(\epsilon_e -\epsilon_g)t}\hat{c}^{\dagger}\hat{\rho}_{S}(t)\hat{c}^{\dagger}\right)\nonumber\\
&&\left.+\hat{\rho}_{S}(t)\left(\hat{c}\hat{c}^{\dagger}-\hat{c}^{\dagger}\hat{c}+e^{\frac{2i}{\hbar}(\epsilon_{e}-\epsilon_{g})t}\hat{c}\hat{c}-e^{-\frac{2i}{\hbar}(\epsilon_{e}-\epsilon_{g})t}\hat{c}^{\dagger}\hat{c}^{\dagger}\right)\right]\nonumber\\
\label{masteqrhs6}
\end{eqnarray}
Finally, substituting this into the right-hand side of the Born-Markov form of the master equation for the open quantum system given by Eq. (\ref{bornmarkovmasteq}) and rearranging terms in this equation, we finally obtain the following form of the master equation describing the time evolution of the ultracold atom gas trapped in an array of $2L+1$ harmonic potentials and coupled to a background BEC that acts as a reservoir of excitations emitted by this trapped ultracold atom gas as they return to their original energy levels after being excited to higher energy levels by multiple laser fields:

\begin{eqnarray}
&&\frac{d}{dt}\hat{\rho}_{S}(t)=-\gamma\left[\left(2\hat{c}\hat{\rho}_{S}(t)\hat{c}^{\dagger}-\left\{\hat{c}^{\dagger}\hat{c},\hat{\rho}_{S}(t)\right\}\right)-\left(2\hat{c}^{\dagger}\hat{\rho}_{S}(t)\hat{c}-\left\{\hat{c}\hat{c}^{\dagger},\hat{\rho}_{S}(t)\right\}\right)\right.\nonumber\\
&&\left.-e^{\frac{2i}{\hbar}(\epsilon_{e}-\epsilon_{g})t}\left[\hat{c}\hat{c},\hat{\rho}_{S}(t)\right]+e^{-\frac{2i}{\hbar}(\epsilon_{e}-\epsilon_{g})t}\left[\hat{c}^{\dagger}\hat{c}^{\dagger},\hat{\rho}_{S}(t)\right]\right]\nonumber\\
\label{masteqnharmonicarraytrappedultracoldatomoqs}
&&\end{eqnarray}

\section{Steady States of Trapped Ultracold Atom Open Quantum System}

Having determined the master equation of the trappd ultracold atom open quantum system given by Eq. (\ref{masteqnharmonicarraytrappedultracoldatomoqs}), let us now determine the system's steady states using this master equation. We first hypothesize that those steady states will have the following forms:

\begin{equation}
\hat{\rho}_{SS,R} = \otimes_{j=1}^{L-1}\left|0\right\rangle\left\langle 0\right|_{j}\otimes\left|N\right\rangle\left\langle N\right|_{L}
\label{ssright}
\end{equation}
\begin{equation}
\hat{\rho}_{SS,L} = \left|N\right\rangle\left\langle N\right|_{1}\otimes_{j=2}^{L}\left|0\right\rangle\left\langle 0\right|_{j}
\label{ssleft}
\end{equation}

Let us now show that these, indeed, are steady states of the system with respect to the position of the harmonic trapping potential in the array. Before we do so, we first define what a steady state of the trapped ultracold atom system with respect to the position of the harmonic trapping potential in the array is. First, we define the particle number operator describing the number of atoms trapped in the $j$th harmonic potential of the array as

\begin{equation}
\hat{N}_{j}=\mathbb{I}_{1}\otimes\mathbb{I}_{2}\otimes\cdots\otimes\hat{a}^{\dagger}_{g,j}\hat{a}_{g,j}\otimes\mathbb{I}_{j+1}\otimes\cdots\otimes\mathbb{I}_{L}
\label{partnumopjthhtp}
\end{equation}
Here, $\mathbb{I}_{n}$ is the identity operator for the $n$th harmonic trapping potential of the system, and $\hat{a}^{\dagger}_{g,j}\hat{a}_{g,j}$ can be written

\begin{equation}
\hat{a}^{\dagger}_{g,j}\hat{a}_{g,j}=\sum_{k=1}^{N}k\left|k\right\rangle\left\langle k\right|_j
\end{equation}

The expectation value for the particle number in the $j$th harmonic potential of the array is then

\begin{equation}
\left\langle\hat{N}_{j}\right\rangle=Tr_{j}\left(\hat{N}_{j}\hat{\rho}\right)
\end{equation}
Now if the density matrix describing the system can be written in block diagonal form as

\begin{equation}
\hat{\rho}=\otimes_{j=1}^{L}\hat{\rho}_{j}=\hat{\rho}_{1}\otimes\hat{\rho}_{2}\otimes\cdots\otimes\hat{\rho}_{j}\otimes\cdots\hat{\rho}_{L}
\end{equation}
where the $j$th block of the density matrix describing states in the $j$th harmonic trapping potential of the array is a square $N\times N$ matrix (where $N$ is the total number of particles in the system) and has the explicit form
\begin{equation}
\hat{\rho}_{j}=\sum_{k,l=1}^{N}(\rho_{kl})_{j}\left|k\right\rangle\left\langle l\right|_{j}
\end{equation}
and $(\rho_{kl})_{j}$ are the components of this $j$th block of the density matrix $\hat{\rho}$ describing the system, then

\begin{equation}
\left\langle\hat{N}_{j}\right\rangle=Tr_{j}\left(\hat{a}^{\dagger}_{g,j}\hat{a}_{g,j}\hat{\rho}\right)=\sum_{k=1}^{N}k(\rho_{kk})_{j}
\label{expvalpartnumjthhtp}
\end{equation}

\subsection{Demonstration that $\hat{\rho}_{SS,R}$ is a position steady state of the trapped ultracold atom open quantum system}

Now let us note that the action of the jump operators $\hat{c}$ and $\hat{c}^{\dagger}$ on these hypothesized steady states will have the following form:

\begin{eqnarray}
&&\hat{c}\hat{\rho}_{SS,R}=\left(\sum_{j=1}^{L-1}\epsilon_{j}(\hat{a}^{\dagger}_{g,j}+\hat{a}^{\dagger}_{g,j+1})(\hat{a}_{g,j}-\hat{a}_{g,j+1})\right)\left(\otimes_{j=1}^{L-1}\left|0\right\rangle\left\langle 0\right|_{j}\otimes\left|N\right\rangle\left\langle N\right|_{L}\right)\nonumber\\
&&=-\epsilon_{L-1}\left(\otimes_{j=1}^{L-1}\left|0\right\rangle\left\langle 0\right|_{j}\otimes N\left|N\right\rangle\left\langle N\right|_{L}+\otimes_{j=1}^{L-2}\left|0\right\rangle\left\langle 0\right|_{j}\otimes\left|1\right\rangle\left\langle 0\right|_{L-1}\otimes\sqrt{N}\left|N-1\right\rangle\left\langle N\right|_{L}\right)\nonumber\\
\label{jumpoprightssright}
\end{eqnarray}
\begin{eqnarray}
&&\hat{\rho}_{SS,R}\hat{c}^{\dagger}=\left(\otimes_{j=1}^{L-1}\left|0\right\rangle\left\langle 0\right|_{j}\otimes\left|N\right\rangle\left\langle N\right|_{L}\right)\left(\sum_{j=1}^{L-1}\epsilon_{j}(\hat{a}^{\dagger}_{g,j}+\hat{a}^{\dagger}_{g,j+1})(\hat{a}_{g,j}-\hat{a}_{g,j+1})\right)\nonumber\\
&&=-\epsilon_{L-1}\left(\otimes_{j=1}^{L-1}\left|0\right\rangle\left\langle 0\right|_{j}\otimes N\left|N\right\rangle\left\langle N\right|_{L}+\otimes_{j=1}^{L-2}\left|0\right\rangle\left\langle 0\right|_{j}\otimes\left|0\right\rangle\left\langle 1\right|_{L-1}\otimes\sqrt{N}\left|N\right\rangle\left\langle N-1\right|_{L}\right)\nonumber\\
\label{jumpopconjleftssright}
\end{eqnarray}
\begin{eqnarray}
&&\hat{c}^{\dagger}\hat{\rho}_{SS,R}=\left(\sum_{j=1}^{L-1}\epsilon_{j}(\hat{a}^{\dagger}_{g,j}-\hat{a}^{\dagger}_{g,j+1})(\hat{a}_{g,j}+\hat{a}_{g,j+1})\right)\left(\otimes_{j=1}^{L-1}\left|0\right\rangle\left\langle 0\right|_{j}\otimes\left|N\right\rangle\left\langle N\right|_{L}\right)\nonumber\\
&&=-\epsilon_{L-1}\left(\otimes_{j=1}^{L-1}\left|0\right\rangle\left\langle 0\right|_{j}\otimes N\left|N\right\rangle\left\langle N\right|_{L}-\otimes_{j=1}^{L-2}\left|0\right\rangle\left\langle 0\right|_{j}\otimes\left|1\right\rangle\left\langle 0\right|_{L-1}\otimes\sqrt{N}\left|N-1\right\rangle\left\langle N\right|_{L}\right)\nonumber\\
\label{jumpopconjrightssright}
\end{eqnarray}
\begin{eqnarray}
&&\hat{\rho}_{SS,R}\hat{c}=\left(\otimes_{j=1}^{L-1}\left|0\right\rangle\left\langle 0\right|_{j}\otimes\left|N\right\rangle\left\langle N\right|_{L}\right)\left(\sum_{j=1}^{L-1}\epsilon_{j}(\hat{a}^{\dagger}_{g,j}+\hat{a}^{\dagger}_{g,j+1})(\hat{a}_{g,j}-\hat{a}_{g,j+1})\right)\nonumber\\
&&=-\epsilon_{L-1}\left(\otimes_{j=1}^{L-1}\left|0\right\rangle\left\langle 0\right|_{j}\otimes N\left|N\right\rangle\left\langle N\right|_{L}-\otimes_{j=1}^{L-2}\left|0\right\rangle\left\langle 0\right|_{j}\otimes\left|0\right\rangle\left\langle 1\right|_{L-1}\otimes\sqrt{N}\left|N\right\rangle\left\langle N-1\right|_{L}\right)\nonumber\\
\label{jumpopleftssright}
\end{eqnarray}
We then substitute $\hat{\rho}_{SS,R}$, given by Eq. (\ref{ssright}), into the master equation describing the dynamics of the trapped ultracold atom open quantum system, given by Eq. (\ref{masteqnharmonicarraytrappedultracoldatomoqs}), and use Eqs. (\ref{jumpoprightssright}), (\ref{jumpopconjleftssright}), (\ref{jumpopconjrightssright}) and (\ref{jumpopleftssright}) as a guide in evaluating the resulting expressions in the master equation. In doing so, we obtain the following:

\begin{eqnarray}
&&\frac{d}{dt}\hat{\rho}_{SS,R}=-\gamma\left(4\epsilon_{L-1}^{2}\left[\otimes_{j=1}^{L-2}\left|0\right\rangle\left\langle 0\right|_{j}\otimes\left|0\right\rangle\left\langle 1\right|_{L-1}\otimes N\sqrt{N}\left|N\right\rangle\left\langle N-1\right|_{L}+\otimes_{j=1}^{L-2}\left|0\right\rangle\left\langle 0\right|_{j}\right.\right.\nonumber\\
&&\left.\otimes\left|0\right\rangle\left\langle 1\right|_{L-1}\otimes N\sqrt{N}\left|N\right\rangle\left\langle N-1\right|_{L}\right]-2\epsilon_{L-1}^{2}\left[\otimes_{j=1}^{L-2}\left|0\right\rangle\left\langle 0\right|_{j}\otimes\left|1\right\rangle\left\langle 0\right|_{L-1}\right.\nonumber\\
&&\otimes\sqrt{N}(N-1)\left|N\right\rangle\left\langle N-1\right|_{L}-\otimes_{j=1}^{L-2}\left|0\right\rangle\left\langle 0\right|_{j}\otimes\left|1\right\rangle\left\langle 0\right|_{L-1}\otimes \sqrt{N}\left|N-1\right\rangle\left\langle N\right|_{L}\nonumber\\
&&-\otimes_{j=1}^{L-2}\left|0\right\rangle\left\langle 0\right|_{j}\otimes\sqrt{2}\left|2\right\rangle\left\langle 0\right|_{L-1}\otimes\sqrt{N(N-1)}\left|N-2\right\rangle\left\langle N\right|_{L}-\otimes_{j=1}^{L-2}\left|0\right\rangle\left\langle 0\right|_{j}\nonumber\\
&&-\otimes_{j=1}^{L-2}\left|0\right\rangle\left\langle 0\right|_{j}\otimes\left|1\right\rangle\left\langle 0\right|_{L-1}\otimes N\sqrt{N}\left|N-1\right\rangle\left\langle N\right|_{L}+\otimes_{j=1}^{L-2}\left|0\right\rangle\left\langle 0\right|_{j}\otimes\left|0\right\rangle\left\langle 1\right|_{L-1}\nonumber\\
&&\otimes\sqrt{N}(N-1)\left|N\right\rangle\left\langle N-1\right|_{L}-\otimes_{j=1}^{L-2}\left|0\right\rangle\left\langle 0\right|_{j}\otimes\left|0\right\rangle\left\langle 1\right|_{L-1}\otimes\sqrt{N}\left|N\right\rangle\left\langle N-1\right|_{L}\nonumber\\
&&-\otimes_{j=1}^{L-2}\left|0\right\rangle\left\langle 0\right|_{j}\otimes\sqrt{2}\left|0\right\rangle\left\langle 2\right|_{L-1}\otimes\sqrt{N(N-1)}\left|N\right\rangle\left\langle N-2\right|_{L}-\otimes_{j=1}^{L-2}\left|0\right\rangle\left\langle 0\right|_{j}\nonumber\\
&&\left.\otimes\left|0\right\rangle\left\langle 1\right|_{L-1}\otimes N\sqrt{N}\left|N\right\rangle\left\langle N-1\right|_{L}\right]+2\cos\left(\frac{2}{\hbar}(\epsilon_{e}-\epsilon_{g})t\right)\epsilon_{L-1}^{2}\left[\otimes_{j=1}^{L-2}\left|0\right\rangle\left\langle 0\right|_{j}\otimes\left|1\right\rangle\left\langle 0\right|_{L-1}\right.\nonumber\\
&&\otimes\sqrt{N}\left|N-1\right\rangle\left\langle N\right|_{L}-\otimes_{j=1}^{L-2}\left|0\right\rangle\left\langle 0\right|_{j}\otimes\sqrt{2}\left|2\right\rangle\left\langle 0\right|_{L-1}\otimes\sqrt{N(N-1)}\left|N-2\right\rangle\left\langle N\right|_{L}\nonumber\\
&&-\otimes_{j=1}^{L-2}\left|0\right\rangle\left\langle 0\right|_{j}\otimes\left|1\right\rangle\left\langle 0\right|_{L-1}\otimes N\sqrt{N}\left|N-1\right\rangle\left\langle N\right|_{L}-\otimes_{j=1}^{L-2}\left|0\right\rangle\left\langle 0\right|_{j}\otimes\left|1\right\rangle\left\langle 0\right|_{L-1}\nonumber\\
&&\otimes\sqrt{N}(N-1)\left|N-1\right\rangle\left\langle N\right|_{L}+\otimes_{j=1}^{L-2}\left|0\right\rangle\left\langle 0\right|_{j}\otimes\left|0\right\rangle\left\langle 1\right|_{L-1}\otimes\sqrt{N}\left|N\right\rangle\left\langle N-1\right|_{L}-\otimes_{j=1}^{L-2}\left|0\right\rangle\left\langle 0\right|_{j}\nonumber\\
&&\otimes\sqrt{2}\left|0\right\rangle\left\langle 2\right|_{L-1}\otimes\sqrt{N(N-1)}\left|N\right\rangle\left\langle N-2\right|_{L}-\otimes_{j=1}^{L-2}\left|0\right\rangle\left\langle 0\right|_{j}\otimes\left|0\right\rangle\left\langle 1\right|_{L-1}\nonumber\\
&&\otimes N\sqrt{N}\left|N\right\rangle\left\langle N-1\right|_{L}-\otimes_{j=1}^{L-2}\left|0\right\rangle\left\langle 0\right|_{j}\otimes\left|0\right\rangle\left\langle 1\right|_{L-1}\nonumber\\
&&\left.\left.\otimes\sqrt{N}(N-1)\left|N\right\rangle\left\langle N-1\right|_{L}\right]\right)\nonumber\\
\label{masteqrhsrightss}
\end{eqnarray}
Notice that the right hand side of the master equation for this state of the system is a zero-diagonal density matrix. Consequently, taking the expectation value of the trapping potential position for the resulting state gives us
\begin{eqnarray}
&&\left\langle\hat{N}\left(-\gamma\left[\left(2\hat{c}\hat{\rho}_{SS,R}\hat{c}^{\dagger}-\left\{\hat{c}^{\dagger}\hat{c},\hat{\rho}_{SS,R}\right\}\right)-\left(2\hat{c}^{\dagger}\hat{\rho}_{SS,R}\hat{c}-\left\{\hat{c}\hat{c}^{\dagger},\hat{\rho}_{SS,R}\right\}\right)\right.\right.\right.\nonumber\\
&&\left.\left.\left.-e^{\frac{2i}{\hbar}(\epsilon_{e}-\epsilon_{g})t}\left[\hat{c}\hat{c},\hat{\rho}_{SS,R}\right]+e^{-\frac{2i}{\hbar}(\epsilon_{e}-\epsilon_{g})t}\left[\hat{c}^{\dagger}\hat{c}^{\dagger},\hat{\rho}_{SS,R}\right]\right]\right)\right\rangle=0
\label{expvalposliouvssr}
\end{eqnarray}

This implies that, if we are to compute for the expectation value of the position of the trapping potential containing this time-evolved state, we obtain

\begin{equation}
\left\langle\frac{d}{dt}\left(\hat{N}\hat{\rho}_{SS,R}\right)\right\rangle=0
\label{expvalpostimeevossr}
\end{equation}
Therefore, the expectation value of the position of the trapping potential containing the state $\hat{\rho}_{SS,R}$, $\left\langle\hat{N}\hat{\rho}_{SS,R}\right\rangle$, remains constant over time, making it a steady state of the system with respect to the position of the trapping potential containing the state, described by the observable $\hat{N}=\sum_{j=1}^{L}\hat{a}^{\dagger}_{g,j}\hat{a}_{g,j}$.

\subsection{Demonstration that $\hat{\rho}_{SS,L}$ is a position steady state of the trapped ultracold atom open quantum system}

We now establish that $\hat{\rho}_{SS,L}$ is a steady state of the trapped ultracold atom open quantum system with respect to the expectation value of the position of the trapping potential. To do so, we first note that the action of the jump operators $\hat{c}$ and $\hat{c}^{\dagger}$ on these hypothesized steady states will have the following form:

\begin{eqnarray}
&&\hat{c}\hat{\rho}_{SS,L}=\left(\sum_{j=1}^{L-1}\epsilon_{j}(\hat{a}^{\dagger}_{g,j}+\hat{a}^{\dagger}_{g,j+1})(\hat{a}_{g,j}-\hat{a}_{g,j+1})\right)\left(\left|N\right\rangle\left\langle N\right|_{1}\otimes_{j=2}^{L}\left|0\right\rangle\left\langle 0\right|_{j}\right)\nonumber\\
&&=\epsilon_{1}\left(N\left|N\right\rangle\left\langle N\right|_{1}\otimes_{j=2}^{L}\left|0\right\rangle\left\langle 0\right|_{j}+\sqrt{N}\left|N-1\right\rangle\left\langle N\right|_{1}\otimes\left|1\right\rangle\left\langle 0\right|_{2}\otimes_{j=3}^{L}\left|0\right\rangle\left\langle 0\right|_{j}\otimes\left|0\right\rangle\left\langle 0\right|_{j}\right)\nonumber\\
\label{jumpoprightssleft}
\end{eqnarray}
\begin{eqnarray}
&&\hat{\rho}_{SS,L}\hat{c}^{\dagger}=\left(\left|N\right\rangle\left\langle N\right|_{1}\otimes_{j=2}^{L}\left|0\right\rangle\left\langle 0\right|_{j}\right)\left(\sum_{j=1}^{L-1}\epsilon_{j}(\hat{a}^{\dagger}_{g,j}+\hat{a}^{\dagger}_{g,j+1})(\hat{a}_{g,j}-\hat{a}_{g,j+1})\right)\nonumber\\
&&=\epsilon_{1}\left(N\left|N\right\rangle\left\langle N\right|_{1}\otimes_{j=2}^{L}\left|0\right\rangle\left\langle 0\right|_{j}-\sqrt{N}\left|N\right\rangle\left\langle N-1\right|_{1}\otimes\left|0\right\rangle\left\langle 1\right|_{2}\otimes_{j=3}^{L}\left|0\right\rangle\left\langle 0\right|_{j}\right)\nonumber\\
\label{jumpopconjleftssleft}
\end{eqnarray}
\begin{eqnarray}
&&\hat{c}^{\dagger}\hat{\rho}_{SS,L}=\left(\sum_{j=1}^{L-1}\epsilon_{j}(\hat{a}^{\dagger}_{g,j}-\hat{a}^{\dagger}_{g,j+1})(\hat{a}_{g,j}+\hat{a}_{g,j+1})\right)\left(\left|N\right\rangle\left\langle N\right|_{1}\otimes_{j=2}^{L}\left|0\right\rangle\left\langle 0\right|_{j}\right)\nonumber\\
&&=\epsilon_{1}\left(N\left|N\right\rangle\left\langle N\right|_{1}\otimes_{j=2}^{L}\left|0\right\rangle\left\langle 0\right|_{j}-\sqrt{N}\left|N-1\right\rangle\left\langle N\right|_{1}\otimes\left|1\right\rangle\left\langle 0\right|_{2}\otimes_{j=3}^{L}\left|0\right\rangle\left\langle 0\right|_{j}\right)\nonumber\\
\label{jumpopconjrightssleft}
\end{eqnarray}
\begin{eqnarray}
&&\hat{\rho}_{SS,L}\hat{c}=\left(\left|N\right\rangle\left\langle N\right|_{1}\otimes_{j=2}^{L}\left|0\right\rangle\left\langle 0\right|_{j}\right)\left(\sum_{j=1}^{L-1}\epsilon_{j}(\hat{a}^{\dagger}_{g,j}+\hat{a}^{\dagger}_{g,j+1})(\hat{a}_{g,j}-\hat{a}_{g,j+1})\right)\nonumber\\
&&=\epsilon_{1}\left(N\left|N\right\rangle\left\langle N\right|_{1}\otimes_{j=2}^{L}\left|0\right\rangle\left\langle 0\right|_{j}+\sqrt{N}\left|N\right\rangle\left\langle N-1\right|_{1}\otimes\left|0\right\rangle\left\langle 1\right|_{2}\otimes_{j=3}^{L}\left|0\right\rangle\left\langle 0\right|_{j}\right)\nonumber\\
\label{jumpopleftssleft}
\end{eqnarray}
We then substitute $\hat{\rho}_{SS,L}$, given by Eq. (\ref{ssleft}), into the master equation describing the dynamics of the trapped ultracold atom open quantum system, given by Eq. (\ref{masteqnharmonicarraytrappedultracoldatomoqs}), and use Eqs. (\ref{jumpoprightssleft}), (\ref{jumpopconjleftssleft}), (\ref{jumpopconjrightssleft}) and (\ref{jumpopleftssleft}) as a guide in evaluating the resulting expressions in the master equation. In doing so, we obtain the following:

\begin{eqnarray}
&&\frac{d}{dt}\hat{\rho}_{SS,L}=-\gamma\left(4\epsilon_{1}^{2}\left[N\sqrt{N}\left|N\right\rangle\left\langle N-1\right|_{1}\otimes\left|0\right\rangle\left\langle 1\right|_{2}\otimes_{j=3}^{L}\left|0\right\rangle\left\langle 0\right|_{j}+N\sqrt{N}\left|N-1\right\rangle\left\langle N\right|_{1}\right.\right.\nonumber\\
&&\left.\otimes\left|1\right\rangle\left\langle 0\right|_{2}\otimes_{j=3}^{L}\left|0\right\rangle\left\langle 0\right|_{j}\right]-2\epsilon_{1}^{2}\left[\sqrt{N}(N-1)\left|N-1\right\rangle\left\langle N\right|_{1}\otimes\left|1\right\rangle\left\langle 0\right|_{2}\otimes_{j=3}^{L}\left|0\right\rangle\left\langle 0\right|_{j}\right.\nonumber\\
&&-N\sqrt{N}\left|N-1\right\rangle\left\langle N\right|_{1}\otimes\left|1\right\rangle\left\langle 0\right|_{2}\otimes_{j=3}^{L}\left|0\right\rangle\left\langle 0\right|_{j}-\sqrt{N(N-1)}\left|N-2\right\rangle\left\langle N\right|_{1}\otimes\sqrt{2}\left|2\right\rangle\left\langle 0\right|_{2}\nonumber\\
&&\otimes_{j=3}^{L}\left|0\right\rangle\left\langle 0\right|_{j}-N\sqrt{N}\left|N-1\right\rangle\left\langle N\right|_{1}\otimes\left|1\right\rangle\left\langle 0\right|_{L-1}\otimes_{j=3}^{L}\left|0\right\rangle\left\langle 0\right|_{j}+\sqrt{N}(N-1)\left|N\right\rangle\left\langle N-1\right|_{1}\nonumber\\
&&\otimes\left|0\right\rangle\left\langle 1\right|_{2}\otimes_{j=3}^{L}\left|0\right\rangle\left\langle 0\right|_{j}-\sqrt{N}\left|N\right\rangle\left\langle N-1\right|_{1}\otimes\left|0\right\rangle\left\langle 1\right|_{2}\otimes_{j=3}^{L}\left|0\right\rangle\left\langle 0\right|_{j}-\sqrt{N(N-1)}\left|N\right\rangle\left\langle N-2\right|_{1}\nonumber\\
&&\left.\otimes\sqrt{2}\left|0\right\rangle\left\langle 2\right|_{2}\otimes_{j=3}^{L}\left|0\right\rangle\left\langle 0\right|_{j}-N\sqrt{N}\left|N\right\rangle\left\langle N-1\right|_{1}\otimes\left|0\right\rangle\left\langle 1\right|_{L-1}\otimes_{j=3}^{L}\left|0\right\rangle\left\langle 0\right|_{j}\right]\nonumber\\
&&-2\cos\left(\frac{2}{\hbar}(\epsilon_{e}-\epsilon_{g})t\right)\epsilon_{1}^{2}\left[\sqrt{N}(N-1)\left|N-1\right\rangle\left\langle N\right|_{1}\otimes\left|1\right\rangle\left\langle 0\right|_{2}\otimes_{j=3}^{L}\left|0\right\rangle\left\langle 0\right|_{j}\right.\nonumber\\
&&-\sqrt{N(N-1)}\left|N-2\right\rangle\left\langle N\right|_{1}\otimes\sqrt{2}\left|2\right\rangle\left\langle 0\right|_{2}\otimes_{j=3}^{L}\left|0\right\rangle\left\langle 0\right|_{j}+N\sqrt{N}\left|N-1\right\rangle\left\langle N\right|_{1}\otimes\left|1\right\rangle\left\langle 0\right|_{2}\nonumber\\
&&\otimes_{j=3}^{L}\left|0\right\rangle\left\langle 0\right|_{j}-\sqrt{N}\left|N-1\right\rangle\left\langle N\right|_{1}\otimes\left|1\right\rangle\left\langle 0\right|_{2}\otimes_{j=3}^{L}\left|0\right\rangle\left\langle 0\right|_{j}+\sqrt{N}(N-1)\left|N\right\rangle\left\langle N-1\right|_{1}\otimes\left|0\right\rangle\left\langle 1\right|_{2}\nonumber\\
&&\otimes_{j=3}^{L}\left|0\right\rangle\left\langle 0\right|_{j}-\sqrt{N}\left|N\right\rangle\left\langle N-1\right|_{1}\otimes\left|0\right\rangle\left\langle 1\right|_{2}\otimes_{j=3}^{L}\left|0\right\rangle\left\langle 0\right|_{j}-\sqrt{N(N-1)}\left|N\right\rangle\left\langle N-2\right|_{1}\nonumber\\
&&\left.\left.\otimes\sqrt{2}\left|0\right\rangle\left\langle 2\right|_{2}\otimes_{j=3}^{L}\left|0\right\rangle\left\langle 0\right|_{j}+N\sqrt{N}\left|N\right\rangle\left\langle N-1\right|_{1}\otimes\left|0\right\rangle\left\langle 1\right|_{2}\otimes_{j=3}^{L}\left|0\right\rangle\left\langle 0\right|_{j}\otimes\left|0\right\rangle\left\langle 1\right|_{L-1}\right]\right)\nonumber\\
\label{masteqrhsleftss}
\end{eqnarray}
Just as with the steady state $\hat{\rho}_{SS,R}$, the right hand side of the master equation for this state of the system is a zero-diagonal density matrix. Consequently, taking the expectation value of the trapping potential position for the resulting state gives us
\begin{eqnarray}
&&\left\langle\hat{N}\left(-\gamma\left[\left(2\hat{c}\hat{\rho}_{SS,L}\hat{c}^{\dagger}-\left\{\hat{c}^{\dagger}\hat{c},\hat{\rho}_{SS,L}\right\}\right)-\left(2\hat{c}^{\dagger}\hat{\rho}_{SS,L}\hat{c}-\left\{\hat{c}\hat{c}^{\dagger},\hat{\rho}_{SS,L}\right\}\right)\right.\right.\right.\nonumber\\
&&\left.\left.\left.-e^{\frac{2i}{\hbar}(\epsilon_{e}-\epsilon_{g})t}\left[\hat{c}\hat{c},\hat{\rho}_{SS,L}\right]+e^{-\frac{2i}{\hbar}(\epsilon_{e}-\epsilon_{g})t}\left[\hat{c}^{\dagger}\hat{c}^{\dagger},\hat{\rho}_{SS,L}\right]\right]\right)\right\rangle=0
\label{expvalposliouvssl}
\end{eqnarray}

This implies that, if we are to compute for the expectation value of the position of the trapping potential containing this time-evolved state, we obtain

\begin{equation}
\left\langle\frac{d}{dt}\left(\hat{N}\hat{\rho}_{SS,L}\right)\right\rangle=0
\label{expvalpostimeevossl}
\end{equation}
Therefore, the expectation value of the position of the trapping potential containing the state $\hat{\rho}_{SS,L}$, $\left\langle\hat{N}\hat{\rho}_{SS,L}\right\rangle$, also remains constant over time, making it another steady state of the system with respect to the position of the trapping potential containing the state, described by the observable $\hat{N}=\sum_{j=1}^{L}\hat{a}^{\dagger}_{g,j}\hat{a}_{g,j}$.

\section{Numerical Results}

Having established the existence of edge steady states for trapped ultracold atoms immersed in a background BEC which undergo a combination of driven and dissipative processes to evolve the system, we now examine, by numerically solving the time evolution equation governing the dynamics of the system, how these steady states emerge over time. In analyzing the time evolution of the system, we focus on the expectation value of the particle number of the time-evolved system at each of the harmonic traps centered at $x_j = (2j-L-2)x_a$, which is defined as

\begin{equation}
N_j (t)=Tr(\hat{N}_{j}\hat{\rho}_{S}(t))
\label{expvalpartnumht}
\end{equation}
Here, $\hat{N}_{j}=\hat{a}^{\dagger}_{g,j}\hat{a}_{g,j}$ is the particle number operator for ultracold atoms at the harmonic potential trap centered at $x_j = (2j-L-2)x_a$.

\begin{figure}[htb]
\includegraphics[width=1.0\columnwidth, height=0.3\textheight]{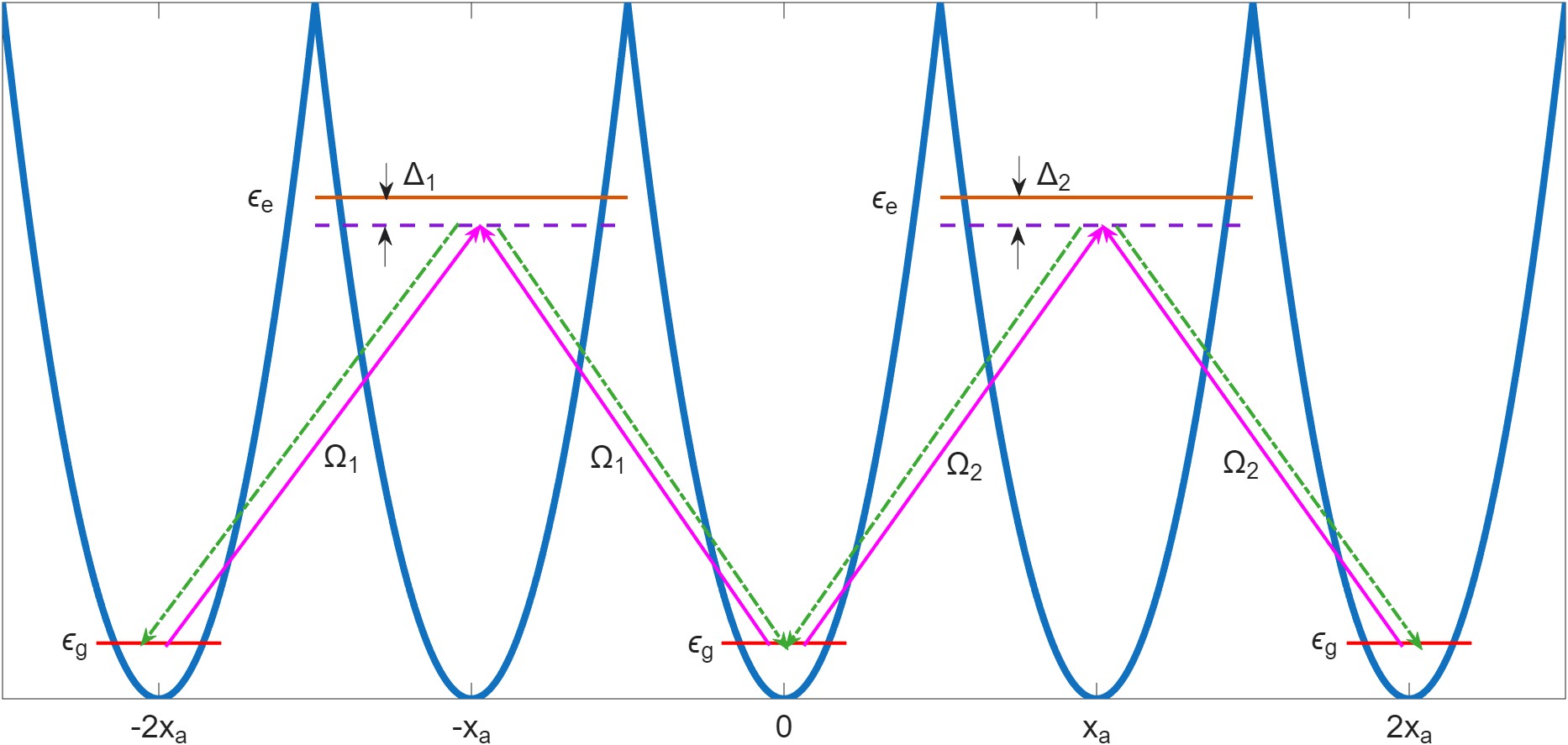}
\caption{\label{fig:5HOTrapArraySchematic} Schematic diagram of a five harmonic oscillator array used to trap the ultracold atoms in this open quantum system. The atoms are initially in the ground state energy level $\epsilon_g$ of the harmonic oscillators centered at $x = 0$ and $x = \pm 2x_a$ (red lines), and are excited by Rabi lasers (pink lines) with Rabi frequencies $\Omega_1$ and $\Omega_2$ to the excited energy level (dashed purple line) located below the higher energy level $\epsilon_e$ of the harmonic potential centered at $x=\pm x_a$ (solid orange line). Emission of an excitation with energy $E_\mathbf{k}$ (dashed green line) into the background BEC will cause the excited atoms to return to the ground state energy level in either of the two neighboring harmonic oscillators centered at $x = -2x_a$ and $x = 0$ if the atoms are excited to the harmonic trap at $x = -x_a$, or in either of the two neighboring harmonic oscillators centered at $x = 0$ and $x = 2x_a$ if the atoms are excited to the harmonic trap at $x = x_a$.}
\end{figure}

In this section, we will consider two particular cases. For the first case, we consider the case wherein the ultracold atoms are trapped in an array of five harmonic potentials. The schematic diagram for this case is given in Fig. \ref{fig:5HOTrapArraySchematic}.

The resulting master equation for this system, which was earlier derived in Ref. (\cite{caballar3}), has the same form as that given by Eq. (\ref{masteqnharmonicarraytrappedultracoldatomoqs}), with the jump operators given by Eq. (\ref{jumpopdef}), with $L = 2$. Explicitly, the jump operator's form is given by 
\begin{equation}
\hat{c}=(\hat{a}^{\dagger}_{g,1}+\hat{a}^{\dagger}_{g,2})(\hat{a}_{g,1}-\hat{a}_{g,2})+\epsilon(\hat{a}^{\dagger}_{g,2}+\hat{a}^{\dagger}_{g,3})(\hat{a}_{g,2}-\hat{a}_{g,3})
\label{3nodejumpop}
\end{equation}

For this system, we assume that its initial state is given by a density matrix with the form
\begin{equation}
\hat{\rho}(t_0) = \left|\phi\right\rangle\left\langle\phi\right| = \left|n_1\right\rangle\left\langle n_1\right|\otimes\left|n_2\right\rangle\left\langle n_2\right|\otimes\left|n_3\right\rangle\left\langle n_3\right|,
\label{initstate3node}
\end{equation}
where $\left|\phi\right\rangle = \left|n_1\right\rangle\otimes\left|n_2\right\rangle\otimes\left|n_3\right\rangle$. We impose the condition that, at the initial instant of time $t_0$, the total number of atoms in the ultracold atom gas is
\begin{equation}
N=n_1+n_2+n_3
\label{convpartnum3node}
\end{equation}
We further assume that this conservation of particle number condition is obeyed by the system at all instants of time, to ensure that the total number of atoms in the ultracold atom gas system remains constant, with no particle loss due to the dissipative interaction between the trapped ultracold atom system and the background BEC that acts as an excitation reservoir. This condition will prove to be crucial in demonstrating the existence of unique edge steady states for this system later on.

\begin{figure}[htb]
\includegraphics[width=1.0\columnwidth, height=0.3\textheight]{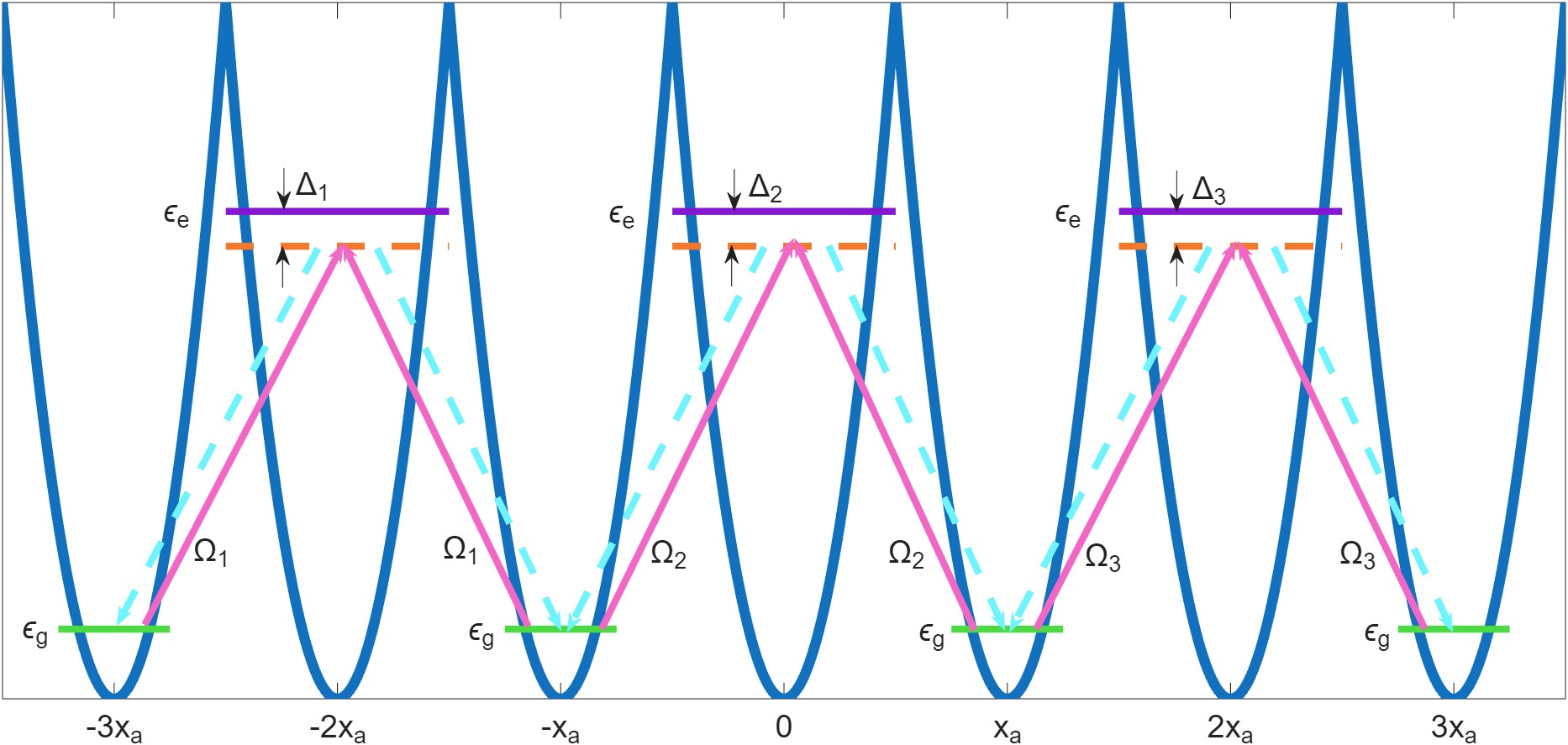}
\caption{\label{fig:7HOTrapArraySchematic} Schematic diagram of a seven harmonic oscillator array used to trap the ultracold atoms in this open quantum system. The atoms are initially in the ground state energy level $\epsilon_g$ of the harmonic oscillators centered at $x = \pm 3x_a$ and $x = \pm x_a$ (green lines), and are excited by Rabi lasers (pink lines) with Rabi frequencies $\Omega_1$, $\Omega_2$ and $\Omega_3$ to the excited energy level (dashed orange line) located below the higher energy level $\epsilon_e$ of the harmonic potential centered at $x=\pm 2x_a$ and $x=0$ (solid purple line). Emission of an excitation with energy $E_\mathbf{k}$ (dashed sky blue line) into the background BEC will cause the excited atoms to return to the ground state energy level in either of the two neighboring harmonic oscillators centered at $x = -3x_a$ and $x = -x_a$ if the atoms are excited to the harmonic trap at $x = -2x_a$, in either of the two neighboring harmonic oscillators centered at $x = -x_a$ and $x = x_a$ if the atoms are excited to the harmonic trap at $x = 0$, or in either of the two neighboring harmonic oscillators centered at $x = x_a$ and $x = 3x_a$ if the atoms are excited to the harmonic trap at $x = 2x_a$.}
\end{figure}

For the second case to be considered in this section, we consider a trapped ultracold atom open quantum system wherein the ultracold atom gas is trapped in an array of seven harmonic potentials, with the schematic diagram for this system given in Fig. \ref{fig:7HOTrapArraySchematic}.

The resulting jump operator that describes the dynamics of this system will, from Eq. (\ref{jumpopdef}) with $L = 3$, then have the following form:

\begin{equation}
\hat{c}=(\hat{a}^{\dagger}_{g,1}+\hat{a}^{\dagger}_{g,2})(\hat{a}_{g,1}-\hat{a}_{g,2})+\epsilon_1(\hat{a}^{\dagger}_{g,2}+\hat{a}^{\dagger}_{g,3})(\hat{a}_{g,2}-\hat{a}_{g,3})+\epsilon_2(\hat{a}^{\dagger}_{g,3}+\hat{a}^{\dagger}_{g,4})(\hat{a}_{g,3}-\hat{a}_{g,4})
\label{4nodejumpop}
\end{equation}

For this system, we assume that its initial state is given by a density matrix with the form
\begin{equation}
\hat{\rho}(t_0) = \left|\phi\right\rangle\left\langle\phi\right| = \left|n_1\right\rangle\left\langle n_1\right|\otimes\left|n_2\right\rangle\left\langle n_2\right|\otimes\left|n_3\right\rangle\left\langle n_3\right|\otimes\left|n_4\right\rangle\left\langle n_4\right|,
\label{initstate4node}
\end{equation}
where $\left|\phi\right\rangle = \left|n_1\right\rangle\otimes\left|n_2\right\rangle\otimes\left|n_3\right\rangle\otimes\left|n_4\right\rangle$. We impose the condition that, at the initial instant of time $t_0$, the total number of atoms in the ultracold atom gas is
\begin{equation}
N=n_1+n_2+n_3+n_4
\label{convpartnum4node}
\end{equation}
As with the five harmonic potential trap system described earlier, we assume that this conservation of particle number condition is obeyed by the system at all instants of time, to ensure that the total number of atoms in the ultracold atom gas system remains constant, with no particle loss due to the dissipative interaction between the trapped ultracold atom system and the background BEC that acts as an excitation reservoir.

For the next two subsections, we assume that $\epsilon_j < 1$ for all $j$ in the jump operators given by Eqs. (\ref{3nodejumpop}) and (\ref{4nodejumpop}) describing the trapped ultracold atom open quantum systems with five and seven harmonic trapping potentials, respectively. The specific values of $\epsilon_j$ and other parameters present in the master equation used to evolve the trapped ultracold atom open quantum systems as given by Eq. (\ref{masteqnharmonicarraytrappedultracoldatomoqs}) are specified in the captions of the figures showing the numerical results of the evolution of the system.

\subsection{Emergence of Edge States when $n_1 << n_j$ and $n_{L+1} << n_j$}

We first consider the case where the number of atoms initially trapped away from the edges of the harmonic trap array are much greater than the number of atoms initially trapped at the edges of the trap, i. e. $n_ j >> n_1, n_j >> n_{L+1}, 1 < j < L+1$. As shown in Figs. (\ref{fig:numexpvalmidtrapmore3node}) and (\ref{fig:numexpvalmidtrapmore4node}) below, the system evolves according to the master equation given by (\ref{masteqnharmonicarraytrappedultracoldatomoqs}) in such a way that left edge steady states will emerge by having $n_1 (t)$ increase over time until it reaches a steady state value $n_1 (t) = N$, where $N$ is the total number of ultracold atoms trapped in the system. On the other hand, the trapped ultracold atom states at the harmonic potentials located away from the left edge of the trap array at $x_j = (2j-L-2)x_a, 1<j\leq L+1$ will vanish, with $n_j (t)$ decreasing over time until $n_j (t) = 0$, irrespective of how many atoms were initially trapped at these harmonic potentials. These results demonstrate the robustness of the steady edge states of the system, analogous to the emergence and robustness of edge states as steady states in topological quantum matter, and demonstrating the topological nature of this driven - dissipative trapped ultracold atom system.

\begin{figure}[htb]
\includegraphics[width=0.5\textwidth, height=0.2\textheight]{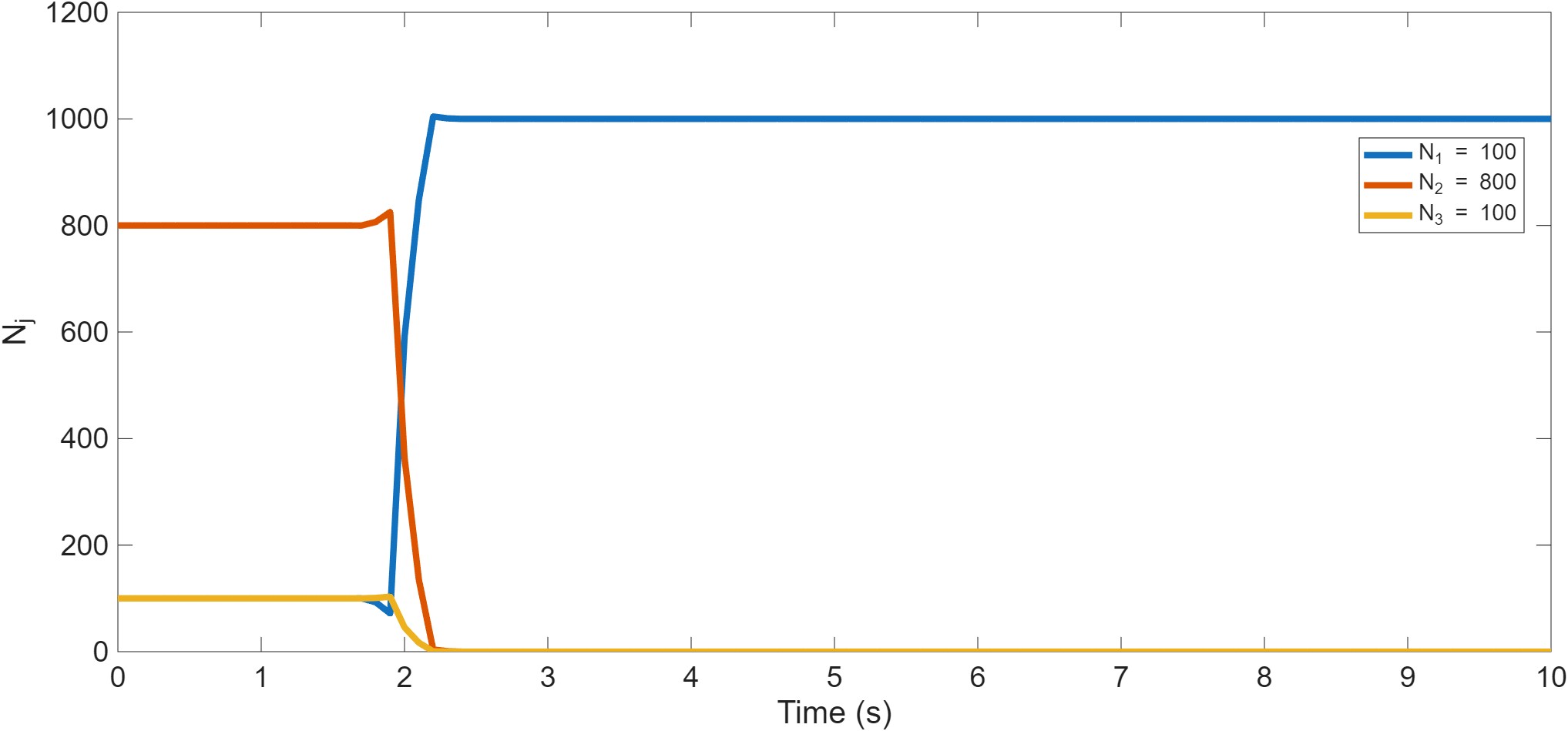}
\includegraphics[width=0.5\textwidth, height=0.2\textheight]{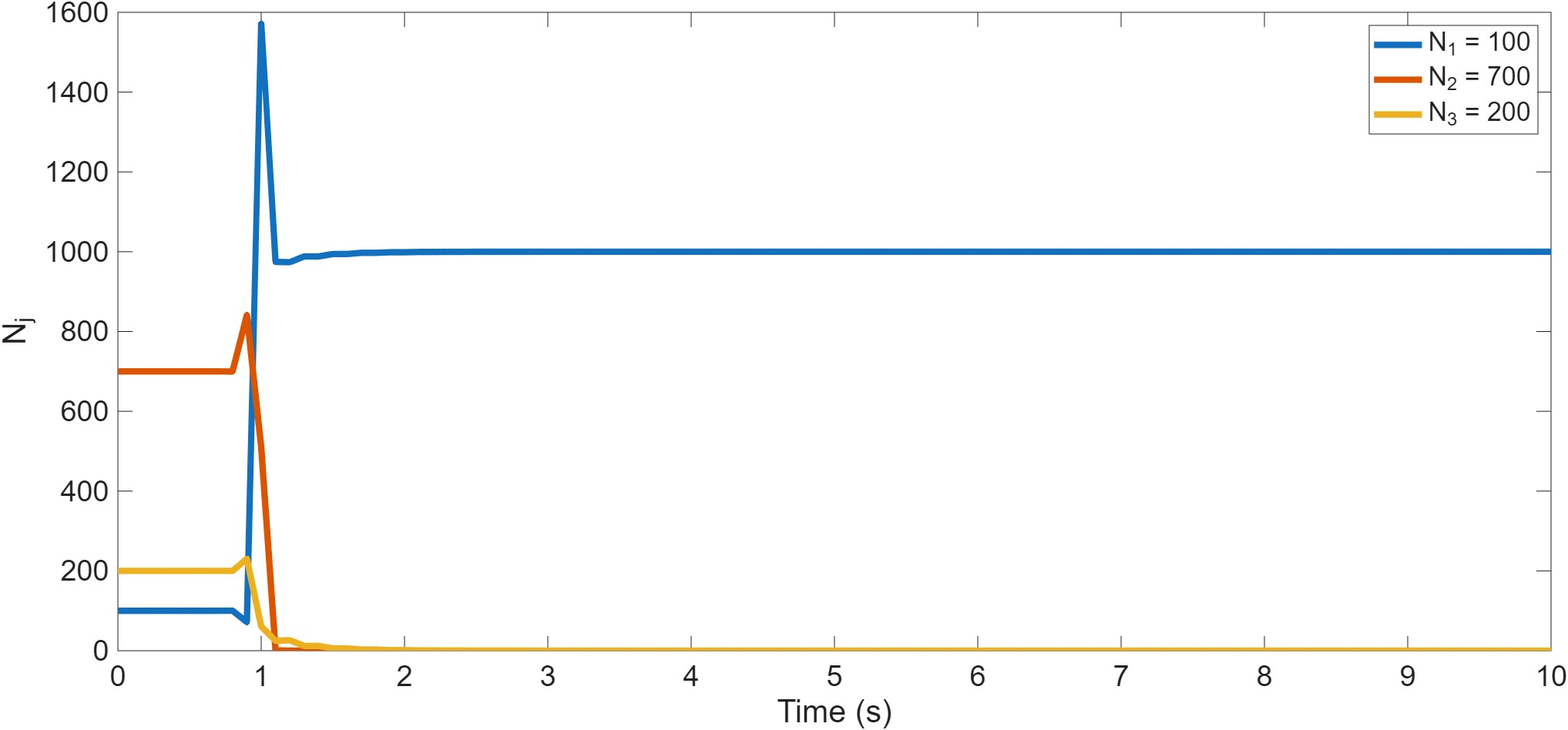}
\includegraphics[width=0.5\textwidth, height=0.2\textheight]{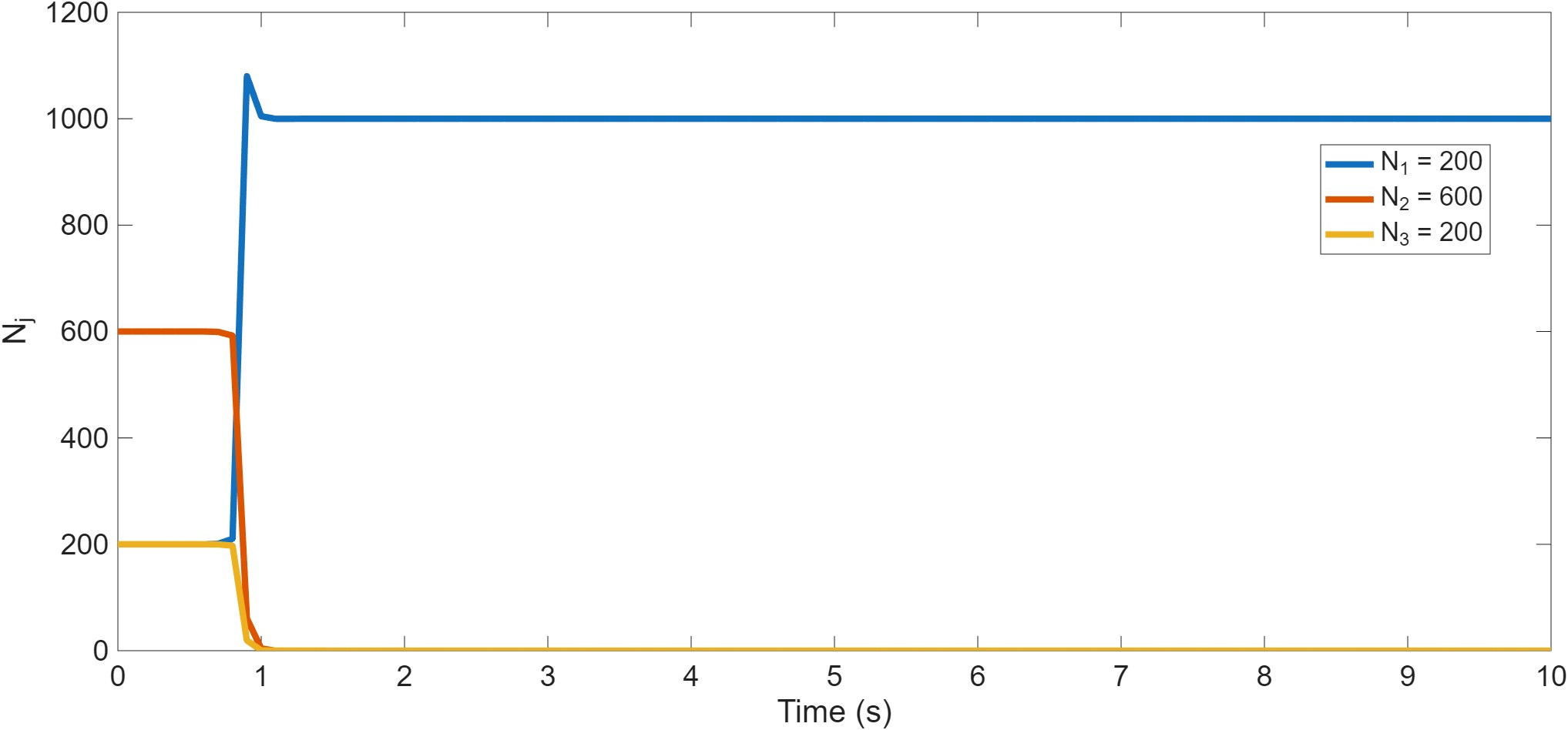}
\includegraphics[width=0.5\textwidth, height=0.2\textheight]{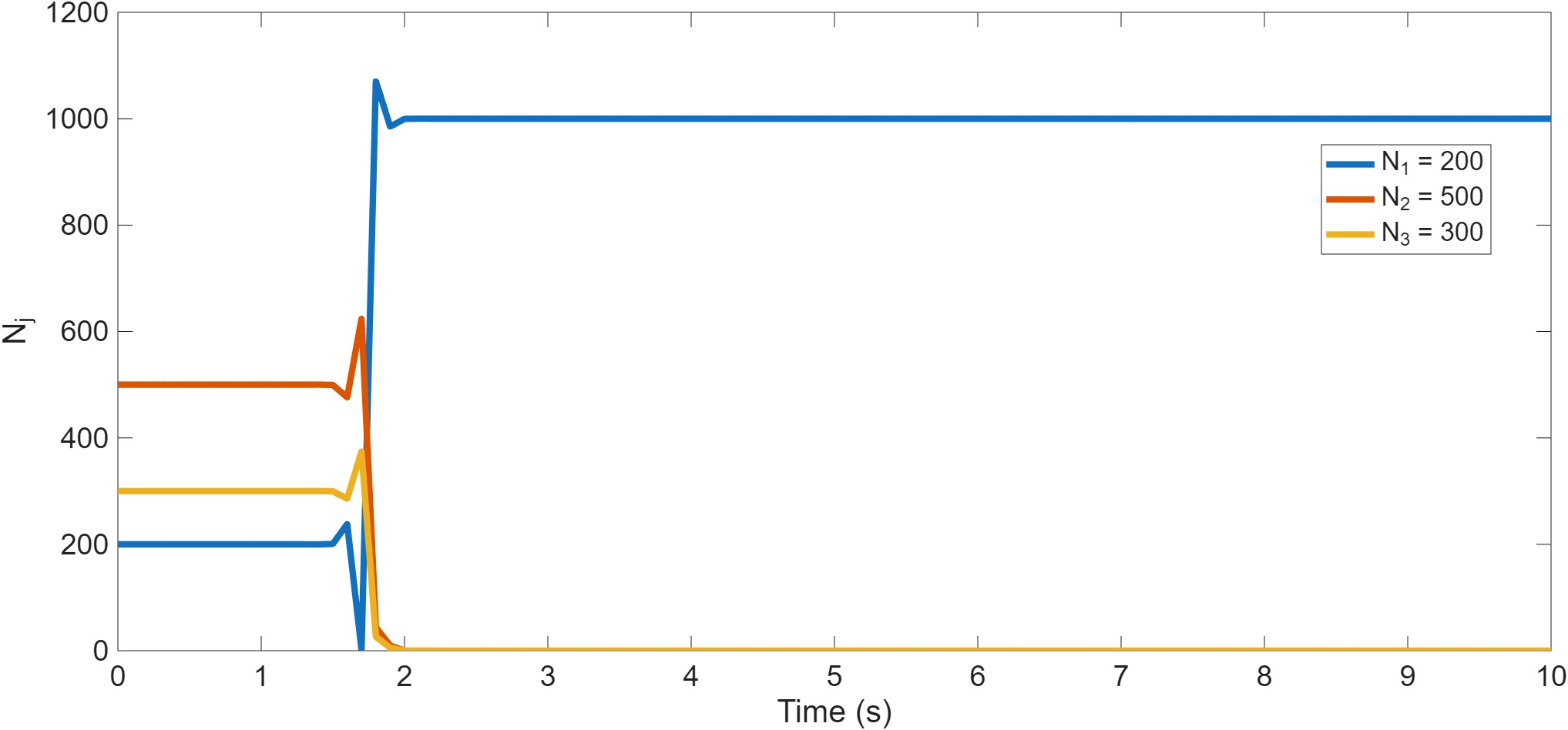}
\caption{\label{fig:numexpvalmidtrapmore3node} Time evolution of the expectation values of the particle number $n_j(t)$ in the harmonic traps located at $x_j = -2x_a$ (for which $n_{1}(0)=N_1$ there), $x_j = 0$ (where $n_{2}(0)=N_2$) and $x_j = 2x_a$ (where $n_{3}(0)=N_3$), with the values of $N_1, N_2, N_3$ indicated for each plot, such that $N_2 >N_1$ and $N_2 > N_3$. Here, $\epsilon = 0.7$, $\varepsilon_e - \varepsilon_g = 1.00\times 10^{-7}$ and $A=0.1$.}
\end{figure}

It is of interest to note that for the 7-harmonic potential array system, it does not matter if $n_2 > n_3$ or vice versa; so long as $n_1 << n_2, n_1 << n_3$ and $n_4 << n_2, n_4 << n_3$, the left edge steady state will always emerge as the trapped ultracold atom system evolves over time. 

\begin{figure}[htb]
\includegraphics[width=0.5\textwidth, height=0.2\textheight]{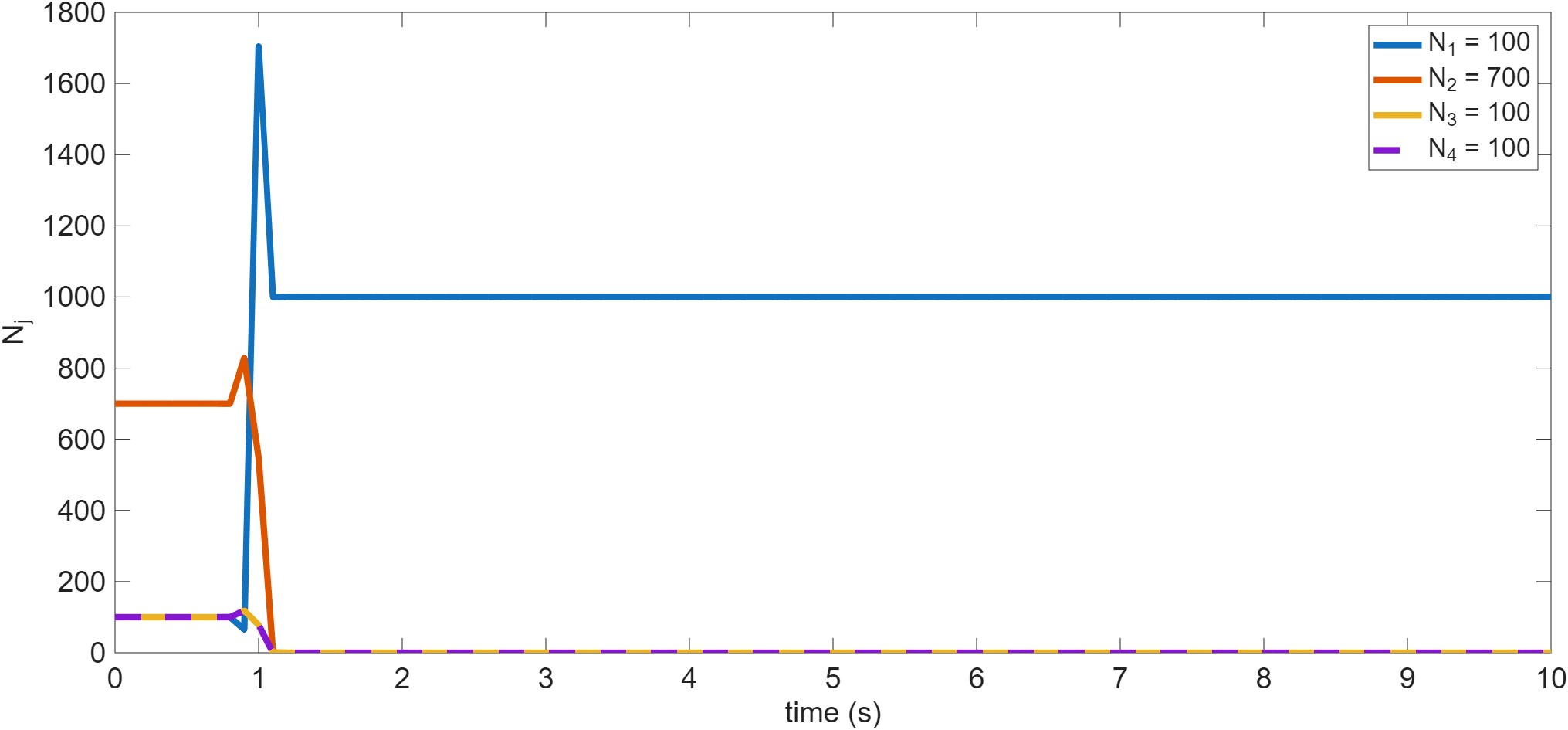}
\includegraphics[width=0.5\textwidth, height=0.2\textheight]{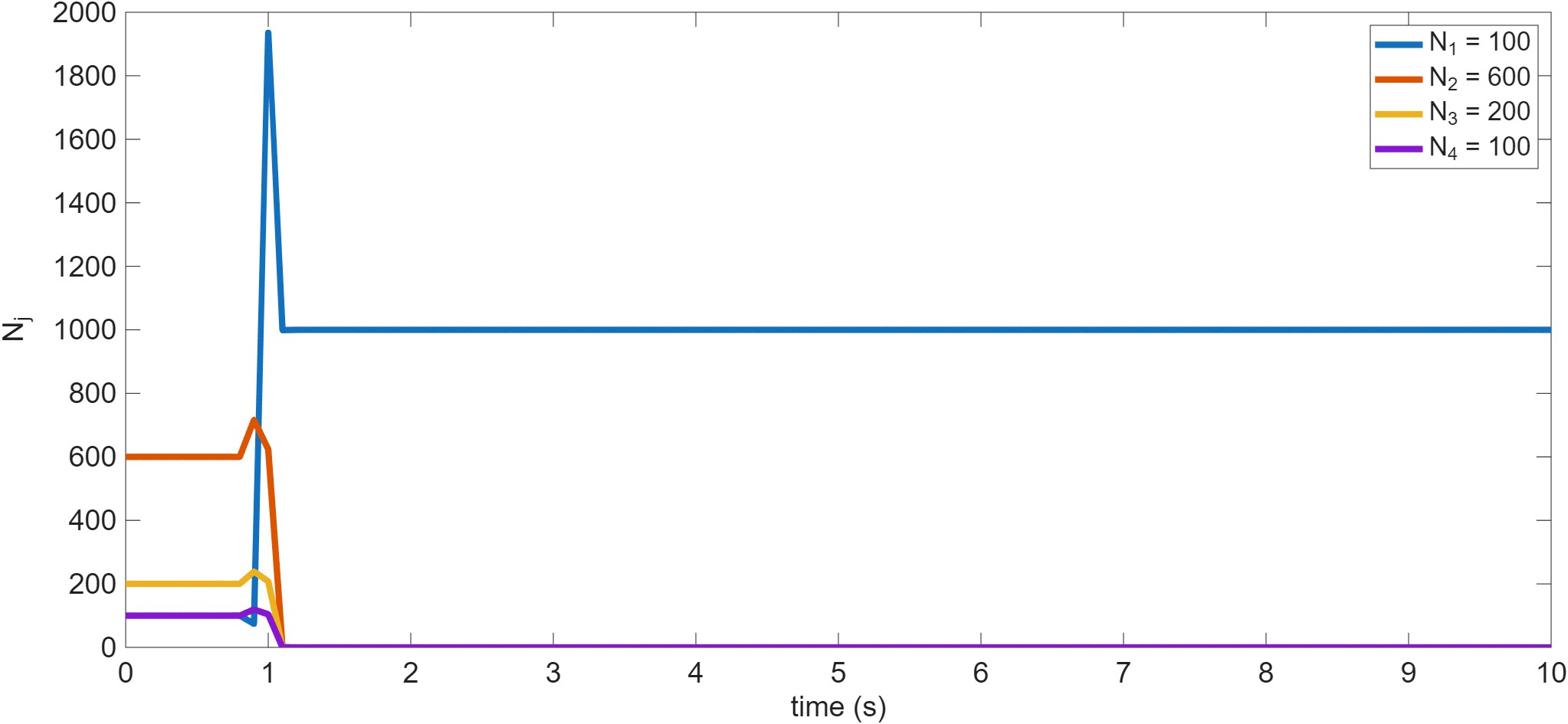}
\includegraphics[width=0.5\textwidth, height=0.2\textheight]{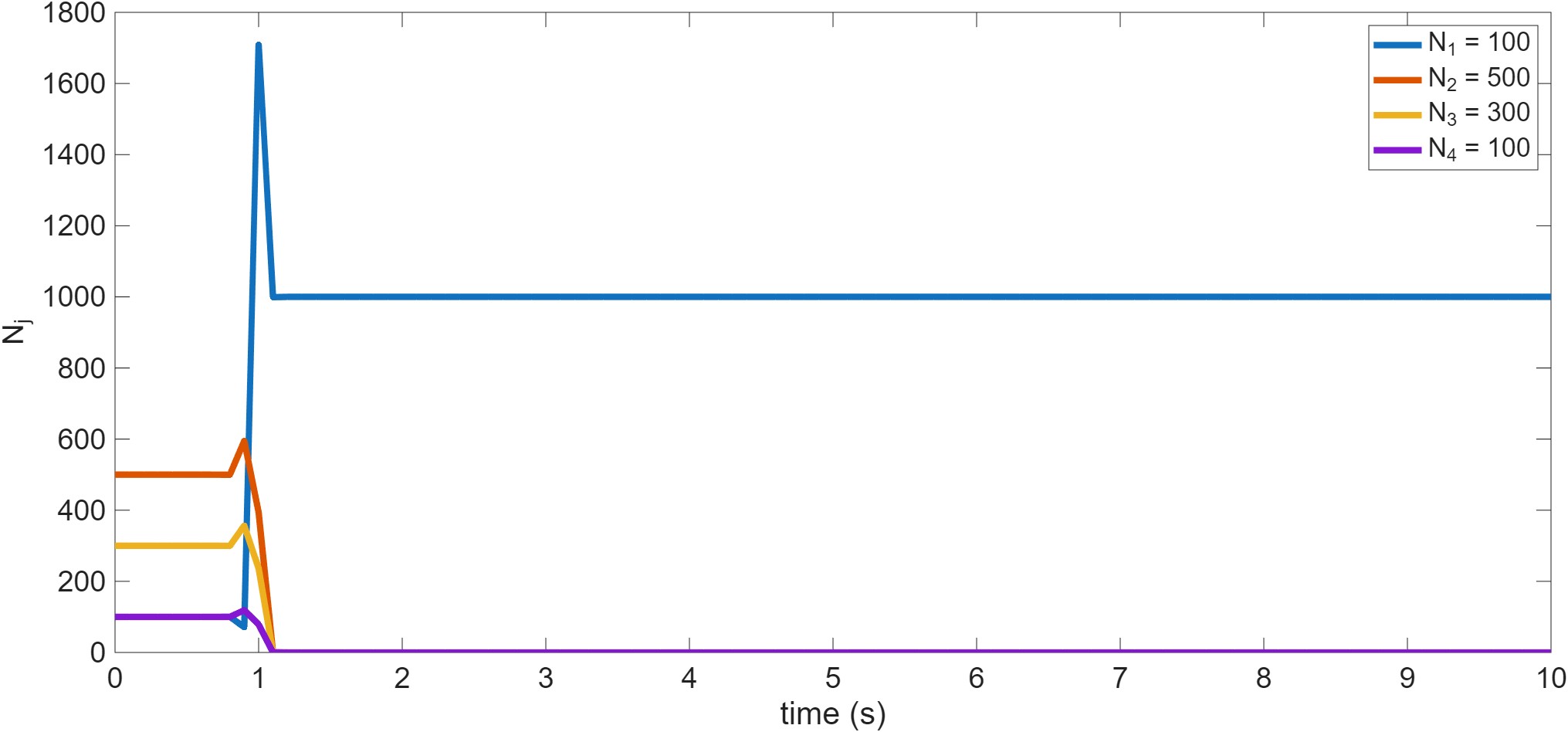}
\includegraphics[width=0.5\textwidth, height=0.2\textheight]{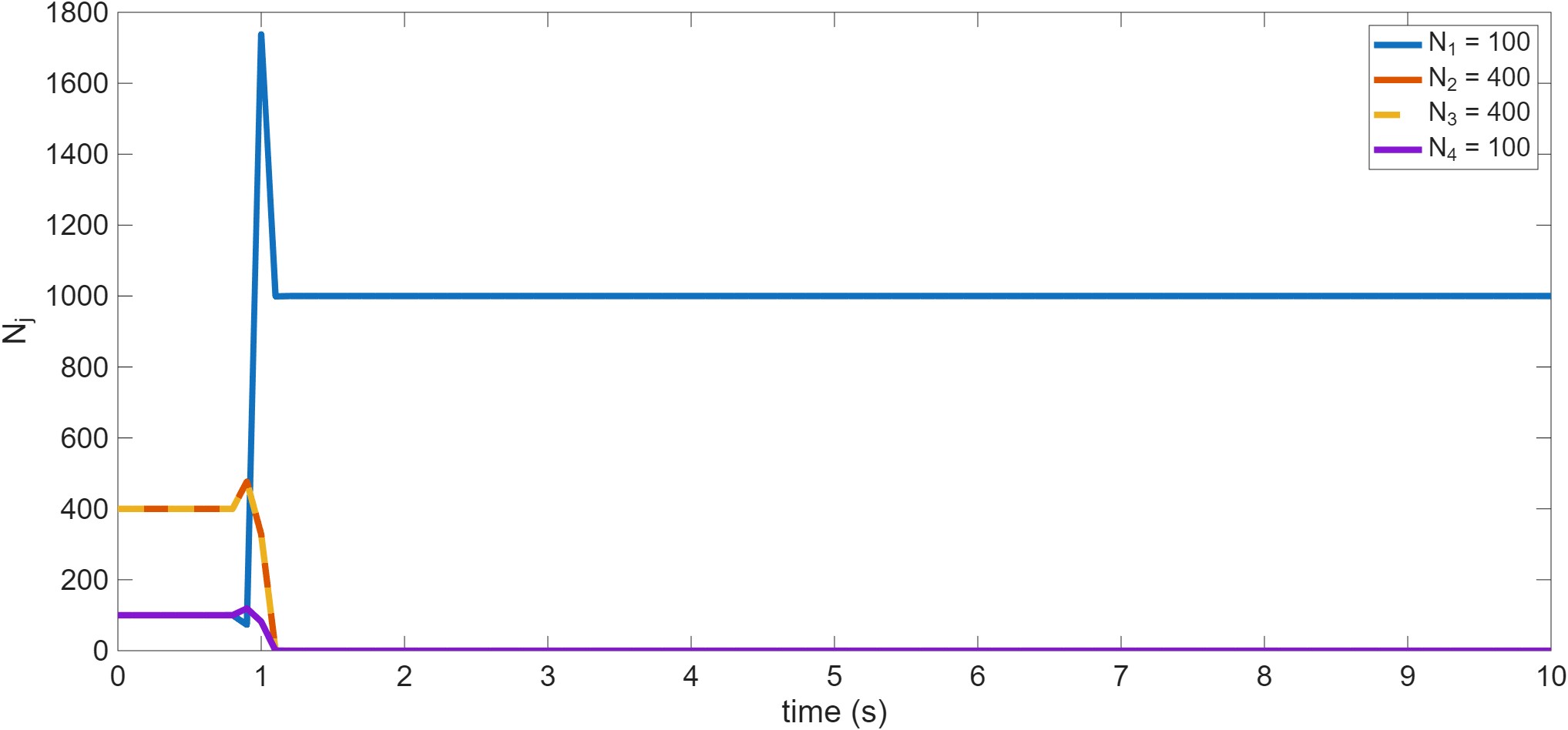}
\caption{\label{fig:numexpvalmidtrapmore4node} Time evolution of the expectation values of the particle number $n_j(t)$ in the harmonic traps located at $x_j = -3x_a$ (for which $n_{1}(0)=N_1$ there), $x_j = -x_a$ (where $n_{2}(0)=N_2$), $x_j = x_a$ (where $n_{3}(0)=N_3$) and $x_j = 3x_a$ (where $n_{4}(0)=N_4$), with the values of $N_1, N_2, N_3, N_4$ indicated for each plot, such that $N_1 = N_4$, $N_2 >N_1$ and $N_3 > N_4$. Here, $\epsilon_1 = 0.7$, $\epsilon_2 = 0.5$, $\varepsilon_e - \varepsilon_g = 1.00\times 10^{-7}$ and $A=0.1$.}
\end{figure}

\subsection{Emergence of Edge States when $n_1$ and $n_{L+1}$ are Varied}

\begin{figure}[htb]
\includegraphics[width=0.5\textwidth, height=0.15\textheight]{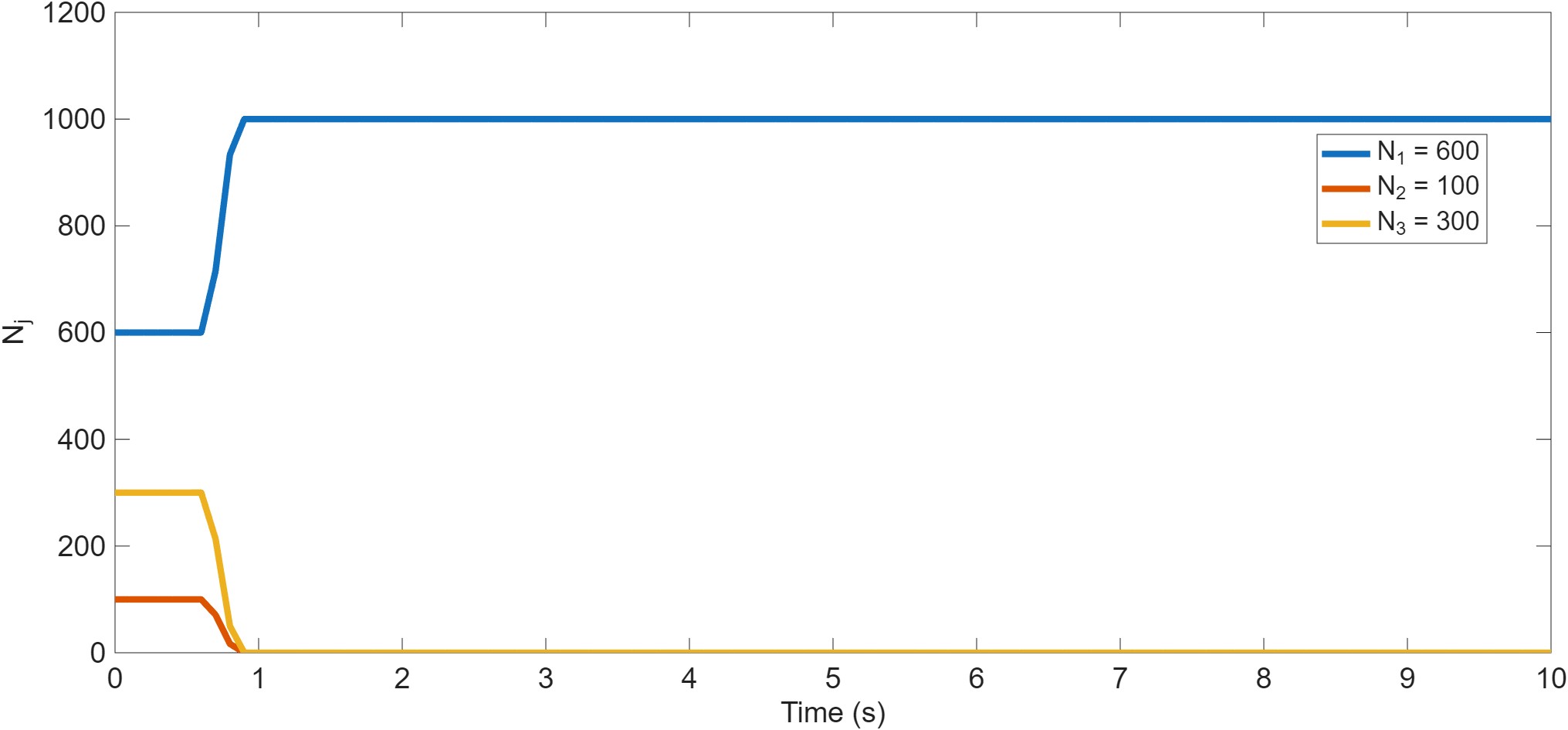}
\includegraphics[width=0.5\textwidth, height=0.15\textheight]{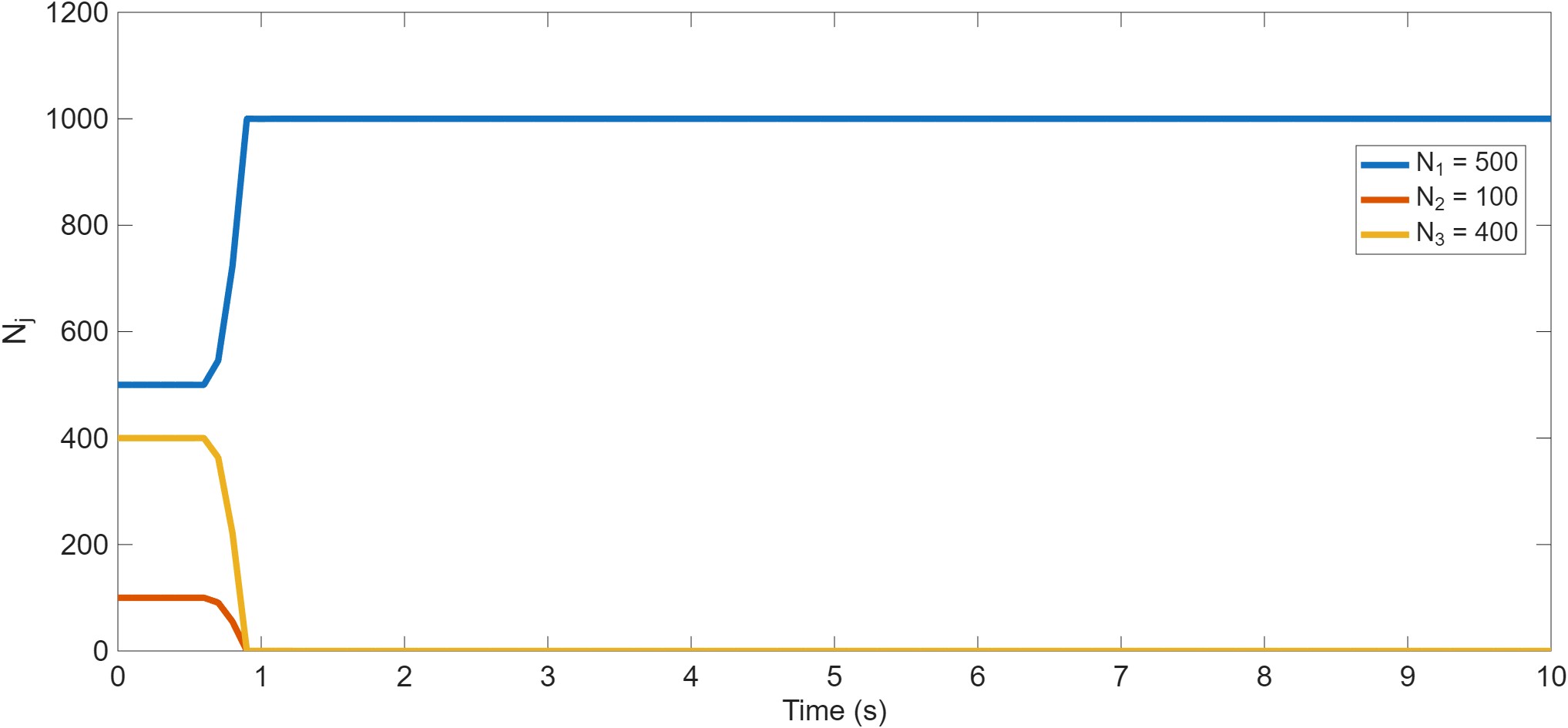}
\includegraphics[width=0.5\textwidth, height=0.15\textheight]{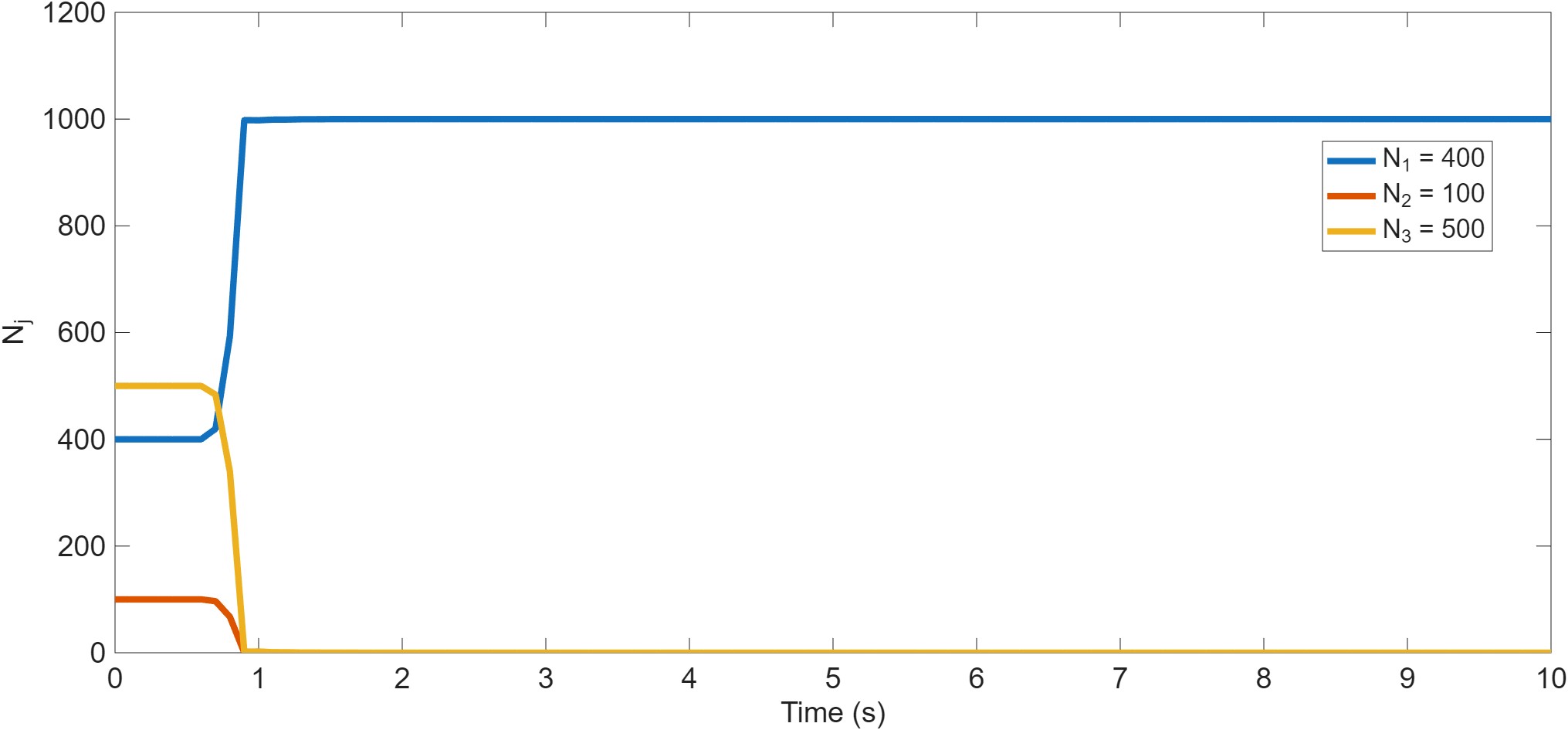}
\includegraphics[width=0.5\textwidth, height=0.15\textheight]{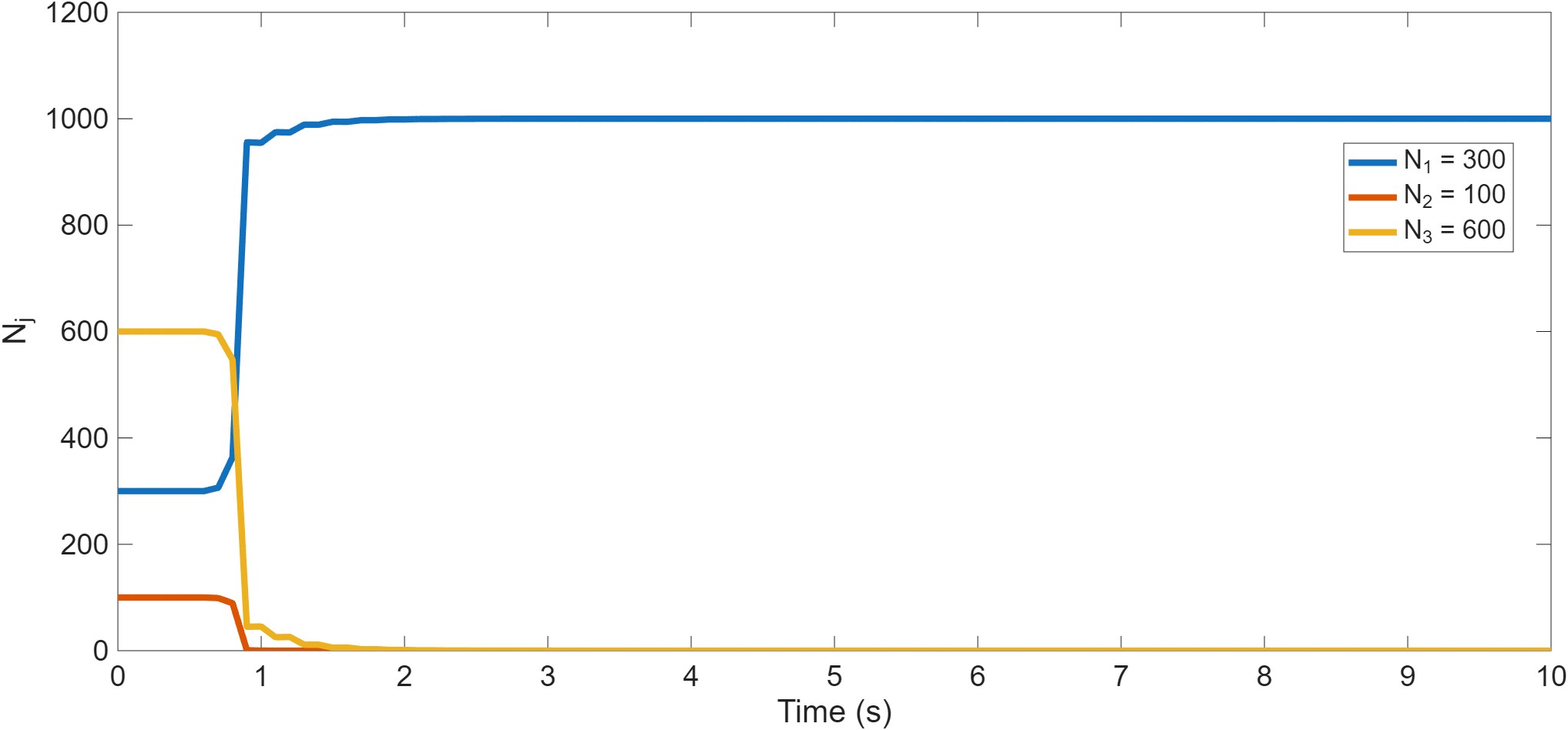}
\includegraphics[width=0.5\textwidth, height=0.15\textheight]{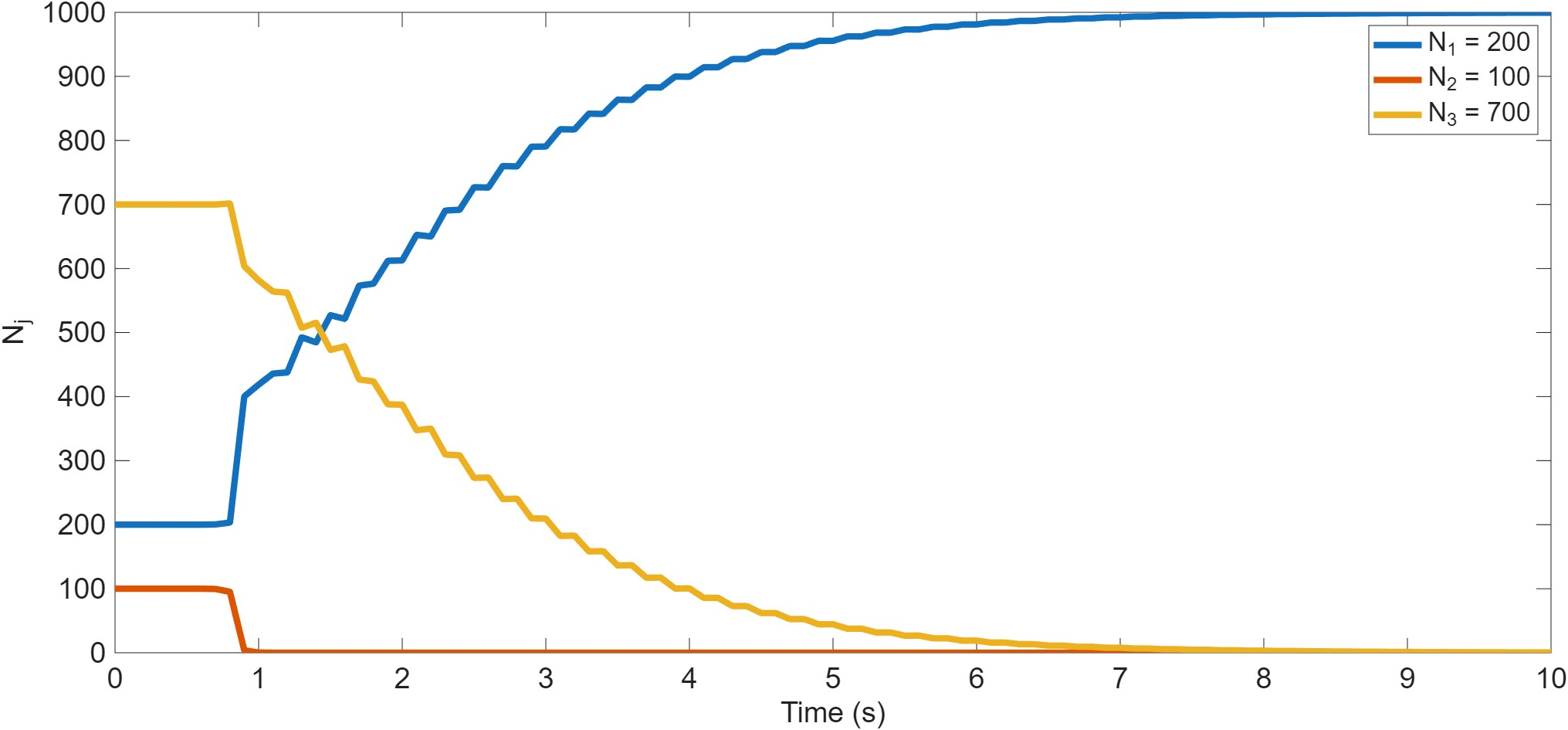}
\includegraphics[width=0.5\textwidth, height=0.15\textheight]{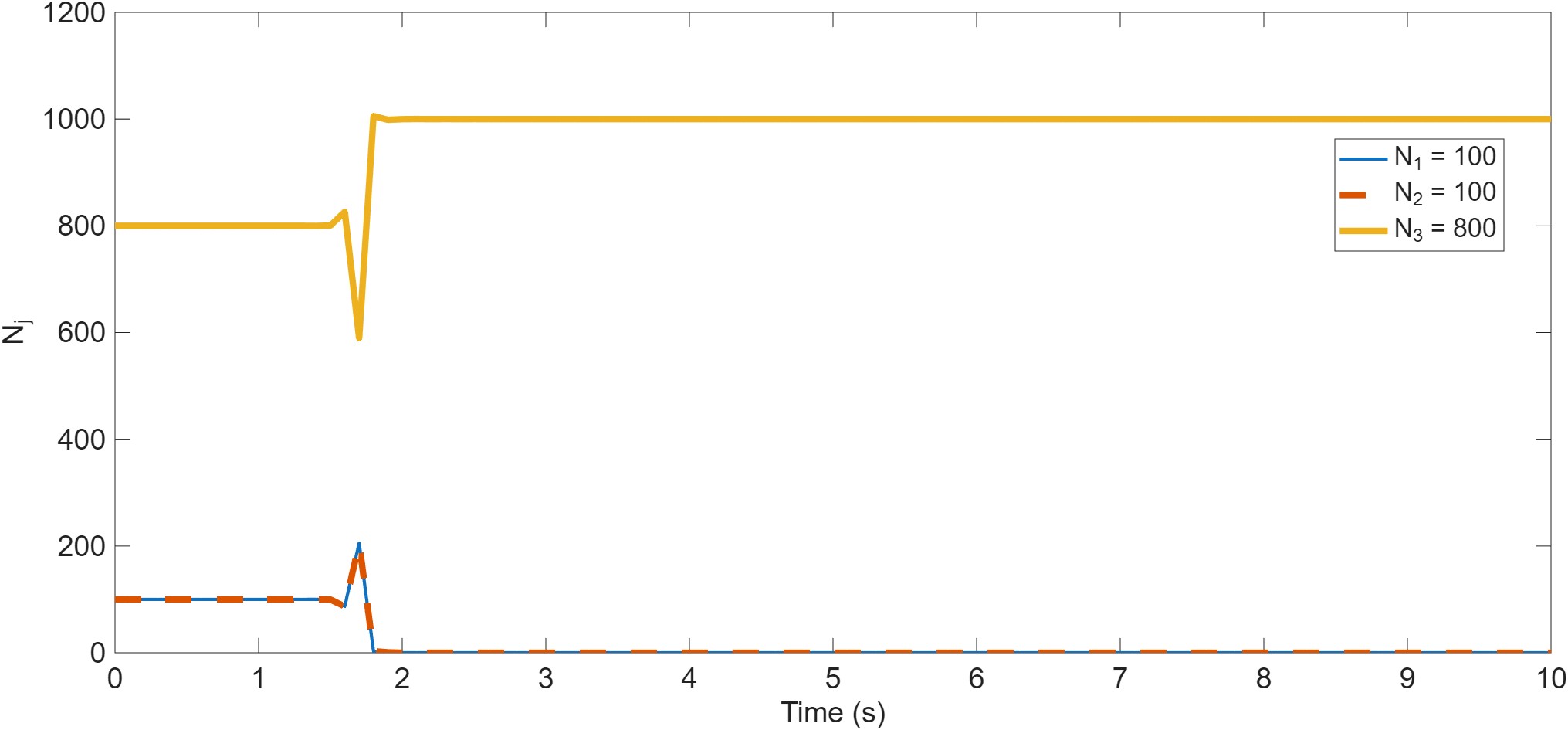}
\caption{\label{fig:numexpvalvaryingN1N33node} Time evolution of the expectation values of the particle number $n_j(t)$ in the harmonic traps located at $x_j = -2x_a$ (for which $n_{1}(0)=N_1$), $x_j = 0$ (where $n_{2}(0)=N_2$) and $x_j = 2x_a$ (where $n_{3}(0)=N_3$), with the values of $N_1, N_2, N_3$ indicated for each plot, such that $N_2 = 100$, $N_1$ decreases from $N_1 = 600$ to $N_1 = 100$ and $N_3$ increases from $N_3 = 300$ to $N_3 = 800$. Here, $\epsilon = 0.7$, $\varepsilon_e - \varepsilon_g = 1.00\times 10^{-7}$ and $A=0.1$.}
\end{figure}

\begin{figure}[htb]
\includegraphics[width=0.5\textwidth, height=0.15\textheight]{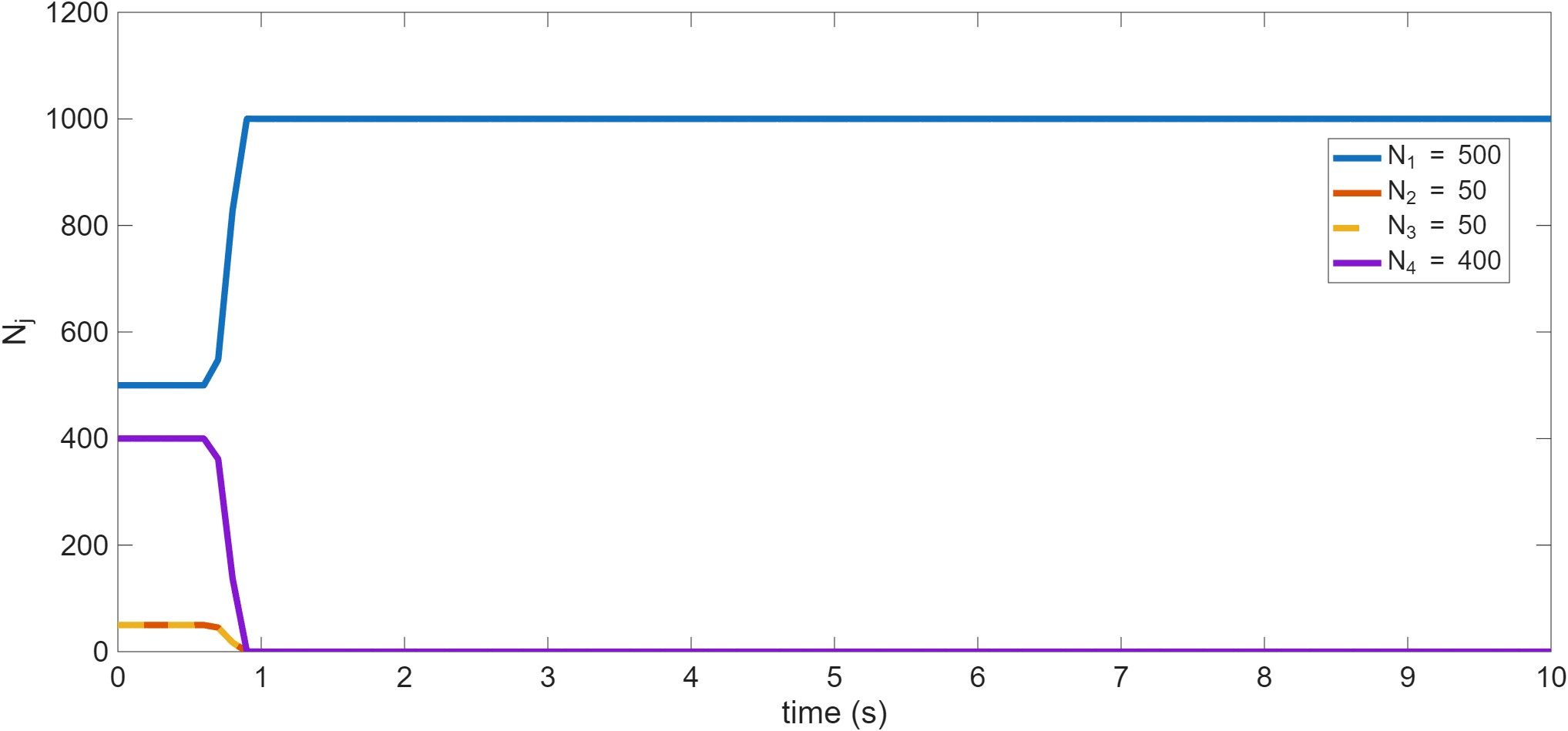}
\includegraphics[width=0.5\textwidth, height=0.15\textheight]{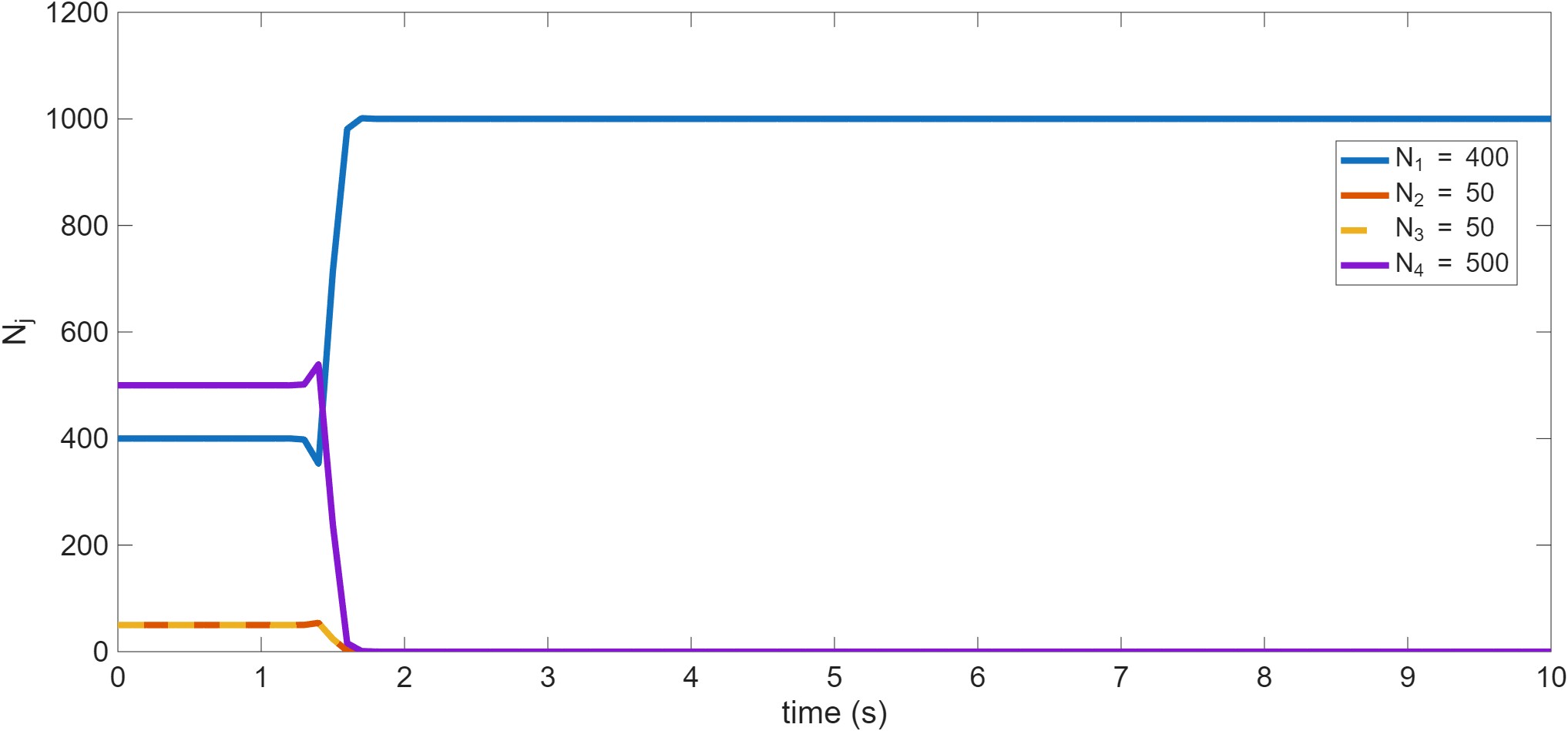}
\includegraphics[width=0.5\textwidth, height=0.15\textheight]{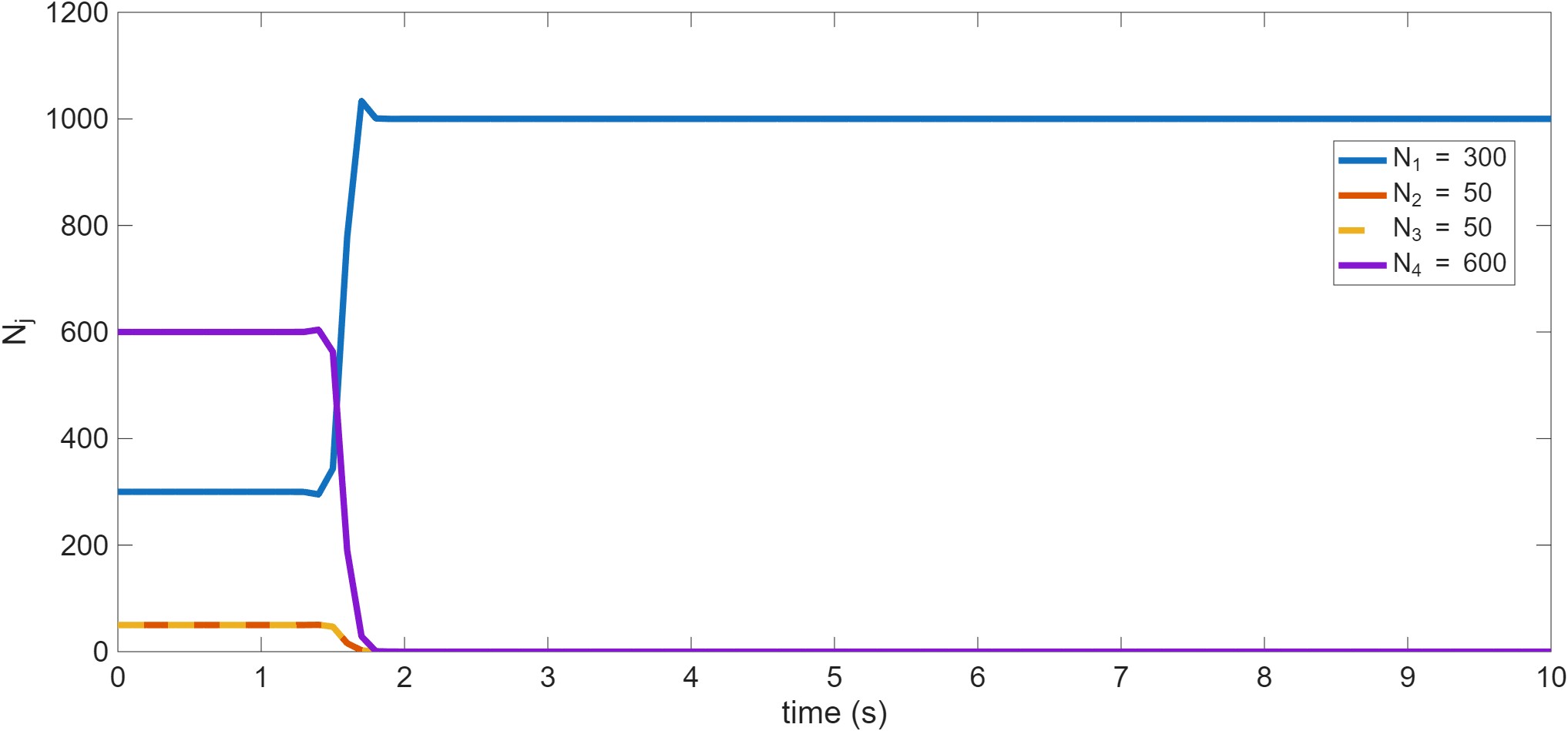}
\includegraphics[width=0.5\textwidth, height=0.15\textheight]{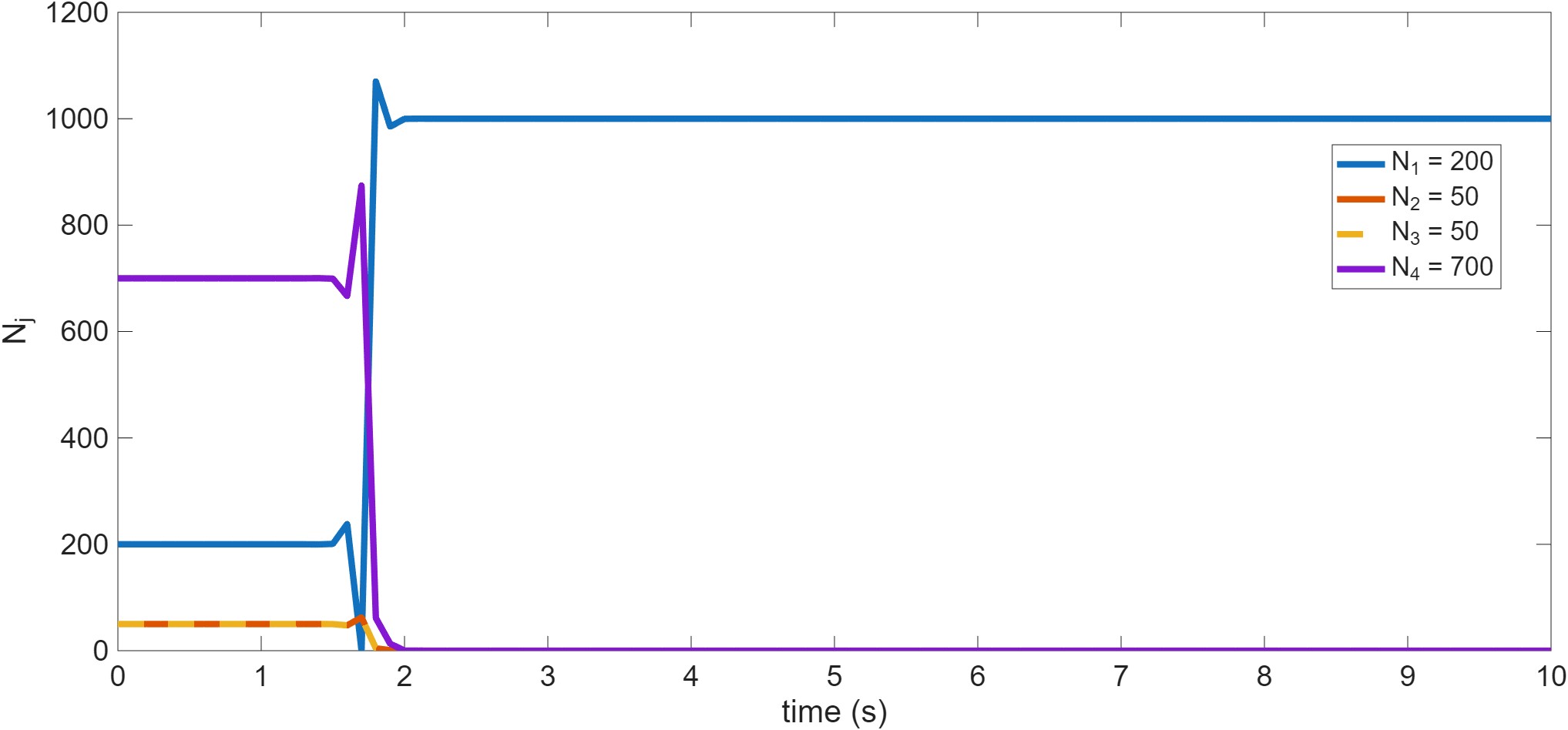}
\includegraphics[width=0.5\textwidth, height=0.15\textheight]{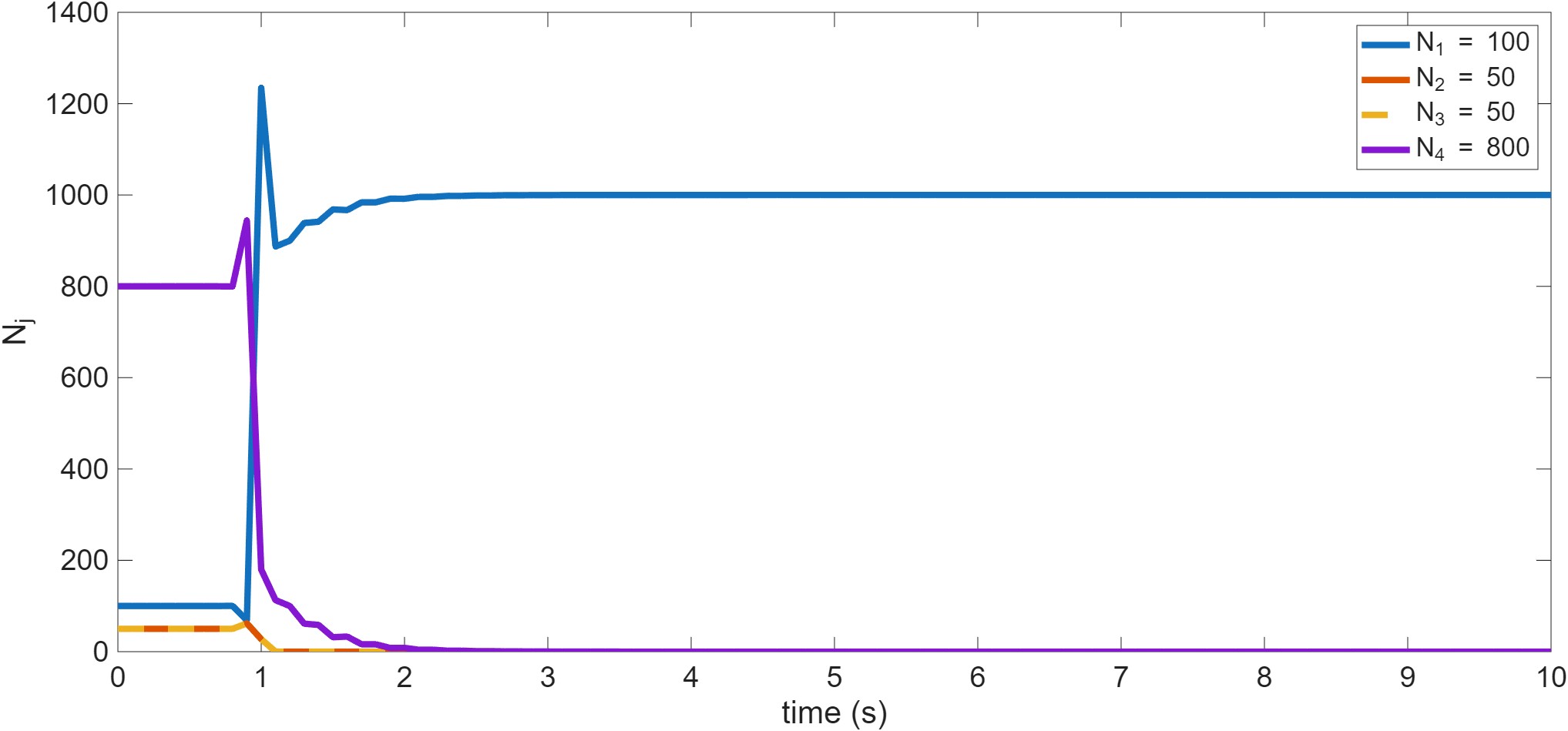}
\includegraphics[width=0.5\textwidth, height=0.15\textheight]{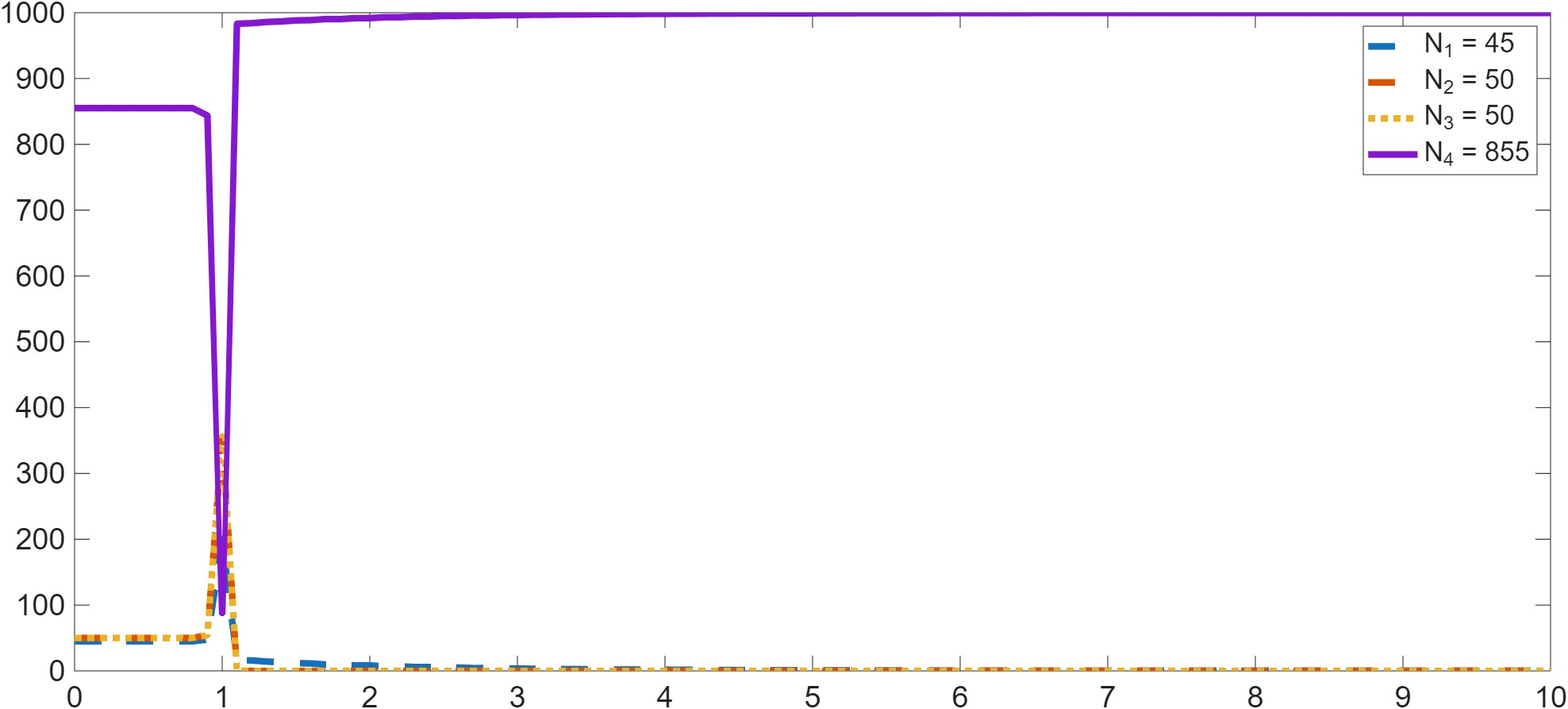}
\caption{\label{fig:numexpvalvaryingN1N34node} Time evolution of the expectation values of the particle number $n_j(t)$ in the harmonic traps located at $x_j = -3x_a$ (for which $n_{1}(0)=N_1$ here), $x_j = -x_a$ (where $n_{2}(0)=N_2$), $x_j = x_a$ (where $n_{3}(0)=N_3$) and $x_j = 3x_a$ (where $n_{4}(0)=N_3$), with the values of $N_1, N_2, N_3, N_4$ indicated for each plot, such that $N_2 = N_3 = 50$, $N_1$ decreases from $N_1 = 600$ to $N_1 = 45$ and $N_4$ increases from $N_3 = 300$ to $N_3 = 855$. Here, $\epsilon_1 = 0.7$, $\epsilon_2 = 0.5$, $\varepsilon_e - \varepsilon_g = 1.00\times 10^{-7}$ and $A=0.1$.}
\end{figure}

Next, we consider what happens when we decrease the number of atoms $n_1$ initially trapped in the harmonic potential at the left edge $x_1 = -x_a$ of the array and simultaneously increase the number of atoms $n_{L+1}$ initially trapped in the harmonic potential at the right edge $x_{L+1} = x_a$ of the array, while keeping the number of atoms $n_j, 1<j<L+1$, initially trapped in the harmonic potentials away from the edges of the array $x_j = (2j-L-2)x_a, 1 < j < L+1$, constant. As shown in Figs. (\ref{fig:numexpvalvaryingN1N33node}) and (\ref{fig:numexpvalvaryingN1N34node}), for a given value of $n_j, 1< j <L+1$ held constant as $n_1$ and $n_{L+1}$ decrease and increase, respectively, the left edge state at $x_1 = -Lx_a$ will persistently emerge even as $n_1$ continously decreases such that $n_{L+1} > n_1$. However, past a value of $n_1$ and $n_{L+1}$ with $n_{L+1} > n_1$, the left edge state will no longer emerge as the steady state of the system, and it will, instead, be the right edge state that emerges as the steady state as the trapped ultracold atom open quantum system evolves over time. For the particular case shown in Fig. (\ref{fig:numexpvalvaryingN1N33node}), for $n_2 = 100$ held constant, the left edge state emerges as the system's steady state for all pairs of values of $(n_1, n_3)$ from $(n_1, n_3) = (600,300)$ up to $(n_1, n_3) = (200, 700)$. However, when $(n_1, n_3) = (100, 800)$, it is the right edge state that emerges as the system's steady state, and will continue to be the system's steady state for all pairs of values $(n_1, n_3) > (100, 800)$ with $n_2 = 100$.    

Figs. (\ref{fig:numexpvalvaryingN1N33node}) and (\ref{fig:numexpvalvaryingN1N34node}) demonstrates a crossover effect, akin to a phase transition, in this trapped ultracold atom open quantum system, with respect to the emergence of the edge steady states. In particular, the figures show that as $n_{L+1}\rightarrow N$, with $n_{L+1}>>n_1$, the steady state for this trapped ultracold atom open quantum system transitions from the left edge state to the right edge state. To investigate this crossover effect further, we consider the case for which $n_3$ is held fixed, while $n_1$ and $n_2$ vary, with $n_3$ having a value close to the total number $N$ of atoms in the ultracold atom gas. As shown in Fig. (\ref{fig:numexpvalvaryingN1N33node}), even if $n_3\approx N$, the steady state that emerges as this trapped ultracold atom open quantum system evolves over time will depend on the value of $n_1$ and $n_2$. In particular, if $n_3>>n_1$, no crossover effect occurs, since $n_3 (t)\rightarrow N, n_1(t),n_2 (t)\rightarrow 0$ as $t\rightarrow\infty$, so that the right edge state emerges as the steady state of this system, with most of the ultracold atoms already trapped in the right edge of the array. However, as $n_1$ increases and $n_2$ simultaneously decreases, with $n_1$ moving closer in magnitude to $n_3$, the crossover effect emerges, with the left edge state emerging as the steady state for this trapped ultracold atom system. 

\begin{figure}[htb]
\includegraphics[width=0.5\textwidth, height=0.2\textheight]{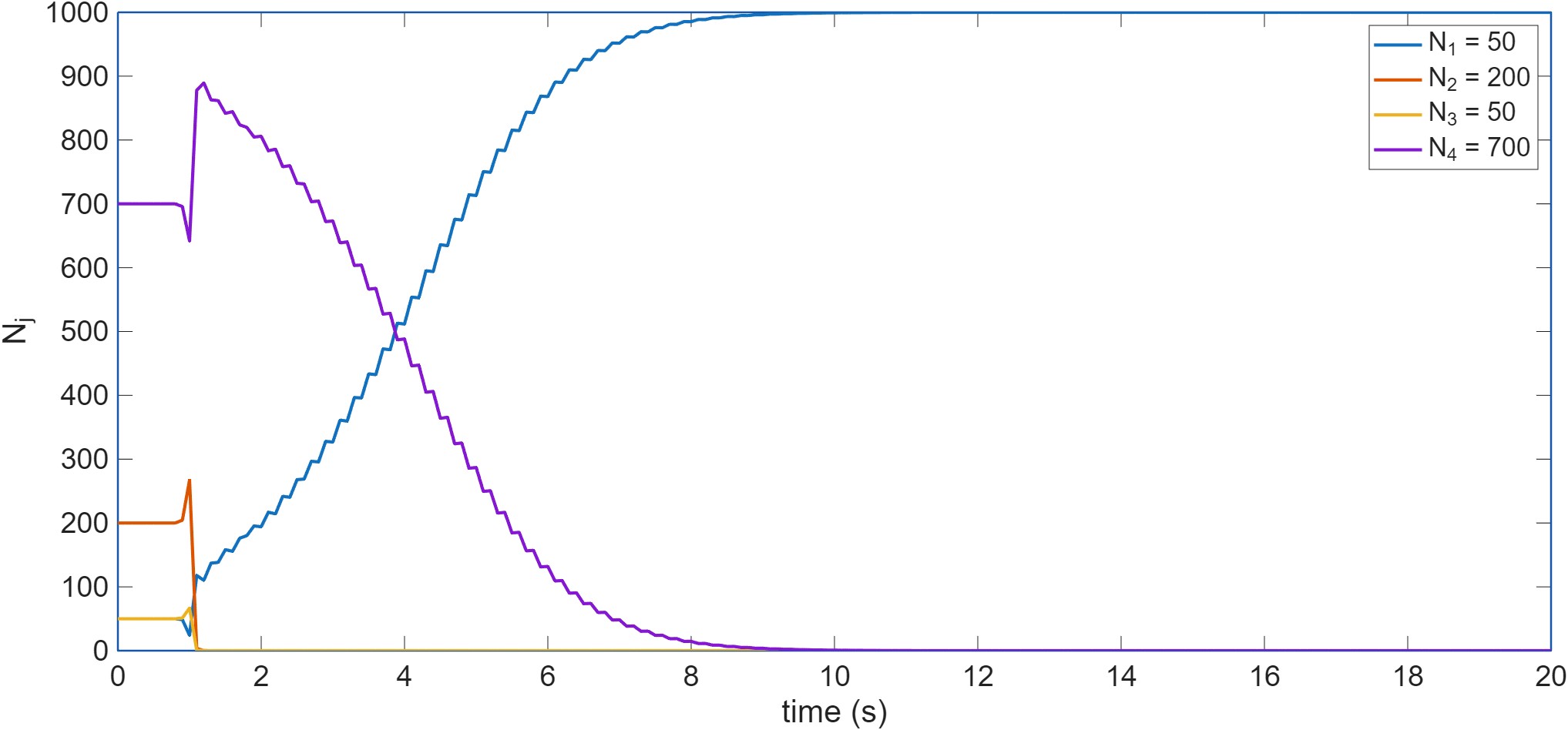}
\includegraphics[width=0.5\textwidth, height=0.2\textheight]{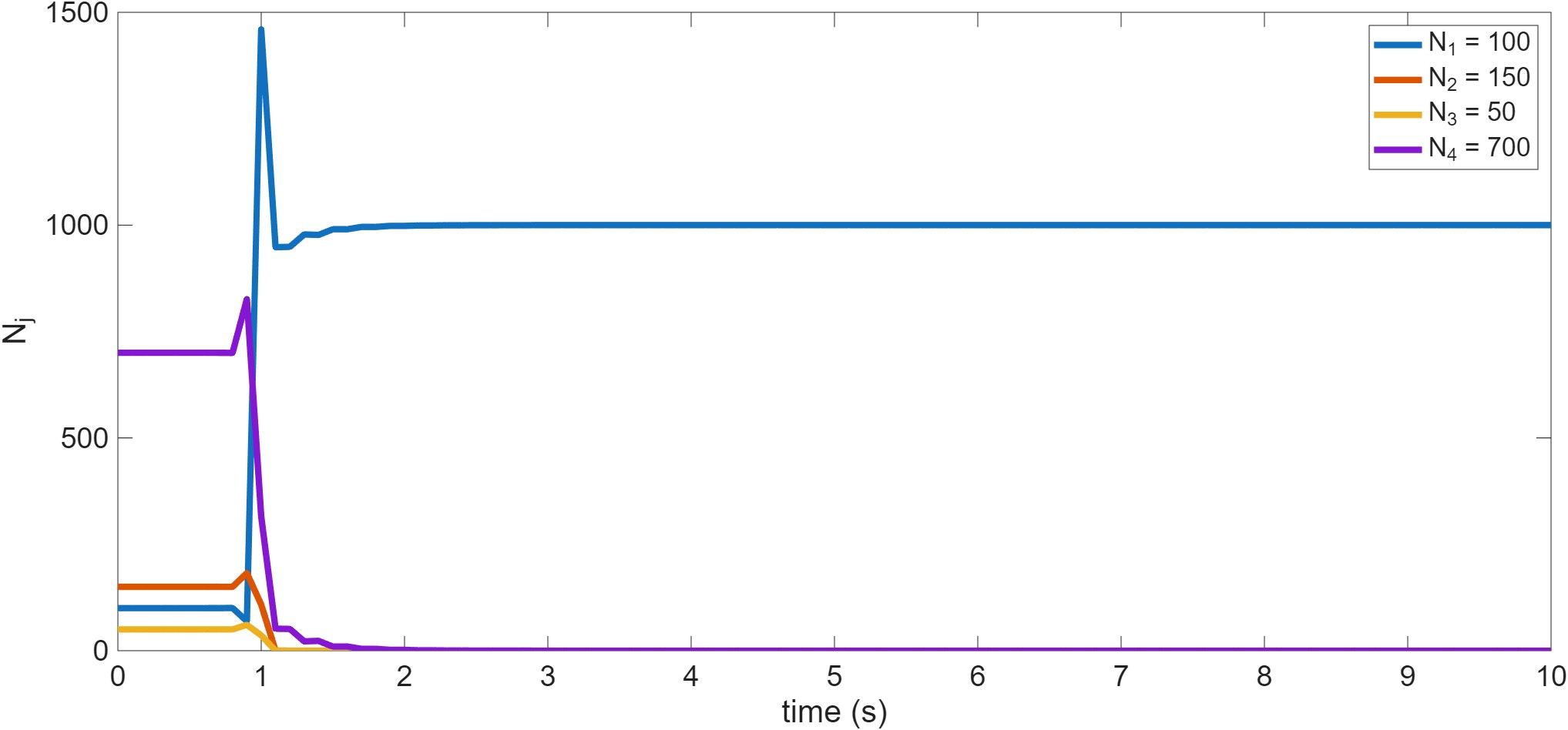}
\includegraphics[width=0.5\textwidth, height=0.2\textheight]{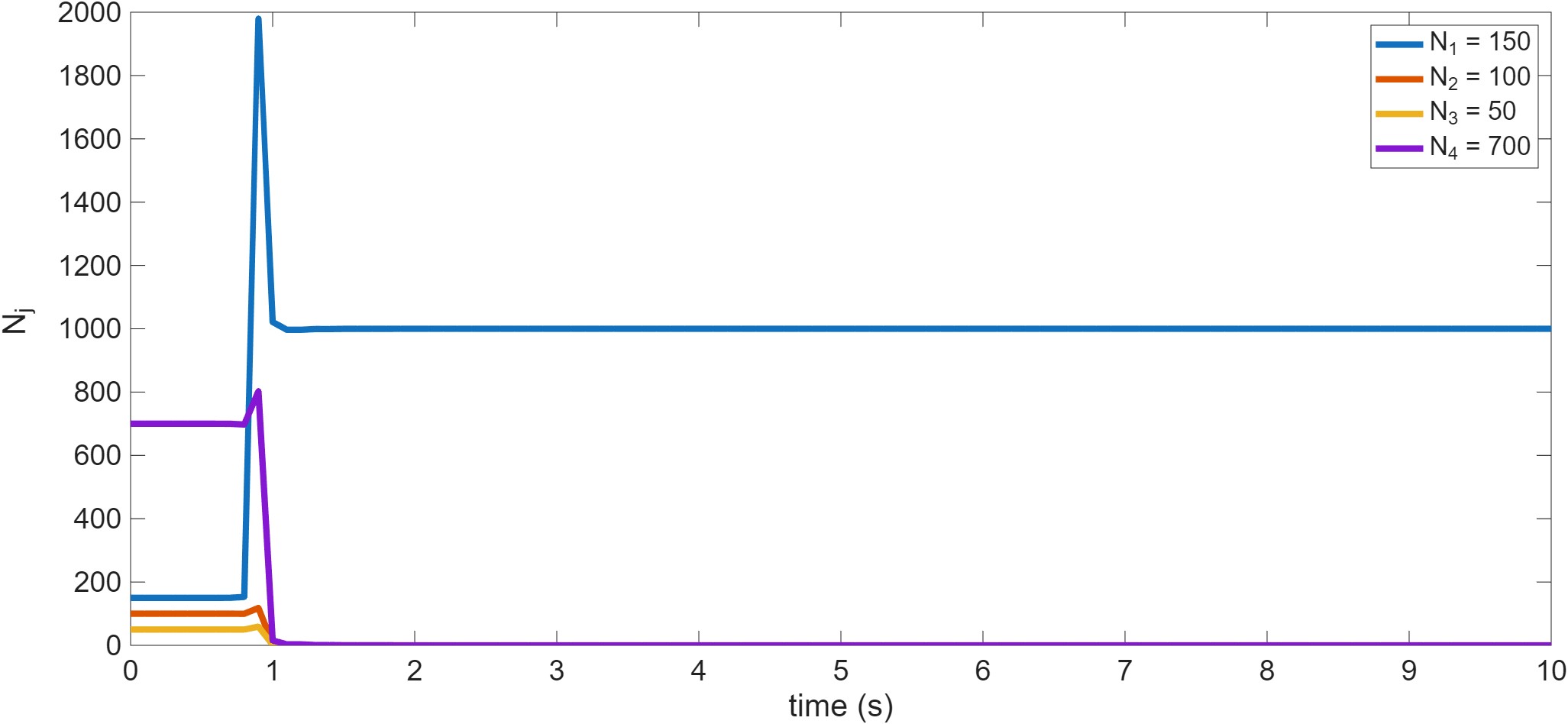}
\includegraphics[width=0.5\textwidth, height=0.2\textheight]{ExpValPartNumN1200N250N350N4700eps1eneg7A1eneg1.jpg}
\caption{\label{fig:numexpvalvaryingN1N24node} Time evolution of the expectation values of the particle number $n_j(t)$ in the harmonic traps located at $x_j = -3x_a$ $n_{1}(0)=N_1$, $x_j = -x_a$ $n_{2}(0)=N_2$, $x_j = x_a$ $n_3 (0) = N_3$ and $x_j = 3x_a$ $n_{3}(0)=N_3$, with the values of $N_1, N_2, N_3, N_4$ indicated for each plot, such that $N_4 = 700$, $N_1$ increases from $N_1 = 50$ to $N_1 = 200$, and $N_2$ decreases from $N_2 = 200$ to $N_2 = 50$. Here, $\epsilon_1 = 0.7$, $\epsilon_2 = 0.5$, $\varepsilon_e - \varepsilon_g = 1.00\times 10^{-7}$ and $A=0.1$.}
\end{figure}

Fig. (\ref{fig:numexpvalvaryingN1N24node}), on the other hand, demonstrates similar behavior for the case wherein we have seven harmonic potentials in the trapping array, with both $n_3$ and $n_4$ held fixed while $n_1$ and $n_2$ vary. For this case, we see that compared to the trapped ultracold atom system with five harmonic potentials in the trapping array, a smaller value of $n_1$ is sufficient to ensure a crossover from the right edge to the left edge state for the steady state of the system. This signifies that the number of harmonic potentials $2L+1$ in the trapping array will also determine the smallest possible value of $n_1$ that will result in the emergence of the crossover effect from the right edge state to the left edge state for the steady state of the trapped ultracold atom open quantum system. More importantly, this result indicates that left edge states are favored as the steady state for this trapped ultracold atom open quantum system, given that the magnitude of the critical value of $n_1$ that ensures the emergence of the crossover from the right edge to the left edge state for the system's steady state decreases as $2L+1$ increases. 

\begin{figure}[htb]
\includegraphics[width=0.5\textwidth, height=0.2\textheight]{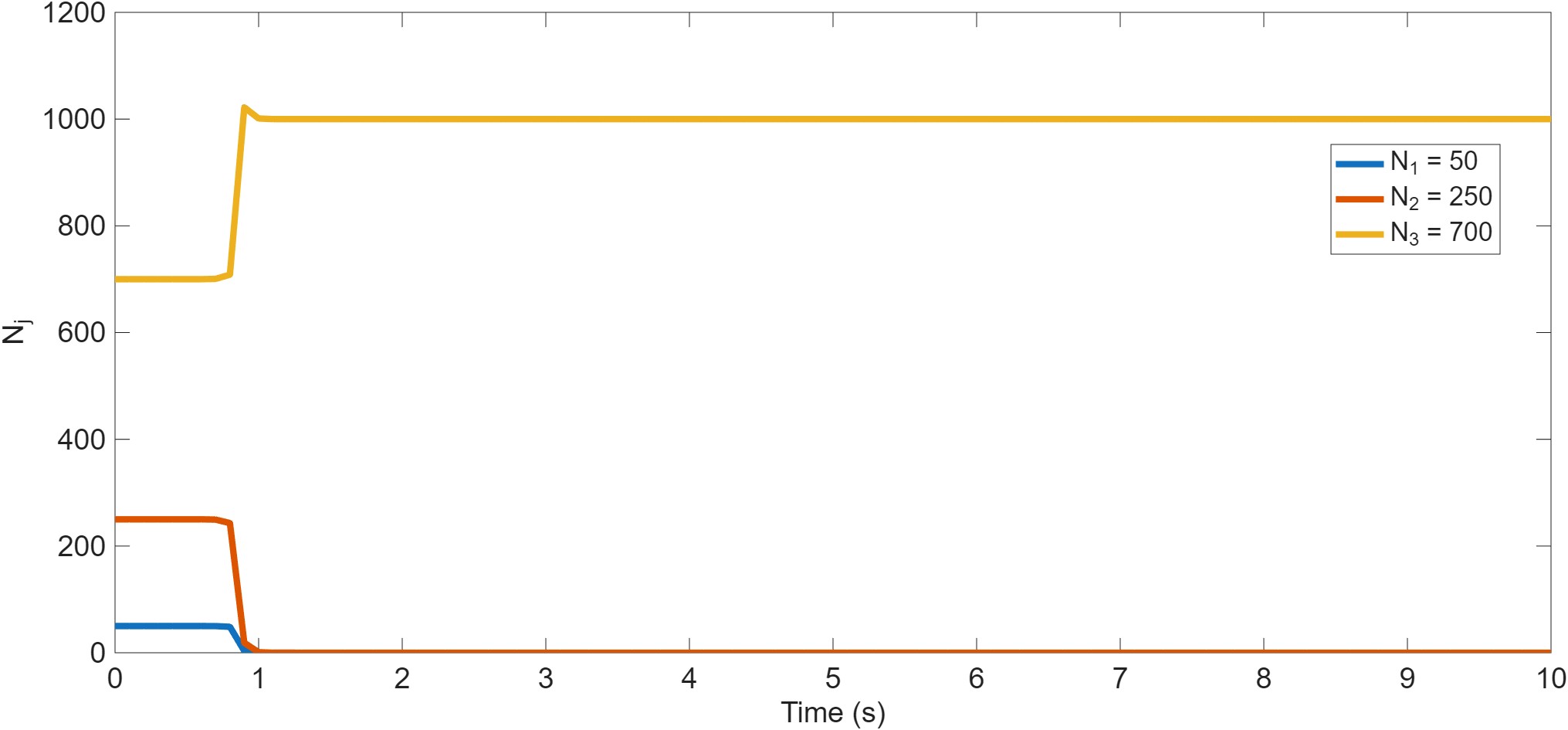}
\includegraphics[width=0.5\textwidth, height=0.2\textheight]{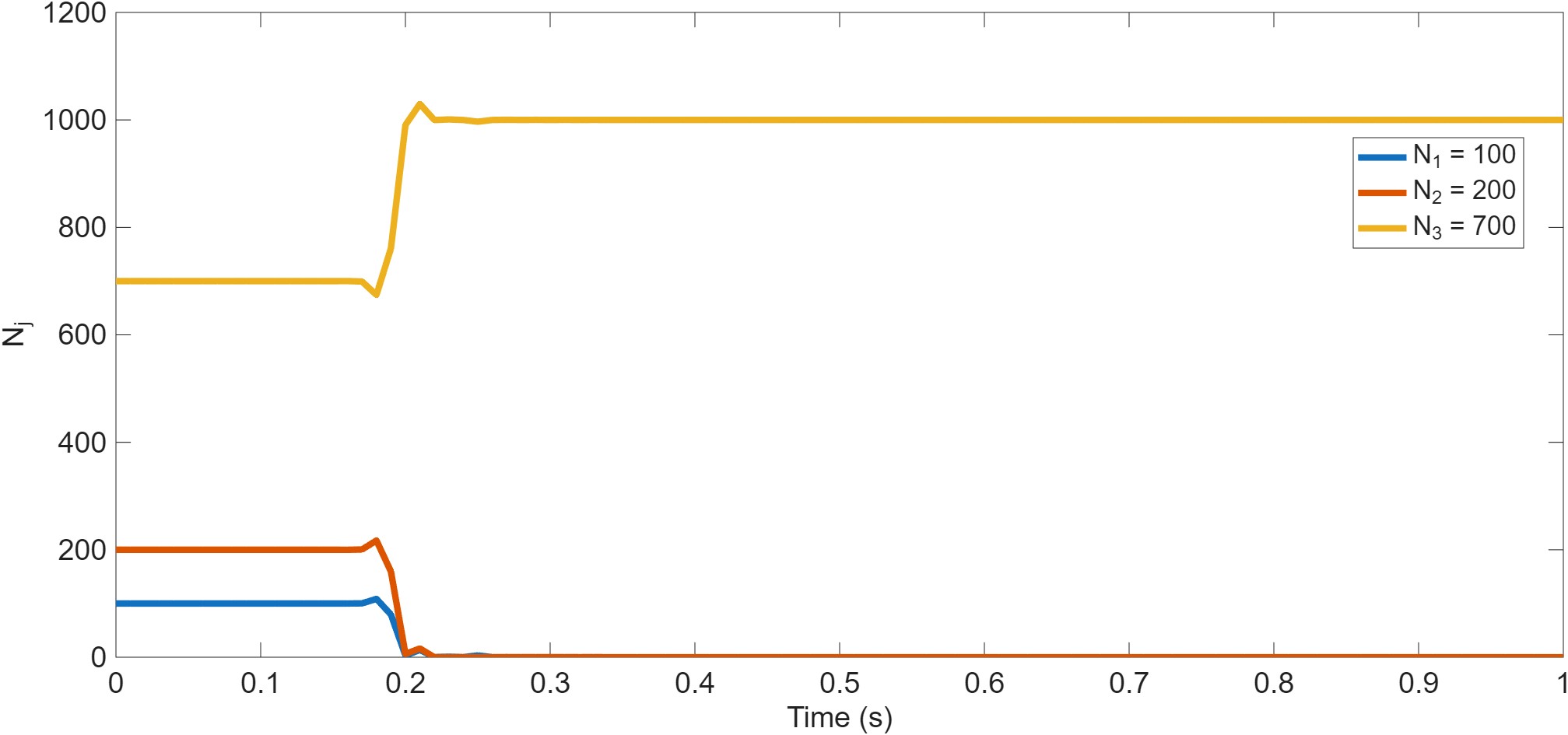}
\includegraphics[width=0.5\textwidth, height=0.2\textheight]{ExpValPartNumN1200N2100N3700eps1eneg7A1eneg1.jpg}
\includegraphics[width=0.5\textwidth, height=0.2\textheight]{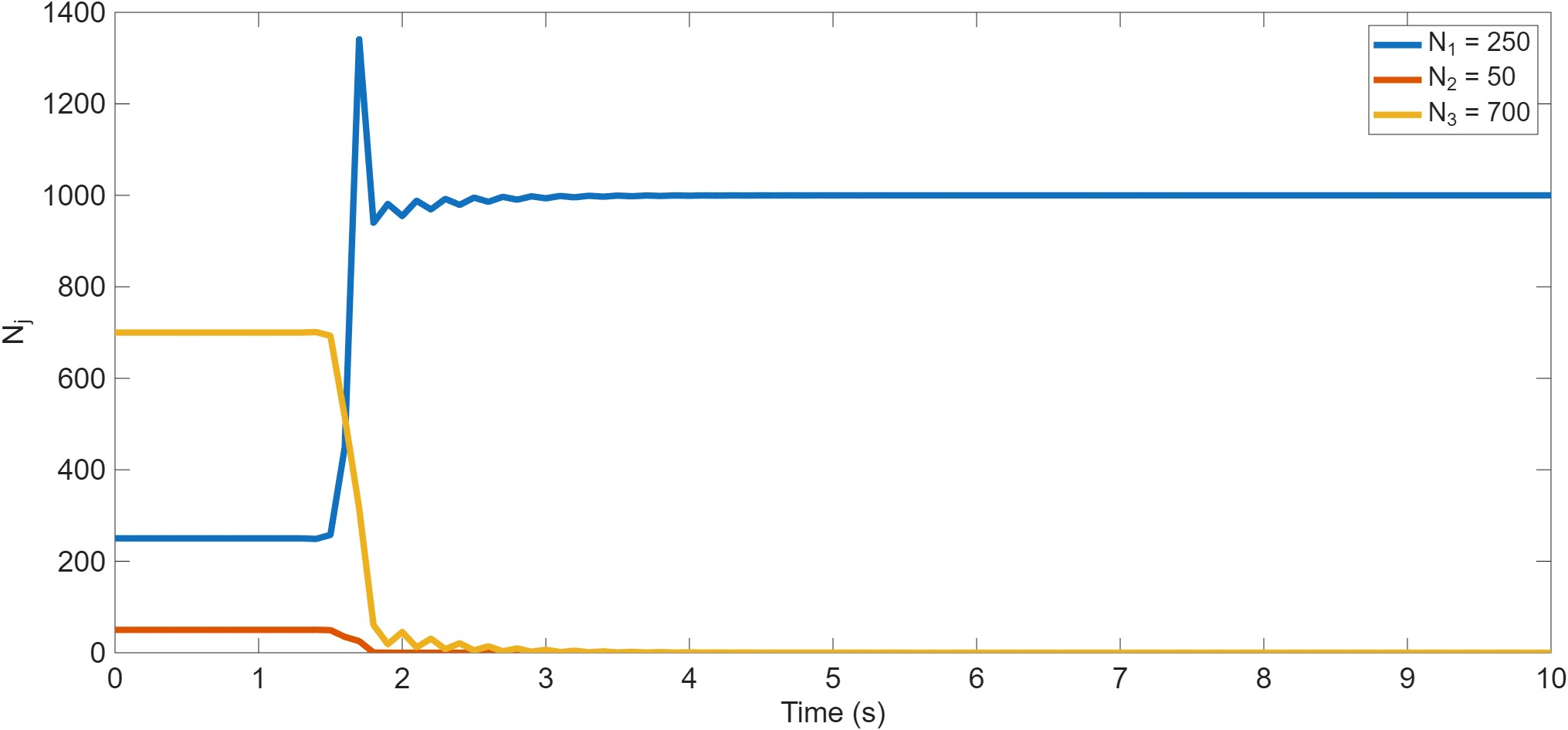}
\caption{\label{fig:numexpvalvaryingN1N23node} Time evolution of the expectation values of the particle number $n_j(t)$ in the harmonic traps located at $x_j = -2x_a$ (for which $n_{1}(0)=N_1$ here), $x_j = 0$ (where $n_{2}(0)=N_2$) and $x_j = 2x_a$ (where $n_{3}(0)=N_3$), with the values of $N_1, N_2, N_3$ indicated for each plot, such that $N_3 = 700$, $N_1$ increases from $N_1 = 50$ to $N_1 = 250$ and $N_2$ decreases from $N_3 = 250$ to $N_3 = 50$. Here, $\epsilon = 0.7$, $\varepsilon_e - \varepsilon_g = 1.00\times 10^{-7}$ and $A=0.1$.}
\end{figure}

\begin{figure}[htb]
\includegraphics[width=1.0\textwidth, height=0.3\textheight]{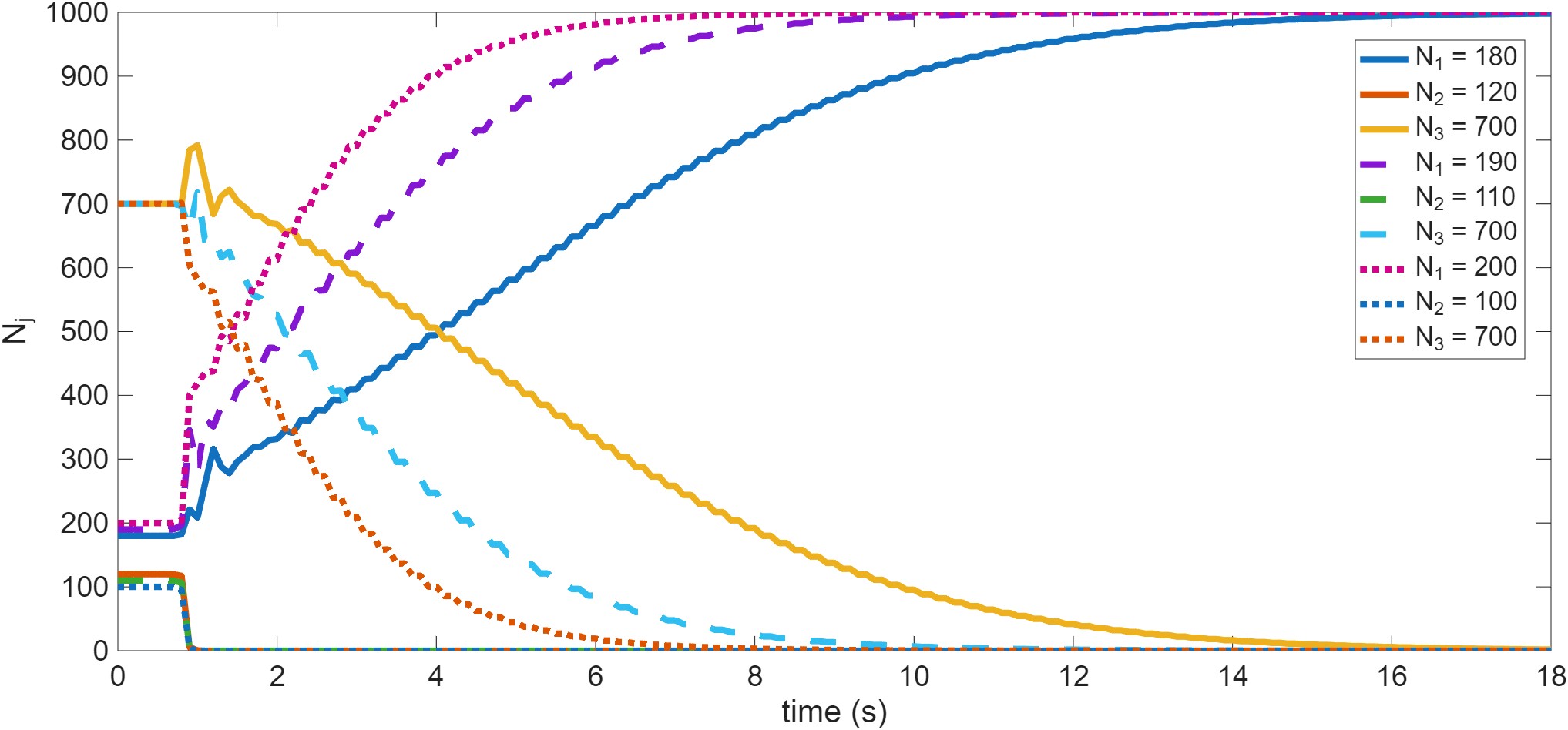}
\caption{\label{fig:numexpvalvaryingN1N23node2} Time evolution of the expectation values of the particle number $n_j(t)$ in the harmonic traps located at $x_j = -2x_a$ (for which $n_{1}(0)=N_1$ here), $x_j = 0$ (where $n_{2}(0)=N_2$) and $x_j = 2x_a$ (where $n_{3}(0)=N_3$), with the values of $N_1, N_2, N_3$ indicated for each plot, such that $N_1$ increases from $N_1 = 180$ to $N_1 = 200$, $N_2$ decreases from $N_2 = 120$ to $N_2 = 100$ and $N_3$ remains fixed at $N_3 = 700$. Here, $\epsilon = 0.7$, $\varepsilon_e - \varepsilon_g = 1.00\times 10^{-7}$ and $A=0.1$.}
\end{figure}

\begin{figure}[htb]
\includegraphics[width=1.0\textwidth, height=0.3\textheight]{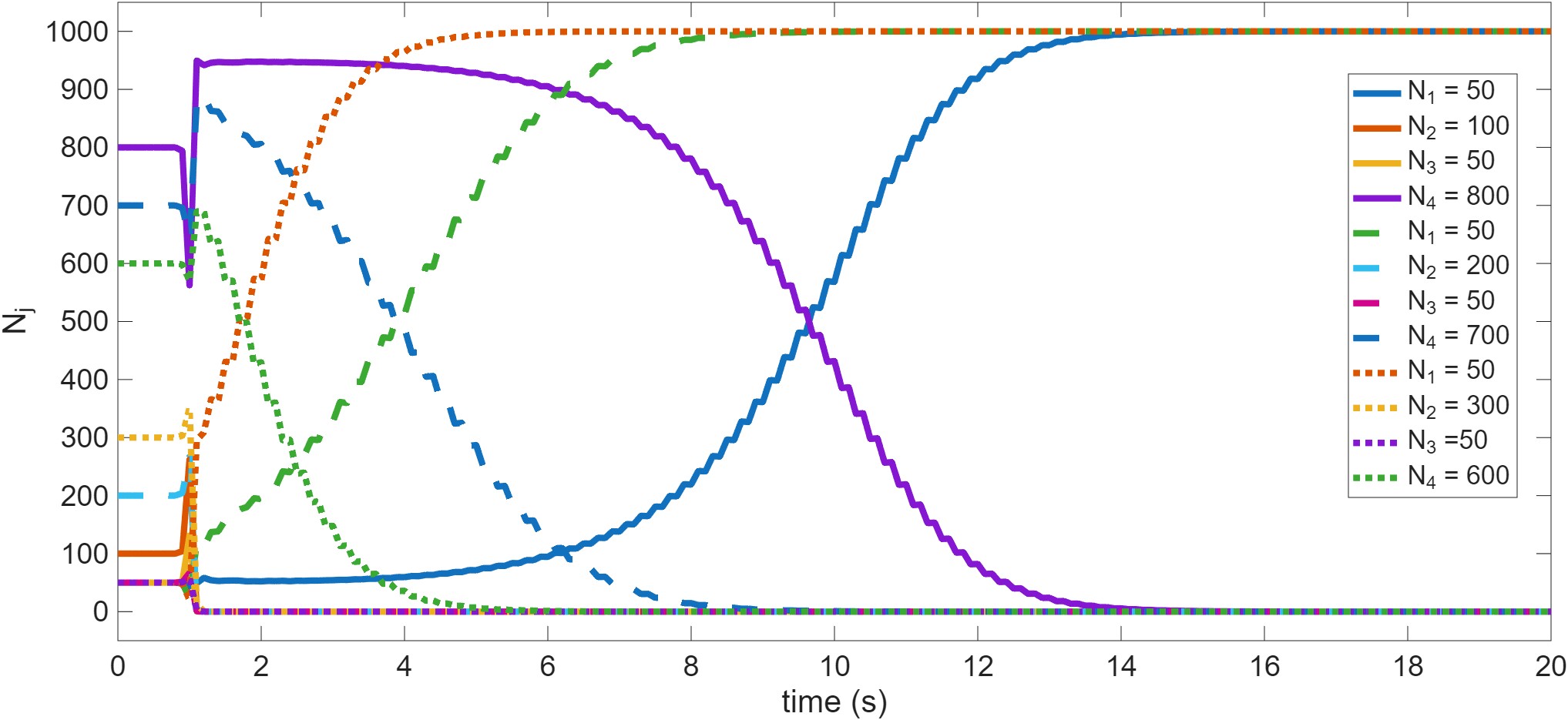}
\caption{\label{fig:numexpvalvaryingN2N44node} Time evolution of the expectation values of the particle number $n_j(t)$ in the harmonic traps located at $x_j = -3x_a$ (for which $n_{1}(0)=N_1$ here), $x_j = -x_a$ (where $n_{2}(0)=N_2$), $x_j = x_a$ (where $n_{3}(0)=N_3$) and $x_j = 3x_a$ (where $n_{4}(0)=N_3$), with the values of $N_1, N_2, N_3, N_4$ indicated for each plot, such that $N_1 = N_3 = 50$, $N_2$ increases from $N_2 = 100$ to $N_2 = 300$ and $N_4$ decreases from $N_4 = 800$ to $N_3 = 600$. Here, $\epsilon_1 = 0.7$, $\epsilon_2 = 0.5$, $\varepsilon_e - \varepsilon_g = 1.00\times 10^{-7}$ and $A=0.1$.}
\end{figure}

An interesting phenomenon can be seen in the time-evolved particle number expectation number values for the 5-harmonic potential trapped ultracold atom open quantum system as shown in Fig. (\ref{fig:numexpvalvaryingN1N23node2}), wherein $n_1$ increases from $n_1 = 180$ to $n_1 = 200$ while $n_2$ decreases from $n_2 = 120$ to $n_2 = 100$, all while $n_3$ remains constant at $n_3 = 700$. The graphs of $n_1 (t)$ and $n_3 (t)$ shown in these figures are, from a topological standpoint, similar to each other, with each successive graph looking like a compressed version of its predecessor. Furthermore, while the points of intersection of $n_1 (t)$ and $n_3 (t)$ occur at an earlier instant of time as $n_1$ increases and $n_3$ correspondingly decreases, the value of $n_j (t)$ for which $n_1 (t) = n_3 (t)$ for these three graphs are approximately equal to each other, being at $n_1 (t) = n_3 (t) = 500$. This value of $n_j (t), j=1,3$, also denotes the crossover point for $n_1 (t)$ and $n_3 (t)$, since, beyond this value of $n_j (t)$, $n_1 (t) > n_3 (t)$. As such, we can then say that these graphs indicate a topological invariant for this trapped ultracold atom open quantum system, namely the crossover point, or the point at which $n_1 (t) \approx n_3 (t)$, which is independent of the particular values of $n_1 $ and $n_3$, the initial number of atoms trapped at the left and right edges of this harmonic potential array. We can see similar behavior in Fig. (\ref{fig:numexpvalvaryingN2N44node}), which depicts the time-evolved particle number expectation values for the 7-harmonic potential trapped ultracold atom open quantum system, for which $n_1 = n_3 = 50$, $n_2$ increases from $n_2 = 100$ to $n_2 = 300$ and $n_4$ decreases from $n_4 = 800$ to $n_3 = 600$. For this case, we see a compression of the graphs of $n_1 (t)$ and $n_4 (t)$ as $n_4$ decreases, with a corresponding decrease in the magnitude of the instant of time $t$ when $n_1 (t) = n_4 (t)$, while the value for which these two particle number expectation values become equal to each other remains the same, being at $n_j (t) = 500, j=1,4$. 

\subsection{Reflection Symmetry of Trapped Ultracold Atom Open Quantum System}

It is of interest to determine the behavior of the system under reflection symmetry. We first note that under the reflection operator, $\hat{R}f(x)=f(-x)=\pm f(x)$ if the function $f(x)$ has either even (positive) or odd (negative) parity. Now if we apply $\hat{R}$ to the overlap integral in Eq. \ref{overlapint} and explicitly consider the limits of integration of the integral, we obtain

\begin{eqnarray}
&&\hat{R}\int_{-\infty}^{+\infty} dx e^{\pm ikx}\phi_{g,j}(x)\phi_{e,j'+1/2}(x)=\left(\frac{m_s \omega_g}{\pi\hbar}\right)^{1/4}\left(\frac{m_s \omega_e}{\pi\hbar}\right)^{1/4}\sqrt{\frac{2m_s \omega_e}{\hbar}}\nonumber\\
&&\times\int_{+\infty}^{-\infty} (-)dx\; e^{\mp ikx}\exp\left(-\frac{m_s \omega_g}{2\hbar}(x-x_j)^2\right)(-(x-x_{j'+1/2}))\exp\left(-\frac{m_s \omega_e}{2\hbar}(x-x_{j'+1/2})^2\right)\nonumber\\
&&\underset{\sigma_g \rightarrow 0}\longrightarrow -2\sigma_{g}^{1/2}\sigma_{e}^{-3/4}\int_{-\infty}^{+\infty} dx\;e^{\pm ikx}\delta(x-x_j)(x-x_{j'+1/2})\exp\left(-\frac{(x-x_{j'+1/2})^2}{2\sigma_e^2}\right)\nonumber\\
&&=-2\sigma_{g}^{1/2}\sigma_{e}^{-3/4}e^{\pm ikx_j}(x_j-x_{j'+1/2})\exp\left(-\frac{(x_j-x_{j'+1/2})^2}{2\sigma_e^2}\right)
\label{overlapintrefl}
\end{eqnarray}

Thus, we see that the overlap integral has odd parity, and consequently, so too will the interaction Hamiltonian describing the trapped ultracold atom open quantum system, as given by Eq. \ref{inthamexp2}. Following the line of reasoning in the previous section describing how the master equation for the trapped ultracold atom open quantum system is obtained, applying the reflection operator to the system then signifies that the jump operator will also have odd parity, i. e.

\begin{equation}
\hat{R}\hat{c}=-\hat{c}=\sum_{j=1}^{L}\epsilon_{j}(\hat{a}_{g,j+1}^{\dagger}+\hat{a}_{g,j}^{\dagger})(\hat{a}_{g,j+1}-\hat{a}_{g,j})
\label{jumpoprefsymm}
\end{equation}

Now looking at the trapped ultracold atom open quantum system as illustrated in Fig. \ref{fig:harmonictraparrayschem}, we can see that, because the system's jump operator has odd parity, the system can demonstrate reflection symmetry if we select the laser frequencies $\Omega_j$ and the corresponding detunings $\Delta_j$ such that $\Omega_j = \Omega_{j+1}$. As an illustration, we consider the case where we have five harmonic potentials trapping a gas of ultracold atoms, illustrated in Fig. \ref{fig:5HOTrapArraySchematic}, such that initially, the trapped ultracold atoms are contained in three harmonic potentials centered at $x = 0$ and $x = \pm 2x_a$. From the figure, we can see that this trapped ultracold atom open quantum system will be symmetric about $x = 0$ if the Rabi frequencies of the Raman lasers exciting the atoms initially trapped at the harmonic potentials centered at $x = 0$ and $x = \pm 2x_a$ are equal to each other, i. e. $\Omega_1 = \Omega_2 = \Omega$. At the same time, the detuning of the excited energy levels at the harmonic potentials centered at $x = \pm x_a$ must be equal to each other, i. e. $\Delta_1 = \Delta_2 = \Delta$. This implies that, from Eq. \ref{3nodejumpop}, which gives the explicit form of the jump operator $\hat{c}$ for this system, the value of $\epsilon$ in the jump operator is $\epsilon = 1$, so that explicitly,

\begin{equation}
\hat{c}=(\hat{a}^{\dagger}_{g,1}+\hat{a}^{\dagger}_{g,2})(\hat{a}_{g,1}-\hat{a}_{g,2})+(\hat{a}^{\dagger}_{g,2}+\hat{a}^{\dagger}_{g,3})(\hat{a}_{g,2}-\hat{a}_{g,3})
\label{3nodejumpoprefsymm}
\end{equation} 

\begin{figure}[htb]
\includegraphics[width=0.5\textwidth, height=0.2\textheight]{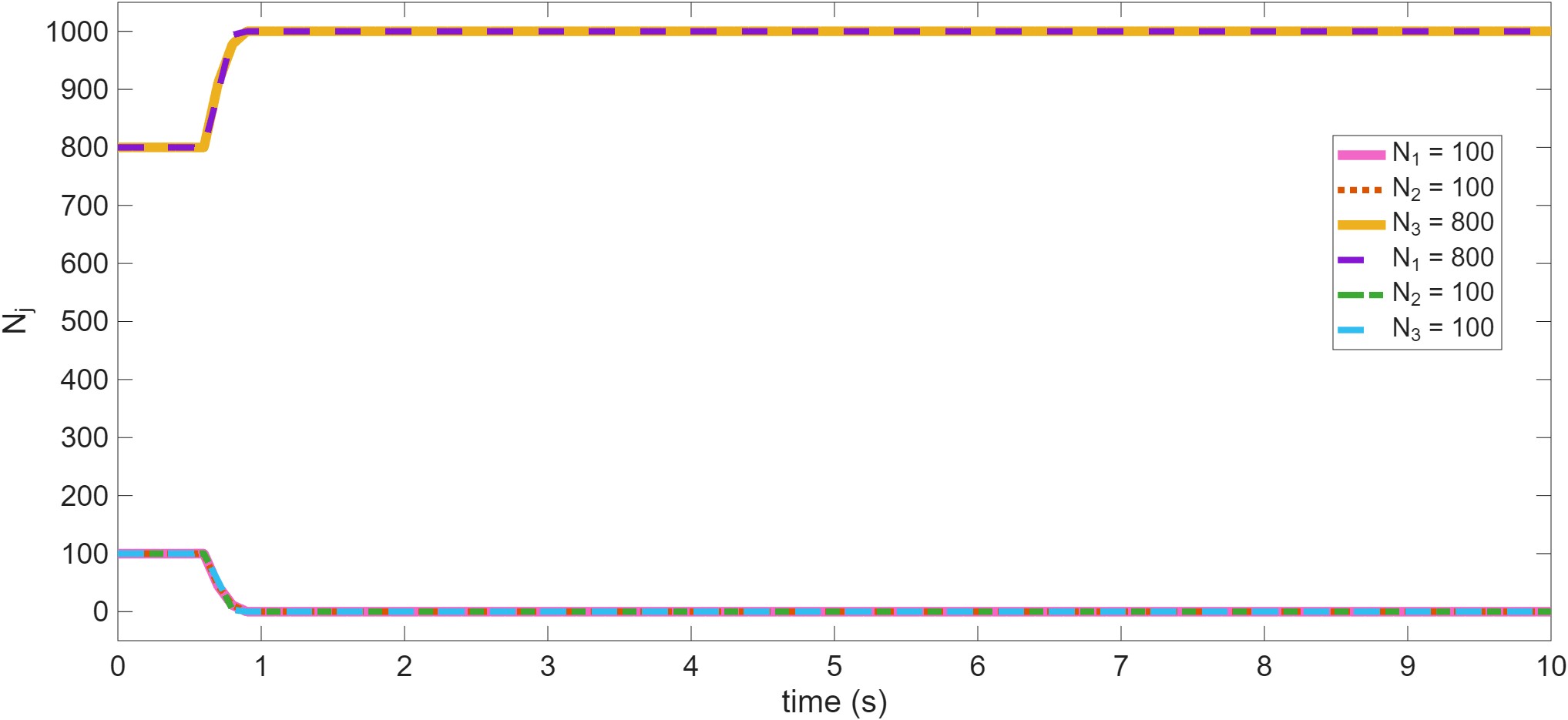}
\includegraphics[width=0.5\textwidth, height=0.2\textheight]{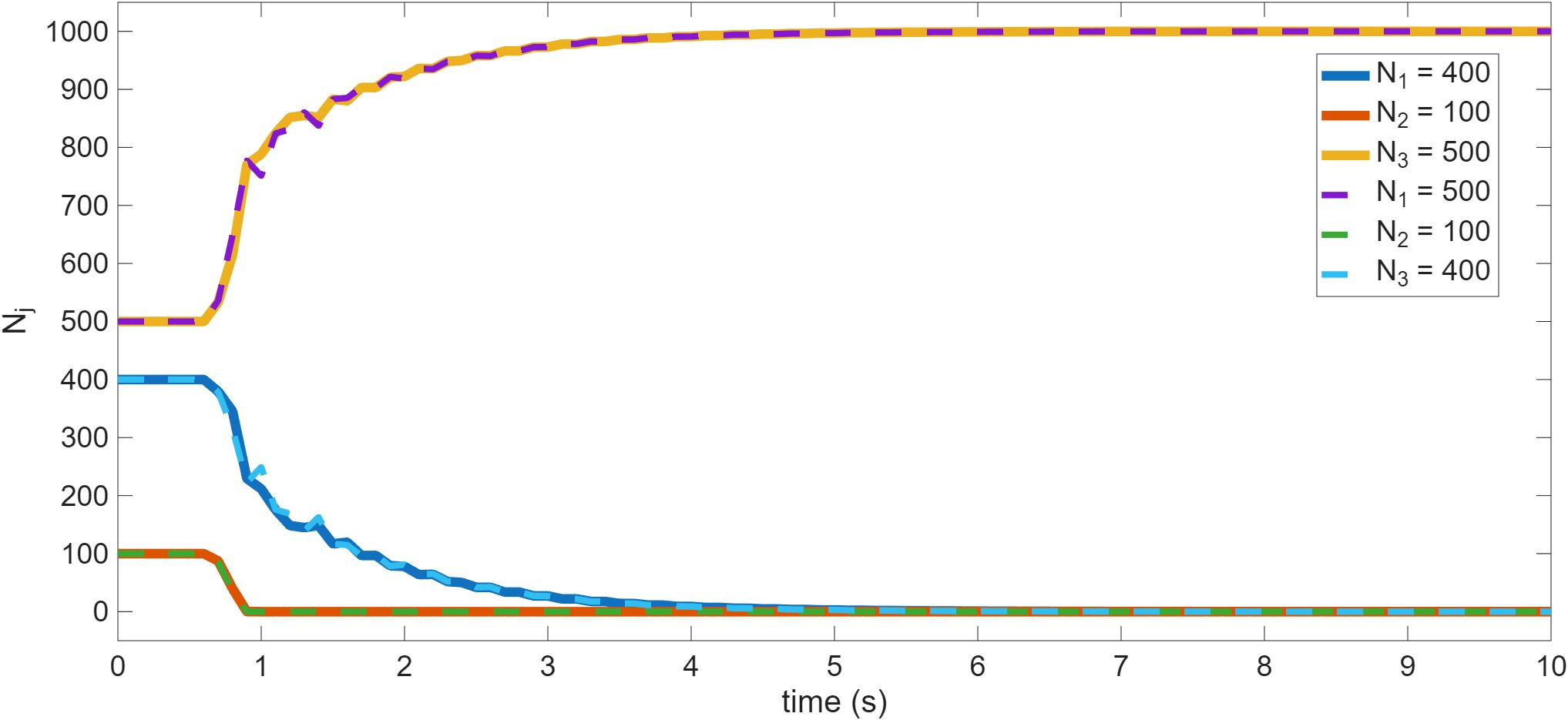}
\includegraphics[width=0.5\textwidth, height=0.2\textheight]{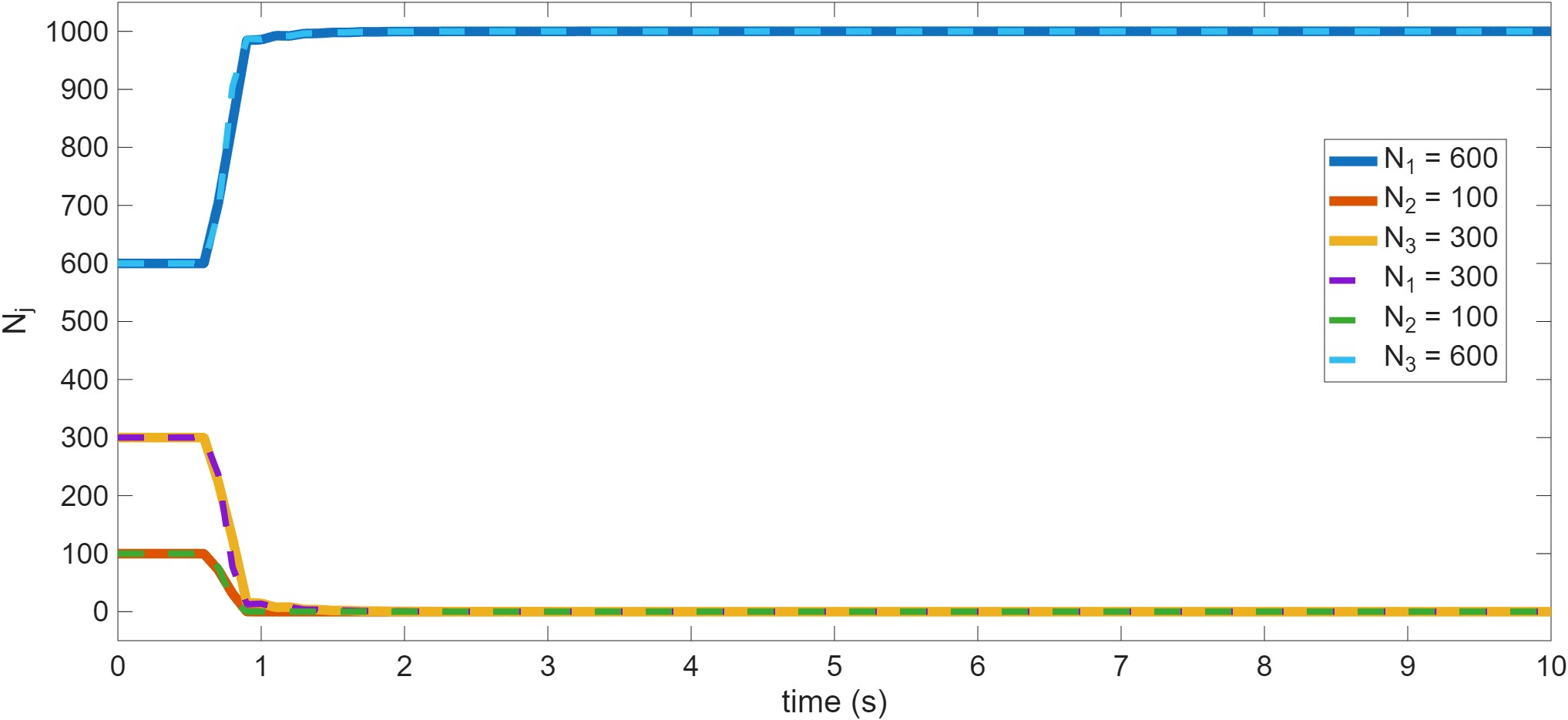}
\includegraphics[width=0.5\textwidth, height=0.2\textheight]{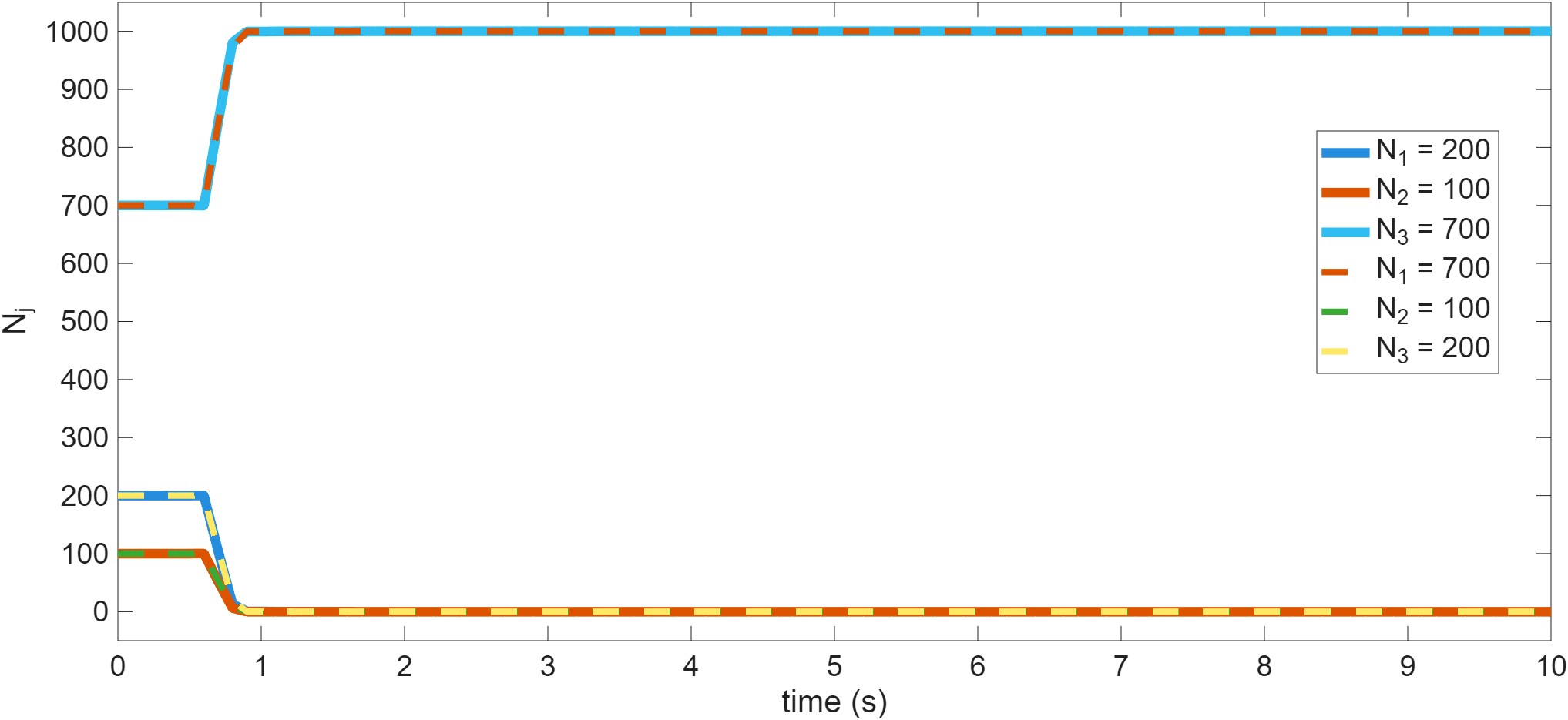}
\caption{\label{fig:numexpvalvaryingN1N33nodesymm} Time evolution of the expectation values of the particle number $n_j(t)$ in the harmonic traps located at $x_j = -2x_a$ (for which $n_{1}(0)=N_1$ here), $x_j = 0$ (where $n_{2}(0)=N_2$) and $x_j = 2x_a$ (where $n_{3}(0)=N_3$), with the values of $N_1, N_2, N_3$ indicated for each plot, such that $N_2 = 100$ and the values of $N_1$ and $N_3$ are inverted. Here, $\epsilon = 1$, $\varepsilon_e - \varepsilon_g = 1.00\times 10^{-7}$ and $A=0.1$.}
\end{figure}

The resulting particle number expectation values at the ground state of the harmonic traps centered at $x = 0, \pm 2x_a$, denoted by $N_j (t)$, are shown for various initial values of $N_j (t)$ in Fig. \ref{fig:numexpvalvaryingN1N33nodesymm}. These expectation values are calculated using the time-evolved density matrix describing the system, with the time evolution due to the master equation given by Eq. \ref{masteqnharmonicarraytrappedultracoldatomoqs} and the jump operators given by Eq. \ref{3nodejumpoprefsymm}. The figure shows that if the trapped ultracold atom open quantum system undergoes a reflection transformation about $x = 0$, i. e. $x_j\rightarrow -x_j$, then so too will the resulting steady-state values of the particle number expectation values at each harmonic trap, i. e. right - edge steady states will become left - edge steady states, and vice versa, under this reflection transformation.

Let us now consider the case where we have seven harmonic potential traps, with the ultracold atoms initially trapped in four of these potentials, which are centered about $x = \pm 3x_a$ and $x = \pm x_a$. It can be seen from Fig. \ref{fig:7HOTrapArraySchematic} that, as with the previous case, this trapped ultracold atom open quantum system has reflection symmetry about the trap centered at $x = 0$ if the Rabi frequencies $\Omega_1$ and detunings $\Delta_1$ of the lasers coupling the ground state energies of the harmonic potentials centered at $x=-3x_a$ and $x=-x_a$ to the excited state energy of the harmonic potential centered at $x = -2x_a$ are equal to the Rabi frequencies $\Omega_3$ and detunings $\Delta_3$ of the lasers coupling the ground state energies of the harmonic potentials centered at $x = x_a$ and $x = 3x_a$ to the excited state energy of the harmonic potential centered at $x = 2x_a$, i. e. $\Omega_1 = \Omega_3$ and $\Delta_1 = \Delta_3$. This implies that, from Eq. \ref{4nodejumpop}, which gives the explicit form of the jump operator $\hat{c}$ for this system, the value of $\epsilon_2$ in the jump operator is $\epsilon_2 = 1$, so that explicitly, with $\epsilon_1 = \epsilon$,

\begin{equation}
\hat{c}=(\hat{a}^{\dagger}_{g,1}+\hat{a}^{\dagger}_{g,2})(\hat{a}_{g,1}-\hat{a}_{g,2})+\epsilon(\hat{a}^{\dagger}_{g,2}+\hat{a}^{\dagger}_{g,3})(\hat{a}_{g,2}-\hat{a}_{g,3})+(\hat{a}^{\dagger}_{g,3}+\hat{a}^{\dagger}_{g,4})(\hat{a}_{g,3}-\hat{a}_{g,4})
\label{4nodejumpoprefsymm}
\end{equation} 

\begin{figure}[htb]
\includegraphics[width=0.5\textwidth, height=0.2\textheight]{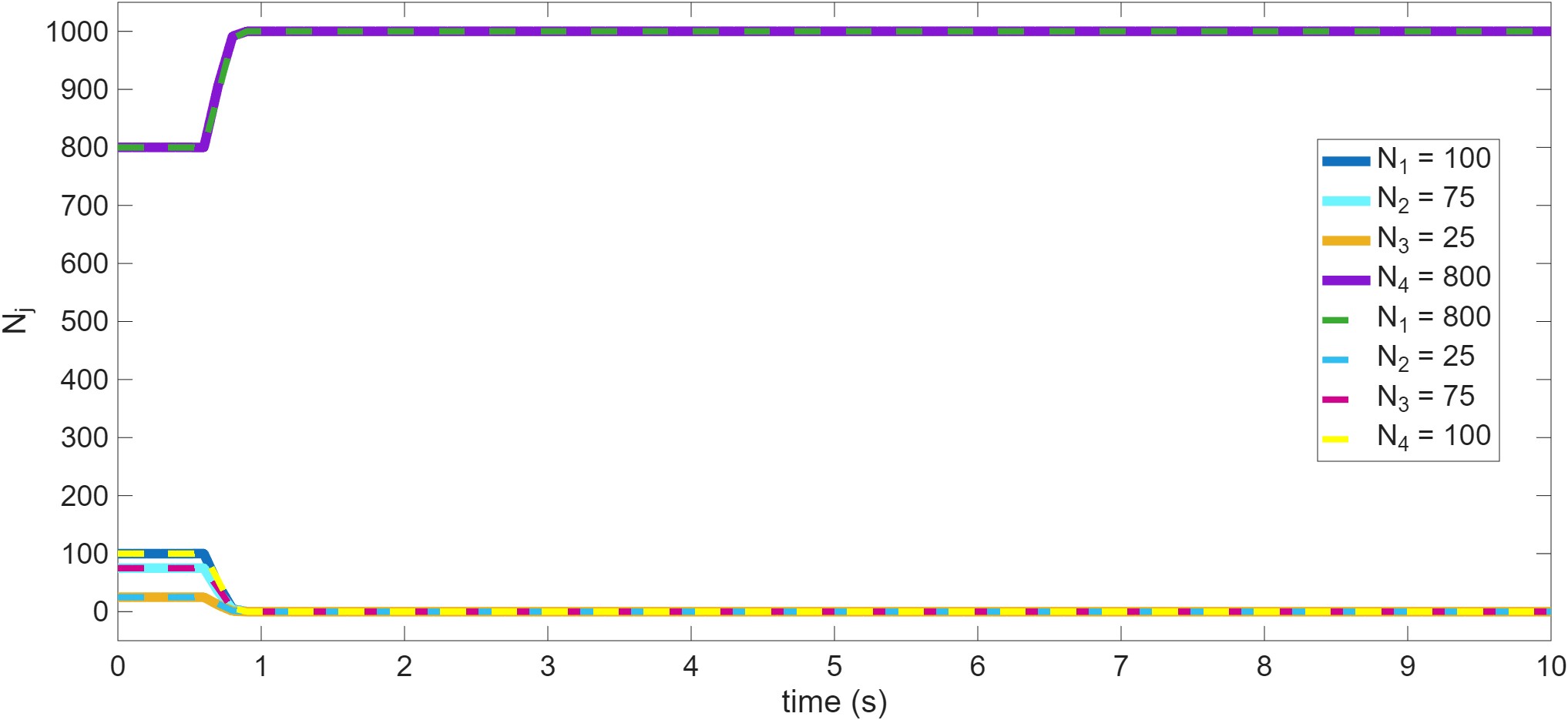}
\includegraphics[width=0.5\textwidth, height=0.2\textheight]{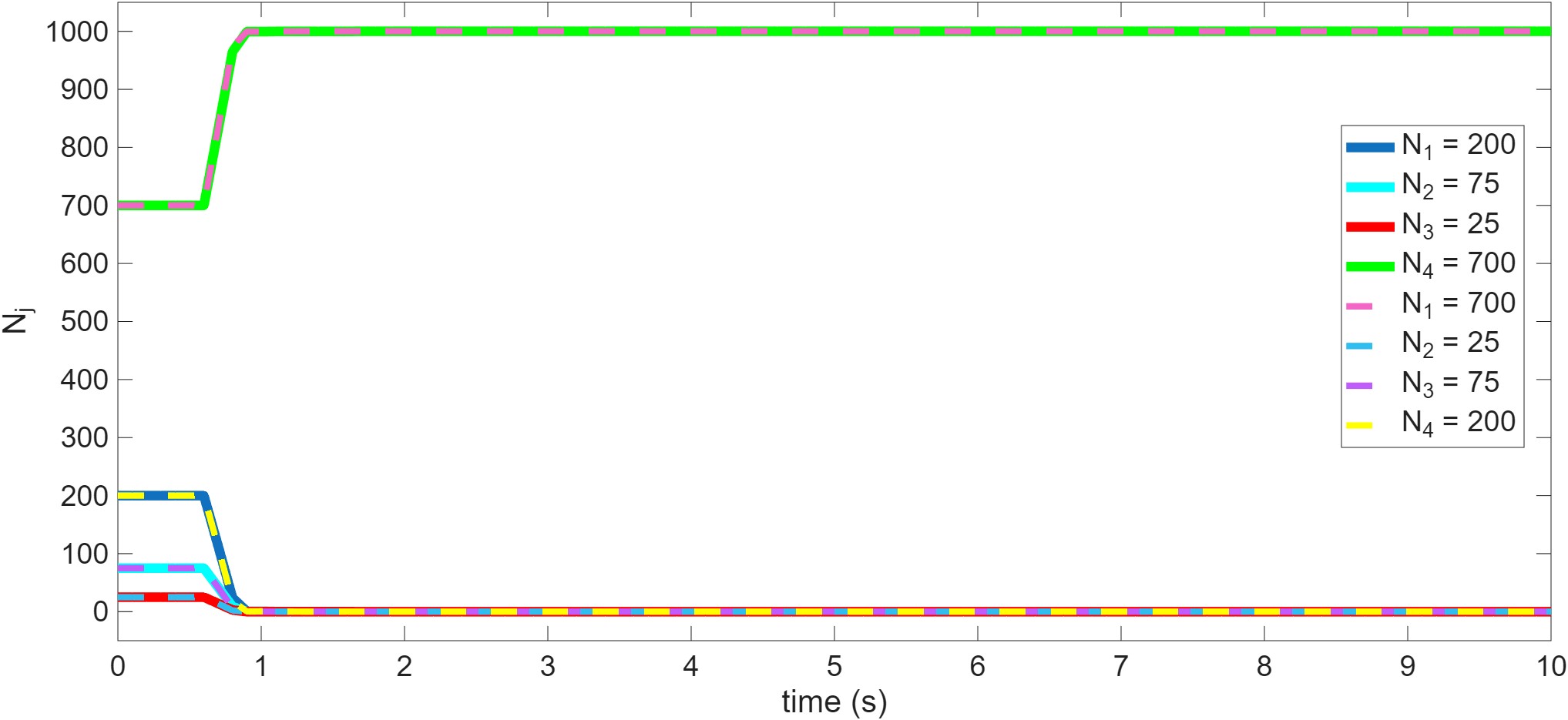}
\includegraphics[width=0.5\textwidth, height=0.2\textheight]{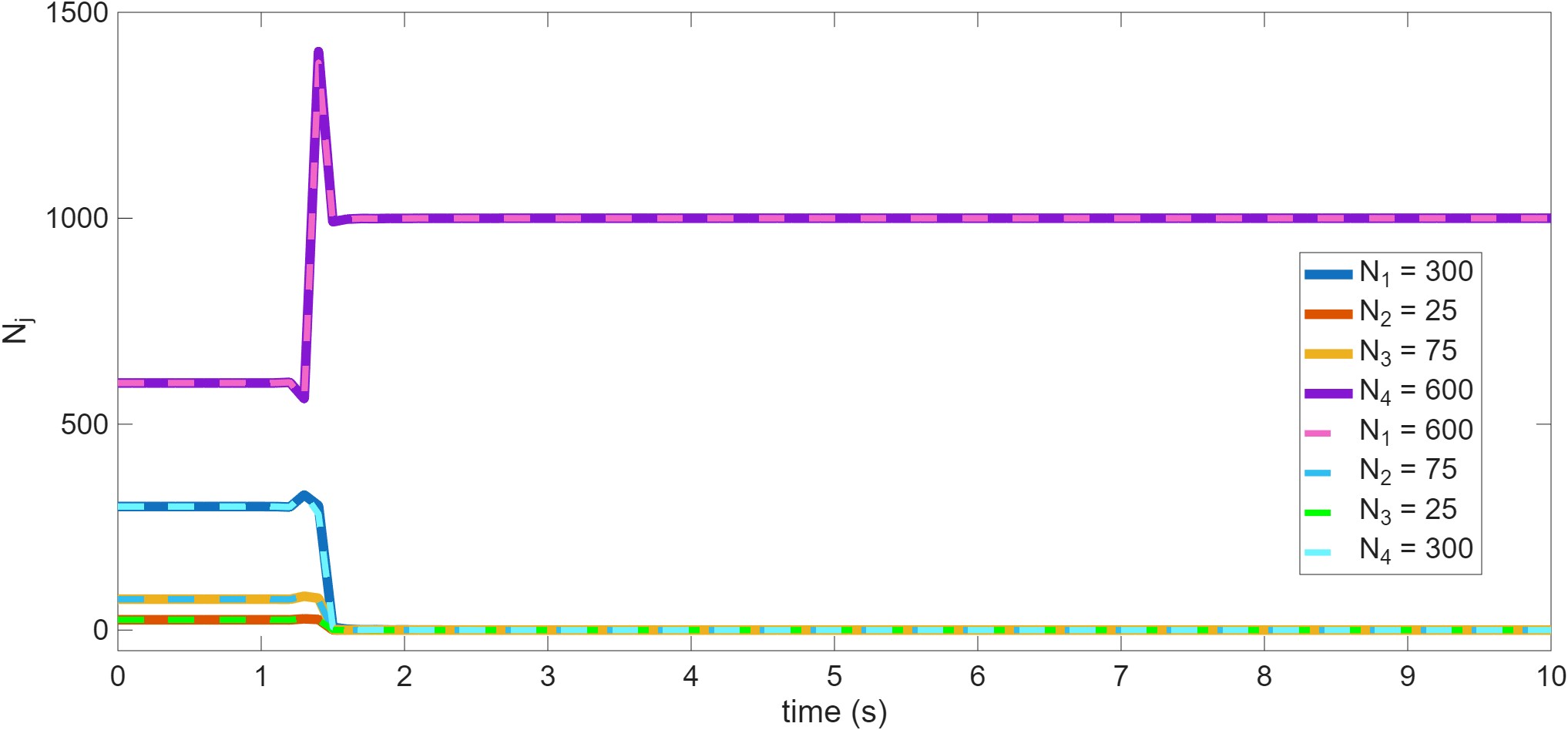}
\includegraphics[width=0.5\textwidth, height=0.2\textheight]{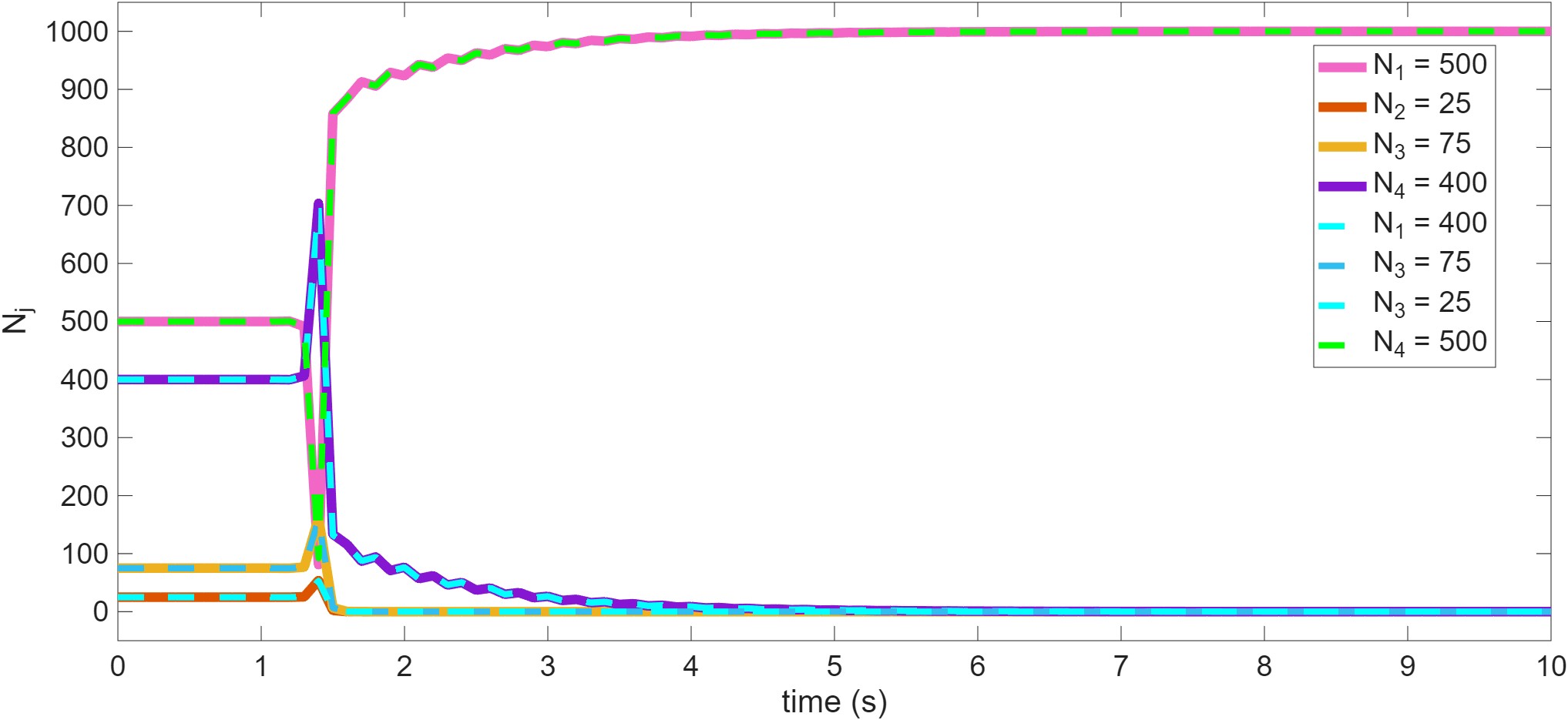}
\caption{\label{fig:numexpvalvaryingN1N44nodesymm} Time evolution of the expectation values of the particle number $n_j(t)$ in the harmonic traps located at $x_j = -3x_a$ (for which $n_{1}(0)=N_1$ here), $x_j = -x_a$ (where $n_{2}(0)=N_2$), $x_j = x_a$ (where $n_{3}(0)=N_2$) and $x_j = 3x_a$ (where $n_{4}(0)=N_4$), with the values of $N_1, N_2, N_3, N_4$ indicated for each plot, such that the values of $N_1$ and $N_3$ are inverted, and the same with the values of $N_2$ and $N_4$. Here, $\epsilon_1 = \epsilon_3 = 1$, $\epsilon_2=0.5$, $\varepsilon_e - \varepsilon_g = 1.00\times 10^{-7}$ and $A=0.1$.}
\end{figure}

The resulting particle number expectation values at the ground state of the harmonic traps centered at $x = 0, \pm 2x_a$, denoted by $N_j (t)$, are shown for various initial values of $N_j (t)$ in Fig. \ref{fig:numexpvalvaryingN1N44nodesymm}. Similar to the five harmonic potential trap case, these expectation values are calculated using the time-evolved density matrix describing the system, with the time evolution due to the master equation given by Eq. \ref{masteqnharmonicarraytrappedultracoldatomoqs} and the jump operators given by Eq. \ref{4nodejumpoprefsymm}. Fig. \ref{fig:numexpvalvaryingN1N44nodesymm} shows that, just as with the five harmonic potential trap case, if the trapped ultracold atom open quantum system undergoes a reflection transformation about $x = 0$, i. e. $x_j\rightarrow -x_j$, then so too will the resulting steady-state values of the particle number expectation values at each harmonic trap, i. e. right - edge steady states will become left - edge steady states, and vice versa, under this reflection transformation, irrespective of the number of atoms trapped in the potentials located at the middle of the array, i. e. at the harmonic potentials centered at $x=\pm x_a$.

These results demonstrate, conclusively, that if there is a symmetric specification of the Rabi frequencies and detunings of the Raman lasers exciting the ultracold atoms trapped at the ground state of the harmonic potentials centered at $x=(2j-L-2)x_a$ and $x=(2j-L)x_a$ to the excited state of the harmonic potential centered at $x=(2j-L-1)x_a$, i. e, $\Omega_1 = \Omega_L$ and $\Delta_1 = \Delta_L$, $\Omega_2 = \Omega_{L-1}$ and $\Delta_2 = \Delta_{L-1}$, etc., then the trapped ultracold atom open quantum system becomes reflection symmetric, and the resulting edge steady states will display reflection symmetry as well, i. e. if the steady state of the trapped ultracold atom open quantum system is a left edge steady state, then under reflection symmetry and a symmetric choice of the system parameters, the edge steady state is transformed into a right edge steady state, and vice versa. 

More importantly, these results demonstrate the dependence of the dynamical behavior of the trapped ultracold atom system on the choice of Rabi frequencies and detunings $\Omega_j$ and $\Delta_j$, respectively, for the system. As mentioned earlier, the values of the parameters $\epsilon_j$ in the jump operators given by Eqs. (\ref{3nodejumpop}) and (\ref{4nodejumpop}) in the master equation given by Eq. (\ref{masteqnharmonicarraytrappedultracoldatomoqs}) are all less than 1. Consequently, for a given set of initial particle number expectation values $\left\{n_j\right\}_{j=1}^{L+1}$ for the harmonic potentials centered at $x=(2j-L-2)x_a$, the resulting particle number steady states favored by these systems are left edge steady states. However, by varying the values of the parameters $\epsilon_j$ in these jump operators such that $\epsilon = 1$ for the five harmonic trapping potential trapped ultracold atom open quantum system and $\epsilon_2 = 1$ for the seven harmonic trapping potential trapped ultracold atom open quantum system, we obtain the reflection - symmetric behavior of the edge steady states emerging in these systems. This signifies that it is possible for us to change the preferred edge steady states by varying the values of the Rabi frequencies and detunings $\Omega_j$ and $\Delta_j$ coupling ground and excited energy levels in adjacent harmonic potentials for the system. 

\section{Discussion of Results}

It can be seen from the numerical results from the previous sections that the edge states, characterized by the number of particles in that state being equal to the total number of atoms in the ultracold atom gas, will emerge as steady states of the trapped ultracold atom open quantum system, irrespective of the number of harmonic potentials used to trap these atoms. At the same time, the manner in which these edge states emerge will depend on the number of atoms initially trapped in the harmonic potentials located at points $x_j = (2j-L-2)x_a, j=1,2,...,L+1, x_a\in\Re$ on the trap array, and the type of edge states that will emerge (left edge states or right edge states) will depend on the number of atoms initially trapped at the edges of the trap array, located at $x_1 = -Lx_a$ and $x_{L+1} = Lx_a$. The emergence of these edge states as unique steady states of the trapped ultracold atom open quantum system, and their robustness despite the reduction of the number of atoms initially trapped in one or both of the edges of the harmonic potential trapping array, is characteristic of the non-Hermitian skin effect in open quantum systems, whereby a large, macroscopic number of eigenstates are localized over time at the boundaries of the system. As mentioned earlier in this paper, this non-Hermitian skin effect is significant in open quantum systems as it is a topological effect that in turn is a manifestation of a non-Hermitian bulk-boundary correspondence in the system, indicating that the spectrum of the Hamiltonian for this system has non-trivial topological features. It would thus be of interest to investigate the spectrum of the Hamiltonian of this trapped ultracold atom open quantum system and determine the topological properties of the eigenvalues of this system's Hamiltonian, and in doing so provide further insight into the rich dynamical behavior underlying this form of driven quantum matter. 

Furthermore, the results demonstrate that, if the system parameters are chosen symmetrically, then the resulting trapped ultracold atom open quantum system becomes reflection symmetric, such that under a reflection transformation, left edge steady states will become right edge steady states, and vice versa. Aside from demonstrating how the system becomes symmetric under reflection, this also demonstrates the dependence of the system's dynamics on its physical parameters, specifically the Rabi frequencies and detunings $\Omega_j$ and $\Delta_j$ of the Raman lasers coupling the ground states of the harmonic potentials centered at $x = (2j-L-2)x_a$ and $x = (2j-L)x_a$ to the excited state of the harmonic potential centered at $x = (2j-L-1)x_a$. By varying the value of these Rabi frequencies and detunings, not only can we alter the symmetry of the system, but we can also alter the nature of the steady states emerging in the system, allowing us to actively select the types of steady states that we would like to emerge in the trapped ultracold atom open quantum system.

It is apparent that the emergence of these unique edge steady states will only depend on the number of atoms initially trapped at each edge of the harmonic potential array. In particular, right edge steady states will emerge if and only if $n_{L+1}>>n_1$, while left edge steady states will emerge if $n_1 \sim O(n_{L+1})$, even if $n_1< n_{L+1}$. However, for a given number $n_{L+1}$ of atoms trapped in the right edge of the harmonic potential array, one can only determine the number of atoms $n_1$ trapped in the left edge of the harmonic potential array that will ensure the emergence of right edge steady states by numerical evolution of the trapped ultracold atom open quantum system according to Eq. (\ref{masteqnharmonicarraytrappedultracoldatomoqs}). More importantly, the results show that if the system parameters are chosen asymmetrically such that the Rabi frequencies and detunings $\Omega_j$ and $\Delta_j$ of the Raman lasers at the right edge of the system is less than the Rabi frequencies and detunings $\Omega_j$ and $\Delta_j$ of the Raman lasers at the left edge of the system, the emergence of left edge steady states are favored by this open quantum system, considering that a greater range of values for $n_1$ will correspond to left edge steady states, even if $n_1 < n_{L+1}$. The results have a wide array of possible applications involving quantum transport of ultracold atoms, dissipative quantum state preparation, and quantum technologies which require localized unique many-body steady states. In particular, because of the existence of edge steady states for this trapped ultracold atom open quantum system, it may possibly be used as a quantum battery (\cite{alicki, campaoli, chang, qiao}) which employs highly efficient charging by using the non-Hermitian skin effect to transport energy for storage at one of its edges. As such, further study of this trapped ultracold atom open quantum system should yield multiple promising fundamental results and applications.

\section*{Acknowledgments}

The author would like to thank the Research Center for the Natural and Applied Sciences (RCNAS) and the College of Science of the University of Santo Tomas for financial, logistical and administrative support during the conduct of this research. The author would also like to thank S-H. Park and S. Verma for insightful discussions preceding this work, and to V. Villegas, M. I. Estayan, K. B. Simbulan, J. A. Albay, P. M. Albano, R. L. Tadle, M. Sabit, M. Bungihan, M. Tan, B. Belmonte, M. N. Confesor and E. Cuansing Jr. for clarificatory discussions during the course of this work. The author is also indebted to the anonymous reviewer of this paper, whose comments greatly improved and focused the content of this paper. As always, the author would like to thank ga T. B. O. Tejada for warm and supportive conversations that helped provide focus and grounding for this work. Palangga ta gid ken ay-ayaten ka la unay ga.

\bibliographystyle{elsarticle-num-names}

\bibliography{SteadyEdgeStatesUltracoldAtomsV7a}

\end{document}